\documentclass[fleqn,usenatbib]{mnras} 

\usepackage{newtxtext,newtxmath}

\usepackage[T1]{fontenc}
\usepackage{ae,aecompl}

\usepackage{graphicx}	
\usepackage{amsmath}	

\usepackage{amssymb}	
\usepackage{multicol}
\usepackage{bm}
\usepackage{booktabs}
\usepackage{cleveref}
\usepackage{natbib}
\usepackage{threeparttable}
\usepackage{adjustbox}

\newcommand{\secpoint}{\mbox{$''\mskip-7.6mu.\,$}}
\newcommand{\angstrom}{\mbox{\normalfont\AA}}

\newcommand{\muv}{$\rm M_{\rm UV}$}

\newcommand{\jwst}{{\it JWST}}
\newcommand{\hst}{{\it HST}}
\newcommand{\oh}{12+\log(\rm O/H)}
\newcommand{\co}{\log(\rm C/O)}
\newcommand{\no}{\log(\rm N/O)}
\newcommand{\oiiihb}{$\rm [OIII]+H\beta$}
\defcitealias{Bian2018}{B18}
\defcitealias{PP04}{PP04}

\title[{Rest-UV spectra at $z>6$}]{Deep rest-UV \jwst{}/NIRSpec spectroscopy of early galaxies: the demographics of CIV and N-emitters in the reionization era}

\author[M. W. Topping]{Michael W. Topping,$^{1}$\thanks{E-mail: michaeltopping@arizona.edu}
Daniel P. Stark$^{1}$,
Peter Senchyna$^{2}$,
Zuyi Chen$^{1}$,
Adi Zitrin$^{3}$,\newauthor
Ryan Endsley$^{4}$,
St\'ephane Charlot$^{5}$,
Lukas J. Furtak$^{3}$,
Michael V. Maseda$^{6}$,
Adele Plat$^{1}$,\newauthor
Renske Smit$^{7}$,
Ramesh Mainali$^{8}$,
Jacopo Chevallard$^{9}$,
Stephen Molyneux$^{7,10}$,\newauthor
Jane R. Rigby$^{11}$
\\
$^{1}$Department of Astronomy / Steward Observatory, University of Arizona, 933 N Cherry Ave, Tucson, AZ 85721\\
$^{2}$The Observatories of the Carnegie Institution for Science, 813 Santa Barbara Street, Pasadena, CA 91101, USA\\
$^{3}$Physics Department, Ben-Gurion University of the Negev, P.O. Box 653, Be’er-Sheva 84105, Israel\\
$^{4}$Department of Astronomy, University of Texas, Austin, TX 78712, USA\\
$^{5}$Sorbonne Universit\'e, CNRS, UMR 7095, Institut d’Astrophysique de Paris, 98 bis bd Arago, 75014 Paris, France\\
$^{6}$Department of Astronomy, University of Wisconsin-Madison, 475 N. Charter St., Madison, WI 53706, USA\\
$^{7}$Astrophysics Research Institute, Liverpool John Moores University, 146 Brownlow Hill, Liverpool L3 5RF, UK\\
$^{8}$Observational Cosmology Lab, Code 665, NASA Goddard Space Flight Center, 8800 Greenbelt Rd., Greenbelt, MD 20771, USA\\
$^{9}$Department of Physics, University of Oxford, Denys Wilkinson Building, Keble Road, Oxford OX1 3RH, UK\\
$^{10}$European Southern Observatory, Karl-Schwarzschild-str. 2, 85748 Garching, Germany\\
$^{11}$Astrophysics Science Division, Code 660, NASA Goddard Space
Flight Center, 8800 Greenbelt Rd., Greenbelt, MD, 20771, USA.
}

\begin{document}

\label{firstpage}
\pagerange{\pageref{firstpage}--XX}
\maketitle

\begin{abstract}
{\it JWST} has recently discovered a subset of reionization era galaxies with ionized gas that is metal poor in oxygen and carbon but heavily-enriched in nitrogen. This abundance pattern is almost never seen in lower redshift galaxies but  is commonly observed in globular cluster stars.  We have recently demonstrated that this peculiar abundance pattern appears in a compact ($\simeq 20$ pc) metal-poor galaxy undergoing a strong burst of star formation. This galaxy was originally selected based on strong CIV emission, indicating a hard radiation field rarely seen locally. In this paper, we present {\it JWST}/NIRSpec observations of another reionization-era galaxy known to power strong CIV emission, the $z=7.04$ gravitationally-lensed galaxy A1703-zd6. The emission line spectrum reveals this is a metal poor galaxy ($12+\log(\rm O/H) = 7.47\pm0.19$) dominated by a young stellar population ($1.6^{+0.5}_{-0.4}$ Myr) that powers a very hard ionizing spectrum (CIV EW = 19.4~\AA, He II EW = 2.2~\AA). The ISM is highly-enriched in nitrogen ($\log(\rm N/O)=-0.6$) with very high electron densities ($8-19\times10^4$ cm$^{-3}$) and extreme ionization conditions rarely seen at lower redshift.  We also find intense CIV emission (EW$\gtrsim20$~\AA) in two new $z\gtrsim 6$ metal poor galaxies.
To put these results in context, we search for UV line emission in a sample of 737 $z\gtrsim 4$ galaxies with NIRSpec spectra, establishing that 40(30)\% of systems with [OIII]+H$\beta$ EW $>2000$~\AA\ have NIV] (CIV) detections with EW$>5$~\AA ($>10$~\AA). 
These results suggest high N/O ratios and hard ionizing sources appear in a brief phase following a burst of star formation in compact high density stellar complexes.

\end{abstract}


%
%
%
%
\section{Introduction} 
\label{sec:intro}

Roughly a decade ago, deep near-infrared spectroscopy began to provide our first window on the rest-frame ultraviolet (UV) spectra of $z\gtrsim 6$ galaxies. This early glimpse revealed strong 
line emission from highly-ionized species of carbon and oxygen, with intensities significantly in excess of those at lower redshifts \citep{Stark2015b, Stark2015a, Stark2017, Laporte2017, Mainali2017, Hutchison2019, Topping2021}. This was  most clearly seen in two gravitationally-lensed galaxies showing strong nebular CIV$\lambda\lambda1548,1550$ emission, RXCJ2248-ID at $z=6.11$ \citep{Mainali2017,Schmidt2017}, and A1703-zd6 at $z=7.045$ \citep{Stark2015b}. The presence of triply ionized carbon requires a supply of 48 eV photons rarely seen in star forming galaxies at lower redshifts. Some argued this may be indicative of a population of very low metallicity stars \citep{Stark2015b,Mainali2017}, whereas others suggested the detections may point to the presence of narrow-line AGN in relatively faint reionization era galaxies \citep{Nakajima2018}. With ground-based instrumentation, our ability to 
delineate the nature of the ionizing sources in these systems was long stunted by limited wavelength coverage ($\lambda_{\rm{rest}} \lesssim$2000~\AA) imposed by our atmosphere.

The launch of {\it JWST} has fundamentally altered our ability to characterize the spectra of reionization era galaxies \citep[e.g.,][]{Schaerer2022b, ArellanoCordova2022, Katz2023, Trump2023, Curtis-lake2022, Fujimoto2023}. 
Early observations have confirmed that hard radiation fields are present in a number of early galaxy spectra (e.g., \citealt{Bunker2023,Tang2023,Castellano2024, Tacchella2023}). Perhaps most striking has been the  the $z=10.60$ galaxy GNz11 reported in \citet{Bunker2023}. In addition to strong high ionization emission lines, the rest-UV spectrum reveals strong NIV]$\lambda\lambda1483,1486$ and NIII]$\lambda1750$ emission, indicating gas that is heavily enriched in nitrogen relative to oxygen ($\log(\rm N/O)=-0.38$) and carbon ($\log(\rm N/C)\sim-0.5$) in spite of being very metal poor ($12+\log(\rm O/H)=7.82$; \citealt{Cameron2023}). This abundance pattern is extremely rare in lower redshift galaxies, suggesting a chemical evolution pathway that is perhaps unique to the reionization era, conceivably requiring a population of very (or super) massive stars ($\gtrsim 10^{2-3}$ 
M$_\odot$; \citealt{Vink2023, Charbonnel2023, Marques-Chaves2024, Nandal2024})  or punctuated star formation histories \citep{Kobayashi2024}.
Similar abundance ratios are seen in many globular cluster stars \citep[e.g.,][]{Gratton2004, Carretta2005, Senchyna2023}, potentially indicating a connection between the stars being formed in GNz11 and those formed in the second generation stellar populations of globular clusters \citep[e.g.,][]{Bastian2018}. The GNz11 spectrum also shows evidence for very high electron densities ($\gtrsim 10^{5}$ cm$^{-3}$), two orders of magnitude above those seen in star forming galaxies at lower redshift \citep[e.g.,][]{Sanders2016},  potentially reflecting conditions associated with a dense broad line region or nuclear star cluster surrounding an AGN \citep{Maiolino2023, Senchyna2023}. 

Whether the peculiar abundance pattern of GNz11 is  common in the reionization era is not known. It is also not clear whether the nitrogen-enhancement is somehow  linked to the presence of the hard ionizing sources and high electron densities that are also seen in GNz11.  But without a larger sample of such galaxies, it is impossible to adequately understand the physical mechanisms driving the origin of the abundance pattern. A natural place to begin investigation of these questions is the two $z\gtrsim 6$ galaxies with CIV detections from ground-based telescopes, both of which are extremely bright in the continuum (J=24.8--25.9) allowing high S/N views of the ionizing sources and gas abundance pattern. While it was clear prior to {\it JWST} that both systems had a population of hard ionizing sources, it was not known whether they also had ionized gas that was nitrogen enhanced or elevated in density similar to GNz11.
In {\it JWST} Cycle 1, we obtained NIRSpec R=1000 spectroscopy targeting both  the $z=6.11$ \citep{Mainali2017,Schmidt2017} and the $z=7.05$ \citep{Stark2015a} CIV emitters (program ID: 2478). The observations were designed to extend the spectroscopic coverage of these sources into the rest-optical for the first time, providing insight into the metal content  of early galaxies with hard radiation fields. We also obtained very deep rest-UV NIRSpec observations, providing complete coverage of the emission lines (and continuum) at $\simeq 1200$-2000~\AA. This far UV spectral window contains a suite of features sensitive to the origin of the radiation field, the electron density, and the level of nitrogen enhancement. 

In the first paper from this program, \citet{Topping2024} presented a detailed investigation of the NIRSpec  observations of RXCJ2248-ID, the $z=6.11$ CIV emitter. The spectrum confirmed the presence of a hard radiation field with strong emission from numerous highly-ionized species (i.e., CIV$\lambda\lambda1548,1550$, He II$\lambda1640$). The rest-optical spectrum was found to be similarly extreme, with very large [OIII]+H$\beta$ EW (3100~\AA) only seen in the upper 2\% of $z\simeq 6-9$ galaxies \citep[e.g.,][]{Endsley2023, Matthee2023}. In spite of the strength of [OIII] and the hydrogen recombination lines, the [OII] doublet is found to be very weak, suggesting an [OIII]/[OII] flux ratio (hereafter O32) of 184, well in excess of nearly any star forming galaxy previously studied at lower redshifts. 
\citet{Topping2024}  report that the various spectral features are consistent with being powered by a young (1.8 Myr) stellar population in a $\lesssim20$ pc region
with ionized gas that is metal poor  ($12+\log(\rm O/H)=7.43$) and extremely dense  (6--31$\times10^4$ cm$^{-3}$). The RXCJ2248-ID spectrum shows emission from NIII] and NIV], revealing a 
nitrogen-enhanced abundance pattern ($\log(\rm N/O)=-0.39$) that is similar to that found in GNz11. However in RXCJ2248-ID no spectral signatures were found that demand the presence of an AGN, suggesting that the high ionization lines may be likely to originate from a low metallicity population of massive stars. 

Based on these results, \citet{Topping2024} suggested that the nitrogen enhancement seen at $z\gtrsim 6$ may be produced in the dense clusters that form during a strong burst of star formation, a phase that many reionization era galaxies go through \citep{Furlanetto2022, Mirocha2023, Dome2023, Endsley2023, Strait2023}. In this framework, the hard ionizing agents would be produced during the burst, and the high electron densities would reflect the extreme dense gas conditions associated with the burst. 
If this picture applies universally, we should also see nitrogen enhancements and high electron densities in the $z=7.05$ galaxy Abell 1703-zd6 (hereafter A1703-zd6), the other CIV emitting primary target in our Cycle 1 program. In this paper, we present the  
deep NIRSpec observations of A1703-zd6.
Our MSA design includes many additional bright lensed $z\gtrsim 6$ galaxies, which we also describe in this paper. Our primary goals are twofold. First we seek to understand more about what drives hard radiation fields in a subset of early galaxies. Second, we seek to understand if the hard radiation fields are linked in some way to spectra like GNz11, with nitrogen enhancements and high electron density. To put these results in context, we will additionally present constraints on the fraction of early galaxies with CIV and NIV] lines in a large database of publicly-available NIRSpec spectra.

The structure of this paper is as follows.
Section~\ref{sec:data} provides an outline of the observations and data analysis, and discusses our measurement methods.
In Section~\ref{sec:results} we describe the spectra of each object in our sample, and we discuss the derived properties of the sample in Section~\ref{sec:sec4}.
Section~\ref{sec:disc} discusses the properties of these sources in the context of the reionization-era galaxy population. 
Finally, we provide a discussion in Section~\ref{sec:discussion}, followed by a summary and brief conclusions in Section~\ref{sec:summary}.
Throughout this paper we assume a cosmology with $\Omega_m = 0.3$, $\Omega_{\Lambda}=0.7$, $H_0=70 \textrm{km s}^{-1}\ \textrm{Mpc}^{-1}$, and adopt solar abundances from \citet[][i.e., $Z_{\odot}=0.014$, $12+\log(\rm O/H)_{\odot}=8.69$]{Asplund2009}. All magnitudes are provided using the AB system \citep{Oke1984}.

\begin{figure}
    \centering
     \includegraphics[width=1.0\linewidth]{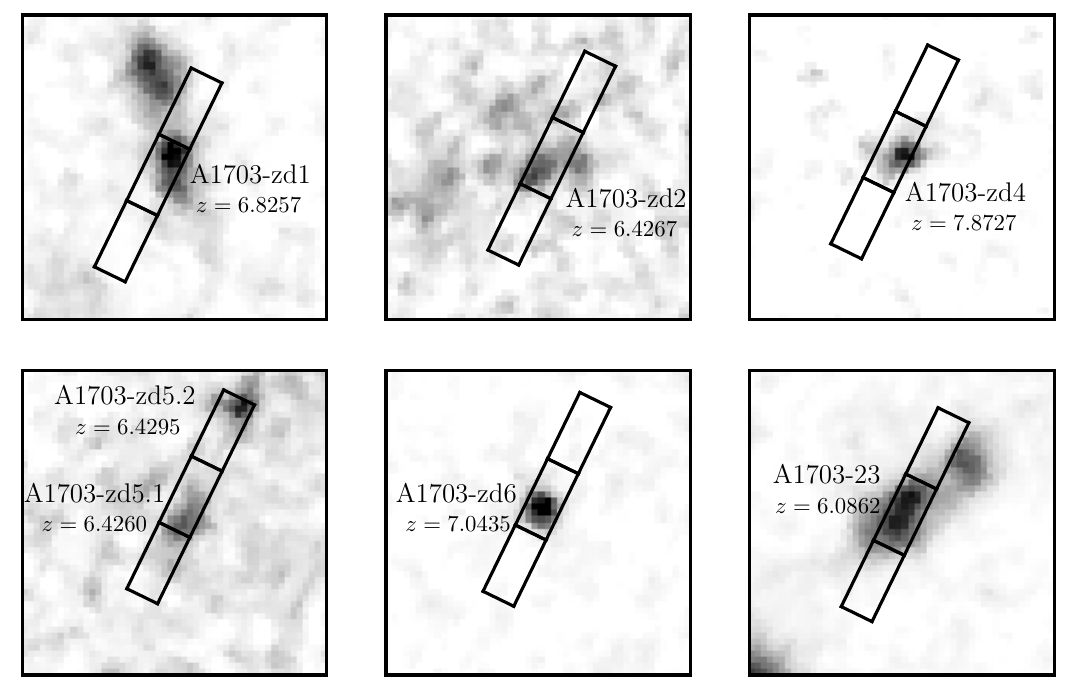}
     \caption{$J_{125}-$band images for each of the objects in our spectroscopic sample. Each postage stamp is $2^{\prime \prime}$ on a side and displayed in an orientation such that north is up and east is to the left. We overlay the position of the NIRSpec shutter in the postage stamp of each object. The listed redshifts were measured from the NIRSpec spectrum.} 
     \label{fig:images}
\end{figure}

%
%
%
%
\section{Data and Measurements}
In this section we provide an overview of the program from which our data derive (Section~\ref{sec:observations}), along with the methods used to measure emission and absorption lines from the spectra. In Section~\ref{sec:properties} we detail the methods used to infer properties of the stellar populations within each galaxy.

\label{sec:data}
\subsection{Observations and Reduction}
\label{sec:observations}
Spectroscopic data were obtained using \jwst{}/NIRSpec in Multi-Object Spectroscopy (MOS) mode targeting two lensing fields, Abell 1703 \citep[RA: 198.7636, Dec: +51.821,][]{Allen1992} and RXCJ2248-4431 \citep[RA: 342.17972, Dec: -44.5330;][]{Guzzo2009}, as part of Cycle 1 program ID 2478 (PI: Stark).
In the Abell 1703 field we obtained data using the G140M/F100LP and G395M/F290LP dispersing modes with total exposure times of 6653 sec. and 1576 sec., respectively, using the NRSIRS2RAPID readout mode.
Following the position angle (PA) assignment of the observations, the MSA pointing centers were defined based on the locations of our highest-priority targets.
For the observations in the Abell 1703 field presented in this analysis, we required that a microshutter slit was placed on A1703-zd6, which was previously spectroscopically confirmed to lie at $z=7.045$ based on Ly$\alpha$ \citep{Schenker2012}, CIV, and OIII] \citep{Stark2015b}.
We additionally required that the location of A1703-zd6 on the MSA would enable continuous wavelength coverage from Ly$\alpha$ to CIII]$\lambda\lambda1907,1909$ (9730\angstrom{}--15530\angstrom{} in the observed frame) in G140M, and [OII]$\lambda\lambda3727,3729$ to [OIII]$\lambda5007$ (29980\angstrom{}--40280\angstrom{} in the observed frame) in G395M.
The final MSA design was configured to fit these constraints and maximize the number of additional high-priority targets, which comprise a sample of $z$ and $i$-drop galaxies selected to lie at $z\simeq6$ (\citealt{Richard2009, Bradley2012, Smit2014}).
Throughout this paper we refer to the additional high-priority targets as A1703-zd2, A1703-zd5.1, A1703-zd5.2, A1703-zd4, and A1703-23. 
Each slit placed upon the targets was composed of three microshutter slitlets, and we obtained data using the standard three-nod pattern, with one exposure for each nod position (Figure~\ref{fig:images}).

\begin{table}
\caption{Sample of galaxies observed with NIRSpec in Abell 1703. H-band magnitudes are from \citet{Bradley2012}, except for A1703-23 for which we use values from \citet{Richard2009}.
}
\begin{adjustbox}{width=1.0\linewidth,center=1\linewidth}
\renewcommand{\arraystretch}{1.2}
\hskip-0.45cm\begin{tabular}{lcccrc}
\toprule
ID & RA & DEC & $\rm H_{160}$ &$\mu~$ & $z_{\rm spec}$ \\
\midrule
 A1703-zd1   &198.747582 &+51.833604 & 24.0 &  $3.1^{+0.3}_{-0.3}$ & 6.8257 \\
 A1703-zd2   &198.777070 &+51.821802 & 24.9 & $19.9^{+7.1}_{-7.0}$ & 6.4267 \\
 A1703-zd5.1 &198.782392 &+51.819382 & 25.7 & $33.6^{+14.0}_{-13.7}$ & 6.4260 \\
 A1703-zd5.2 &198.782206 &+51.819568 & 25.3 & $33.5^{+17.8}_{-16.9}$ & 6.4295 \\
 A1703-23    &198.755951 &+51.807251 & 23.8 &  $3.0^{+0.2}_{-0.2}$ & 6.0862 \\
 A1703-zd4   &198.780017 &+51.839995 & 25.4 &  $2.0^{+0.1}_{-0.1}$ & 7.8727 \\
 A1703-zd6   &198.754233 &+51.834663 & 25.9 &  $3.9^{+0.1}_{-0.2}$ & 7.0435 \\
\bottomrule
\end{tabular}
\end{adjustbox}
\label{tab:demographics}
\end{table}

\begin{table*}
\caption{Derived and measured quantities for galaxies observed with NIRSpec in Abell 1703. Reported stellar population properties are derived from \textsc{beagle} SED fitting, and have been corrected for the effects of gravitational lensing (see Section~\ref{sec:data}).\\
$^a$ CIV feature displays significant P-Cygni profile (see Section~\ref{sec:zd1}).
}
\begin{adjustbox}{width=1.0\textwidth,center=1\textwidth}
\renewcommand{\arraystretch}{1.2}
\hskip-0.45cm\begin{tabular}{lcccrccl}
\toprule
ID &  $z_{\rm spec}$ & $\log(\rm M/M_{\odot})$  & $\log(\rm SFR/M_{\odot}yr^{-1})$ & Age [Myr] &$\tau_{\rm V}$ & $\rm [OIII]+H\beta~EW~[\angstrom{}]$ & $\rm CIV~EW~[\angstrom{}]$ \\
\midrule
 A1703-zd1   & 6.8257 & $9.05\pm0.22$ & $1.48^{+0.36}_{-0.32}$ &  $37^{+27}_{-16}$    & $0.08^{+0.01}_{-0.03}$ &$1290\pm90$ & \quad$-^a$ \\
 A1703-zd2   & 6.4267 & $7.52\pm0.47$ & $0.25^{+0.59}_{-0.53}$ &  $17^{+18}_{-7}$     & $0.04^{+0.06}_{-0.03}$ & $1579\pm450$& $\quad31.6\pm7.0$ \\
 A1703-zd5.1 & 6.4260 & $7.74\pm0.77$ &$-0.34^{+1.06}_{-1.37}$ &  $70^{+117}_{-55}$   & $0.16^{+0.12}_{-0.13}$ & $443\pm162$ & $<4.3$ \\
 A1703-zd5.2 & 6.4295 & $8.17\pm0.71$ & $0.67^{+0.95}_{-1.06}$ &  $23^{+45}_{-17}$    & $0.02^{+0.08}_{-0.01}$ & $1137\pm210$& $\quad10.5\pm1.0$ \\
 A1703-23    & 6.0862 & $9.54\pm0.16$ & $1.40^{+0.29}_{-0.21}$ & $150^{+33}_{-61}$    & $0.17^{+0.04}_{-0.04}$ & $584\pm34$ & $<1.9\hfill$ \\
 A1703-zd4   & 7.8727 & $8.50\pm0.27$ & $1.06^{+0.52}_{-0.45}$ &  $23^{+27}_{-13}$    & $0.07^{+0.04}_{-0.04}$ & $1895\pm520$ & $<4.9\hfill$ \\
 A1703-zd6   & 7.0435 & $7.70\pm0.24$ & $1.49^{+0.29}_{-0.26}$ & $1.6^{+0.5}_{-0.4}$  & $0.01^{+0.01}_{-0.01}$ & $4116\pm690$& $\quad19.4\pm1.4$ \\
\bottomrule

\end{tabular}
\end{adjustbox}
\label{tab:properties}
\end{table*}

We reduced the spectroscopic dataset using a pipeline composed of standard STScI tools \citep{Bushouse2024}\footnote{\url{https://github.com/spacetelescope/jwst}} in addition to custom-made routines.
The initial processing of the raw uncalibrated (\texttt{*\_uncal.fits}) 
frames included a bias and dark current correction, as well as ramp fitting to the groups yielding the mean count rate in each pixel.
During this stage, jumps in consecutive groups resulting from cosmic rays are identified and masked, including significant events such as `snowballs' and `showers'.
The resulting full-frame 2D images were corrected for $1/f$ noise using the \textsc{nsclean} \citep{Rauscher2023} package, and the 2D spectrum traced out by each object was cut out from all of the full-frame exposures.
We applied a flat-field correction, applied a wavelength solution and absolute photometric calibration using the updated Calibration Reference Data System (CRDS) context to the cutout spectra for each exposure of every object.
We subtracted the background from each exposure following the nodding pattern described above.
The background-subtracted exposures were then interpolated onto a common wavelength grid.
Each nodded exposure of the targets were then background subtracted and combined including the rejection of pixels that have been flagged in previous reduction stages.
These 2D exposures for each object were then interpolated onto a common wavelength grid and combined to produce the final spectrum.
We fit the spatial profile of each 2D spectrum, and used the fitting information to perform an optimal extraction \citep{Horne1986} yielding the final 1D spectra.

We corrected the emission from each lensed image for the effects of magnification.
Our lens model builds on the model published by \citet{Zitrin2010}, which used the Light-Traces-Mass approach and included a set of 16 multiply imaged systems to constrain the model, most of which had spectroscopic redshifts and were previously known \citep[see][]{Richard2009, Limousin2008}. Here we rerun the model using a parametric lens modeling code, namely a revision of the \citet{Zitrin2015} parametric pipeline, which has been extensively used in recent JWST results \citep[e.g.,][]{Pascale2022, Pascale2024, Furtak2023, Meena2023}.  The model consists of two principal components: cluster galaxies and larger-scale dark matter halos. Cluster galaxies are modeled as double Pseudo Isothermal Ellipsoids \citep[dPIEs;][]{Eliadottir2007}, scaled by their luminosity following common scaling relations \citep[e.g.,][]{Jullo2007}. Two dark matter haloes are incorporated, each modeled as a pseudo isothermal elliptical mass distribution \citep[PIEMD; e.g.,][]{Keeton2001}.  One halo is initially centred on the brightest cluster galaxy (BCG) but is allowed to roam around the BCG center, with the exact position optimized in the minimization procedure. A second halo is set on a second bright cluster galaxy north-north-west of the BCG, at [13:15:03.1829, +51:49:56.629]. The mass of the BCG and this second bright galaxy, as well as a third, central galaxy at [13:15:07.8290, +51:48:57.955], are modelled independently of the scaling relation, allowing some more freedom. In the minimization the redshifts of the photometric-redshift systems, i.e. those lacking a spectroscopic redshift, are left free to be optimized as well. The minimization is performed in the source plane via a long MCMC and the final model has a (lens-plane) image reproduction rms of 0\secpoint8.

\begin{figure}
    \centering
     \includegraphics[width=1.0\linewidth]{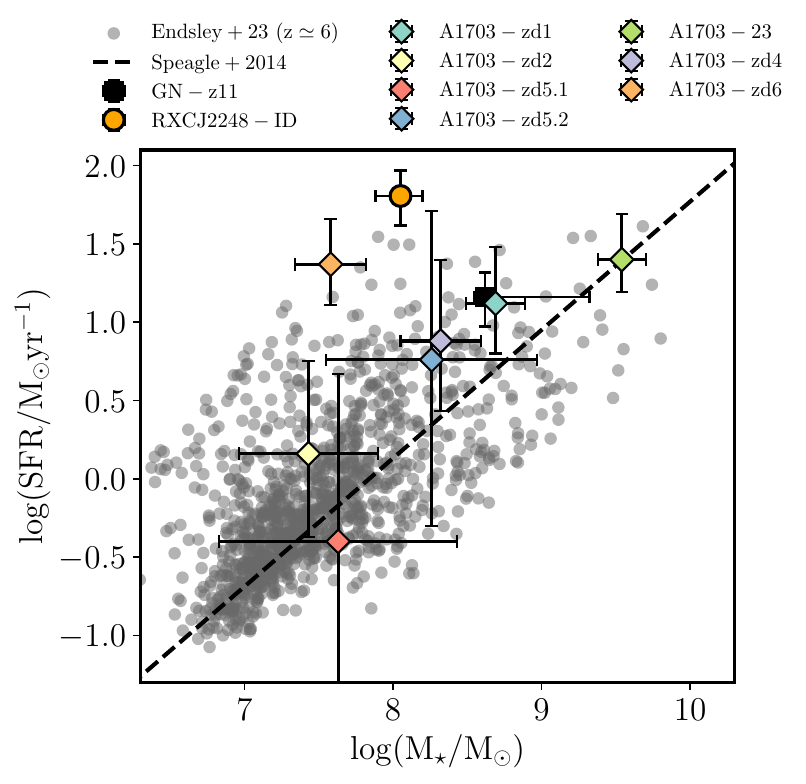}
     \caption{Star-formation rate versus stellar mass for the two primary targets discussed in this work (violet and orange points) inferred from the \textsc{beagle} SED fitting. We compare to the values inferred for GN-z11 \citep[black square;][]{Bunker2023} and RXCJ2248-ID \citep[orange circle;][]{Topping2024}, in addition to a population of star-forming galaxies at $z\sim6-9$ identified by \citet{Endsley2023} (grey circles). For reference, we display the `main sequence' derived by \citet{Speagle2014} calculated at $z=6$. Our primary targets lie at significantly elevated SFR relative to the main sequence and the population of early star-forming galaxies at fixed stellar mass. } 
     \label{fig:mainsequence}
\end{figure}

\begin{figure*}
    \centering
     \includegraphics[width=1.0\linewidth]{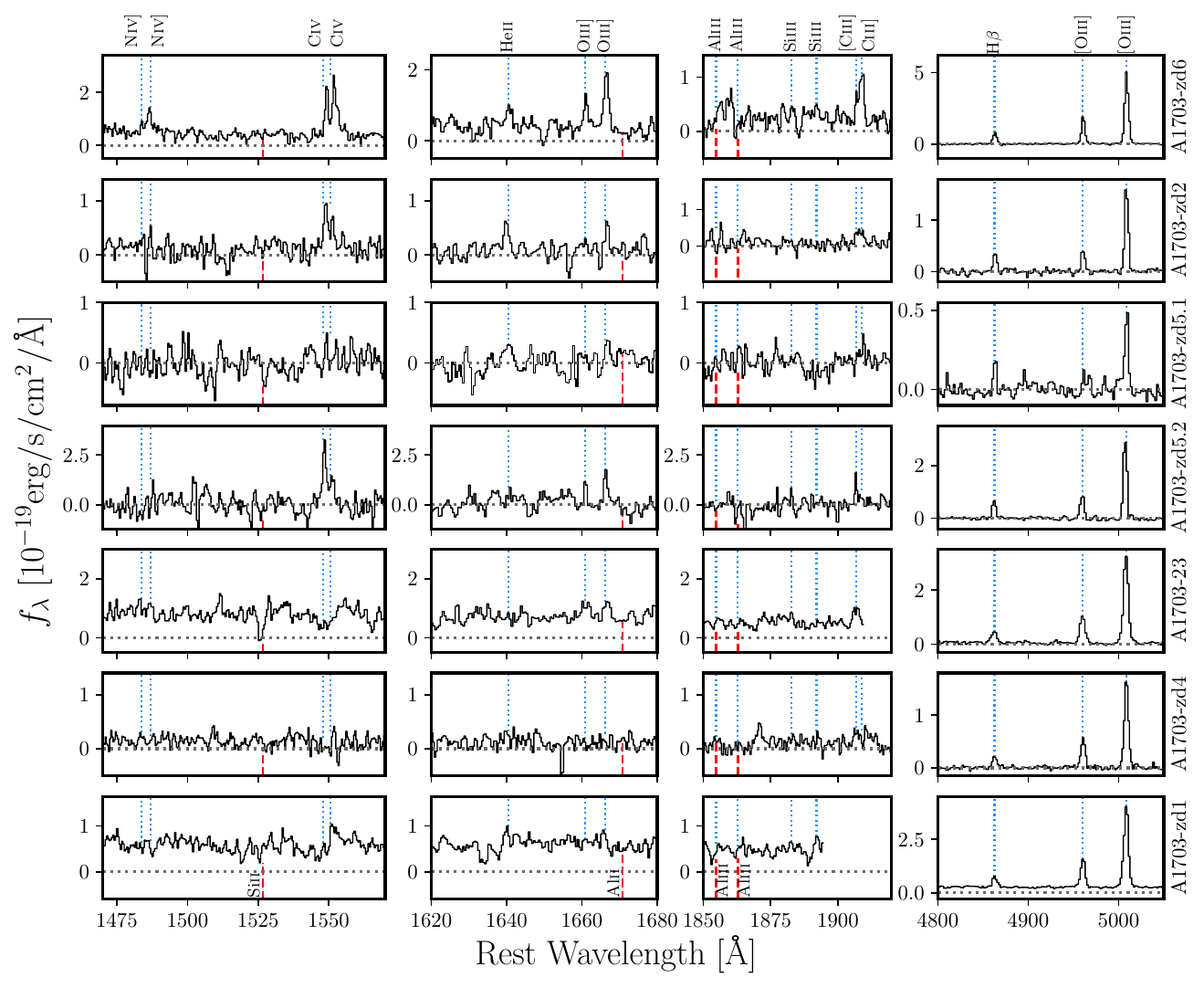}
     \caption{Presentation of the rest-frame UV and optical spectra for the sample of $z>6$ galaxies in Abell 1703. Each row corresponds to one object, which is labeled on the side of the right-most panel. Each column is a cutout of the rest-UV spectrum around the wavelengths where prominent features are detected.  Vacuum wavelengths are indicated by vertical blue dotted lines rising above each spectrum, which are also labeled at the top of the figure. Transitions typically seen in absorption are indicated by vertical red dashed lines below each spectrum. These absorption lines are labeled in the bottom row of the figure.} 
     \label{fig:UVspectrum}
\end{figure*}

Emission line fluxes, centroids, and widths were measured by fitting Gaussian profiles to the observed features.
For emission lines that are isolated, we fit a model of a single feature, while emission lines in close proximity in wavelength were fit simultaneously.
All of the emission line fitting was provided an initial estimate of the redshift, and its centroid was allowed to vary within one resolution element for emission lines that trace the systemic redshift, and within 500 km/s for resonance and absorption lines. In the case of multiple simultaneous line fits, the best-fit redshift was allowed to vary, however the spacing between the lines was fixed.
In some cases, such as for [OII]$\lambda\lambda3727,3729$ and NIII]$\lambda1750$, the lines are significantly blended, and line fluxes were measured by integrating over the combined profiles.
The continuum level for each line was measured by fitting a linear component to spectral windows on either side of the line.
We obtained uncertainties on each property (flux, centroid, FWHM, and EW) by first creating 5000 mock spectra that were derived by perturbing the observed spectrum by its corresponding error spectrum.
We repeated the above fitting procedure on each of the mock spectra, and assign uncertainties based on the inner 68th percentile of the recovered distributions of each property.

\subsection{Stellar Populations from SED Fitting}
\label{sec:properties}

Before analyzing the emission line properties, we investigate the basic physical properties of our sample using synthesized broadband photometry from the spectra, following a similar approach to that presented in \citet{Topping2024}.
For consistency with previous works, we derive photometry based on the \jwst{}/NIRCam filter throughputs and fit the resulting broadband fluxes using SED models.
The A1703 sample was observed using the G140M and G395M gratings, which completely encompass the \jwst{}/NIRCam F115W, F150W, F335M, F356W, F410M, and F444W filters. As such, we utilized synthesized photometry in only these filters for the SED fitting, with the redshift of each object fixed to its spectroscopic value.
Briefly, we fit the resulting photometry of both objects using BayEsian Analysis of GaLaxy sEds \citep[BEAGLE][]{Chevallard2016}.
These models implement updated models of \citet{Bruzual2003}, which include nebular emission using the prescription detailed in \citet{Gutkin2016}.
We assume a constant star-formation history (CSFH) with a minimum age of 1 Myr, and adopt a uniform prior on metallicity (ionization parameter) within the range $\log(Z/Z_{\odot})\in [-2.24, 0.3]$ ($\log(U)\in [-4, -1]$).
Finally, we assume an attenuation law following that of the SMC \citep{Pei1992}, and a fixed dust-to-metal ratio of $\xi_d=0.3$. Throughout this analysis, we adopt galaxy properties (e.g., stellar mass, SFR) that have been corrected for gravitational lensing using the magnification factors provided in Table~\ref{tab:properties}.
By assuming a CSFH, the inferred stellar population parameters are representative of the most recent episode of star formation, which will dominate the emission in the rest-frame UV and optical.
This modelling does not account for the possibility that each galaxy hosts a significant older stellar population, which could lead to a significant underestimate of properties such as the stellar mass and SFR \citep[e.g.,][]{Whitler2023a}. However, this possibility will not significantly impact any of our conclusions.

We present the model properties that reproduce the SEDs  in Table~\ref{tab:properties}.
The A1703 $z>6$ sample spans a factor of $\sim170$ stellar mass and $\sim70$ in SFR,  ranging from
$\log(\rm M/ M_{\odot})=7.70-9.54$ and $\log(\rm SFR/ M_{\odot}yr^{-1})=-0.34-1.49$, respectively.
In Figure~\ref{fig:mainsequence} we show the SFRs of our sample as a function of their stellar mass along with the larger sample of $z\simeq6-9$ LBGs from  \citet{Endsley2023}.
Our sample roughly extends over the full range of SFRs spanned by the broader $z\gtrsim6$ population at $\log(\rm M/ M_{\odot})\gtrsim8$.
When assuming a CSFH, we find the youngest galaxy has a CSFH age of just $1.6$ Myr, while the oldest system has an age of $150$ Myr. This suggests that our spectra are sampling galaxies dominated by a very young stellar population in addition to those that have may have a significant contribution from more evolved stars.
Additionally, the best-fit SEDs imply that our sample has a low average dust content, which is consistent with the typically blue UV slopes observed for galaxies during this epoch \citep[e.g.,][]{Topping2024b, Cullen2023}.
However, the emission of some objects in our sample (e.g., A1703-zd4, A1703-23) is best modelled by SEDs that include a non-zero amount of attenuation.
In the following section, we describe the spectrum of each galaxy in our sample.

\begin{figure}
    \centering
     \includegraphics[width=1.0\linewidth]{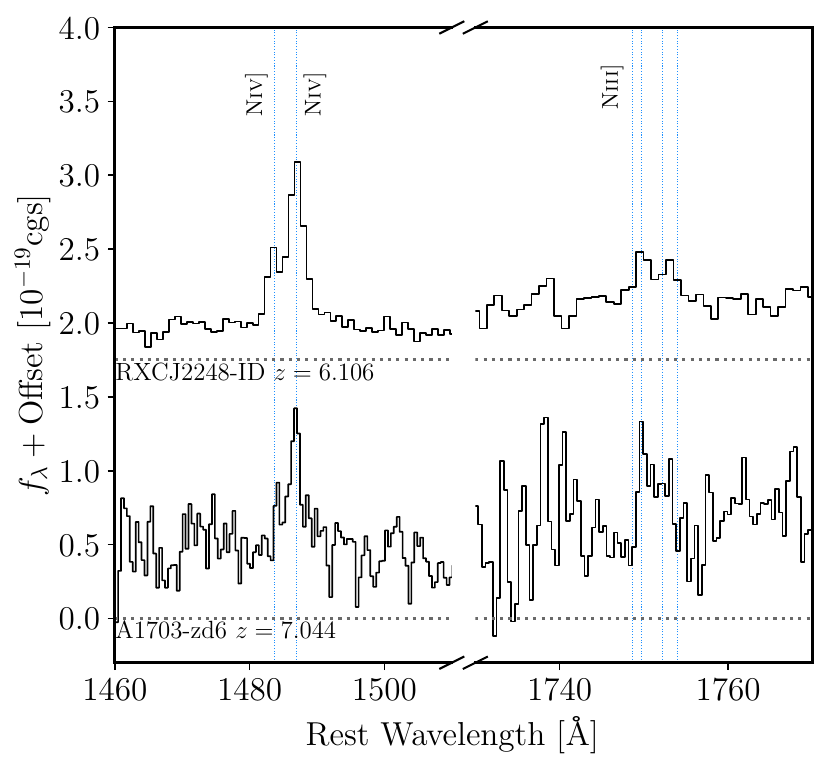}
     \caption{Comparison of  nitrogen lines detected in deep NIRSpec G140M spectra of RXCJ2248-ID \citep{Topping2024} and A1703-zd6 (this work) at $z=6.106$ and $z=7.044$, respectively. Both spectra have been shifted into the rest frame. The vacuum wavelengths of the NIV] doublet and NIII] multiplet members are indicated by vertical blue lines. The line of zero flux for each spectrum is shown as a horizontal dotted line.} 
     \label{fig:nitrogenlines}
\end{figure}

\begin{figure*}
    \centering
     \includegraphics[width=1.0\linewidth]{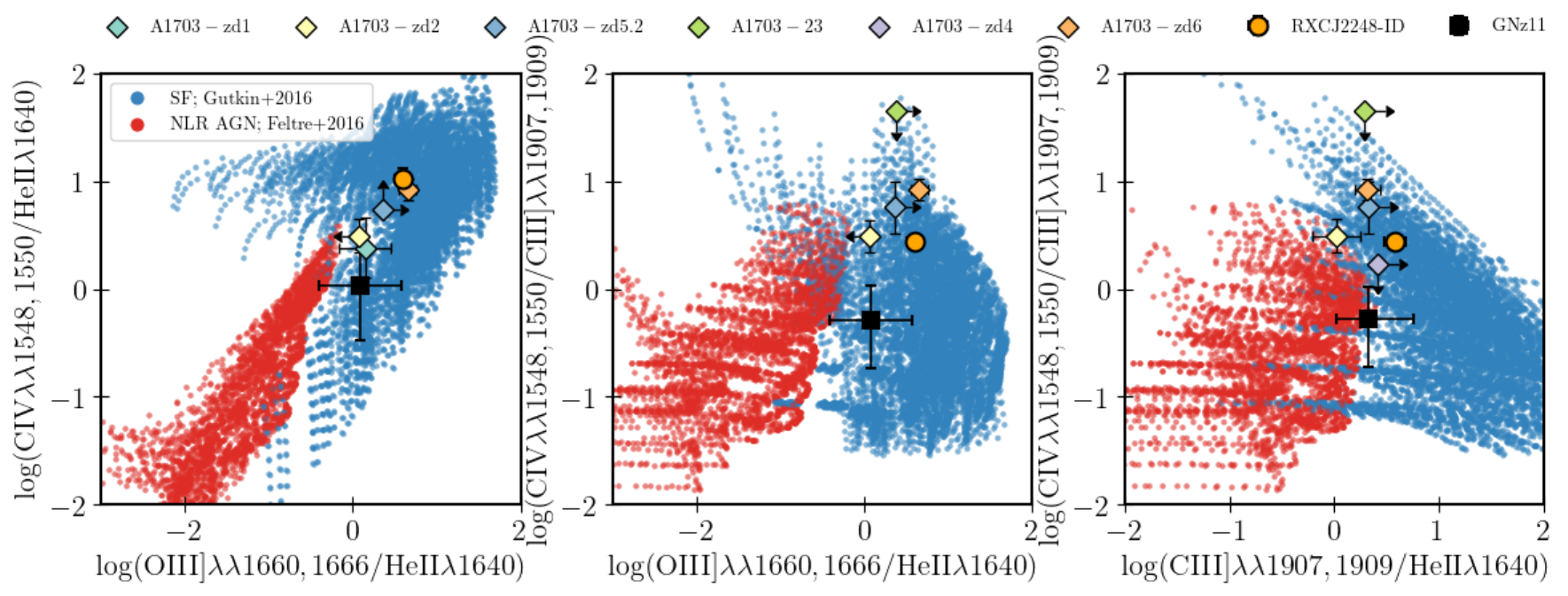}
     \caption{Rest-UV line ratios compared to photoionization models of HII regions within star-forming galaxies (blue points; \citealt{Gutkin2016}), and models where a narrow line AGN dominates the ionizing radiation field (red points; \citealt{Feltre2016}). Galaxies in our A1703 spectroscopic sample are displayed as the colored diamonds. For comparison, we plot the rest-UV line ratios of RXCJ2248-ID \citep{Topping2024} and GNz11 \citep{Bunker2023} as the orange circle and black square, respectively. Line ratios indicated as limits are provided at the $3\sigma$ level.} 
     \label{fig:UVlineratios}
\end{figure*}

%
%
%
%
\section{Spectroscopic Properties}
\label{sec:results}

In this section we describe the spectroscopic properties of each object in our sample. Our focus  will be on emission and absorption line measurements (i.e., EWs, line ratios), building an empirical physical picture for each individual galaxy. We will then build on and synthesize these results in \S4, calculating gas-phase properties (i.e., electron densities, abundance patterns) using the empirical measurements reported in this section.

Figure~\ref{fig:UVspectrum} presents  spectra of the A1703 galaxies in the rest-UV and rest-optical, at wavelengths where lines are commonly detected.
Line fluxes and EWs measured following the methods outlined in Section~\ref{sec:observations} are provided in Table~\ref{tab:UVlines} for the G140M spectra, and Table~\ref{tab:opticallines} for the G395M spectra. Multiple emission lines are detected in each galaxy. We also detect absorption lines in the two  brightest continuum sources (A1703-zd1 and A1703-23).

\subsection{A1703-zd6}
\label{sec:zd6}
A1703-zd6 was  identified  as a bright ($\rm H=25.9$) $z$-band dropout, moderately lensed ($\mu=3.9^{+0.1}_{-0.2}$) by the foreground Abell 1703 cluster \citep{Bradley2012}. 
\citet{Schenker2012} confirmed the redshift ($z=7.045$) via detection of 
strong Ly$\alpha$ emission ($\rm{EW}=65\pm12$\angstrom{}) with Keck/NIRSpec. Motivated by the discovery of high ionization UV lines in metal poor $z\simeq 2-3$ galaxies (e.g. \citealt{Erb2010, Christensen2012, Stark2014}), a Keck/MOSFIRE J-band spectrum was subsequently obtained covering rest-frame 1433-1680~\AA. The observations revealed very strong nebular CIV emission at 1.2458 $\mu$m ($\lambda$1548) and 1.2474$\mu$m ($\lambda$1550), indicating the presence of a hard ionizing spectrum rarely seen in lower redshift star forming galaxies \citep{Stark2015b}. The  R=1000 NIRSpec spectrum presented in this paper covers the rest-UV (9700-18900~\AA) and 
rest-optical (28700-52700~\AA), providing a new window on the ionizing agents and ionized gas in A1703-zd6. Below we will discuss the lines measured in the spectrum and compare to NIRSpec observations of RXCJ2248-ID  \citep{Topping2024}, the other $z\gtrsim 6$ CIV emitting galaxy known prior to {\it JWST}.

The rest-optical spectrum (Figure~\ref{fig:UVspectrum}) 
shows very prominent Balmer lines (H$\beta$, H$\gamma$, H$\delta$, H$\epsilon$) and strong [OIII] nebular ($\lambda\lambda$4959,5007) and auroral ($\lambda$4363)  emission, yielding a systemic redshift of $z=7.0435$.
The H$\beta$ EW ($423.8$~\AA) is not only at the high end of values seen at 
$z\gtrsim 6$ \citep[e.g.][]{Roberts-Borsani2024} but is also larger than that found in the vast majority of extreme emission line galaxies found at lower redshifts (e.g., \citealt{Maseda2018, Tang2019}). The [OIII]$\lambda\lambda4959,5007$ EW is similarly elevated ($3692$~\AA), matched by a small number of known $z\gtrsim 6$ galaxies in existing {\it JWST} surveys \citep{Tang2023, llerena2023, Boyett2024}. Indeed, the [OIII]+H$\beta$ EW ($4116$~\AA) is in the upper 0.5\% of values seen in the reionization-era population \citep{Endsley2023, Matthee2023}.  The stellar population synthesis models described in \S2.2  reproduce the strong optical lines, provided the age of the stellar population dominating the light is very young ($1.6$ Myr) and the ionization parameter very large ($\log(U) = -1.7\pm0.3$). In this picture, 
the strong CIV emission is powered by a hard radiation field associated with an extremely young stellar population formed in a recent burst of star formation (Figure~\ref{fig:mainsequence}).

In contrast to the strong nebular emission described above,  the A1703-zd6 spectrum does not reveal detection of the [OII] doublet. The lower ionization state gas (if present) appears to not be intense in its emission. The ratio of [OIII] and [OII] fluxes in A1703-zd6  is found to be $>$36 ($3\sigma$), well above the typical values at $z\gtrsim 6$ (median $\rm O32 \simeq 10$ in existing spectroscopic samples; \citealt{Tang2023, Cameron2023, Sanders2023}). Such a  large O32 value is  qualitatively consistent with the well-known relationship 
between [OIII] EW and O32 \citep{Tang2019,Sanders2020}, whereby extreme ionization conditions (large O32) tends to be found in galaxies with very large [OIII] EW. The large flux ratio of [Ne III] and [OII] ($>3.8$; hereafter Ne3O2) presents more evidence for extreme ionization conditions.
The lower limits on the O32 and Ne3O2  ratios in A1703-zd6 are consistent with those measured in RXCJ2248-ID \citep{Topping2024}. In RXCJ2248-ID, it was shown that the extremely large O32 value ($184$) was partially driven by elevated electron densities ($\sim 10^5$ cm$^{-3}$) which result in collisional de-excitation of [OII]. At these high densities, we also expect  anomalously strong He I$\lambda$5876 emission, owing to elevated collisional excitation.  A1703-zd6 presents a strong He I $\lambda$5876 line, with a  flux 26\% that of H$\beta$.  We will demonstrate that the densities of A1703-zd6 are indeed very large in \S~\ref{sec:density}.

The UV continuum slope of the synthesized SED of A1703-zd6 is very blue ($\beta=-2.6$), comparable to galaxies with similar luminosities at $z\gtrsim 6$ \citep{Topping2024b, Cullen2023, Morales2024}. Such blue colors suggest very little reddening from dust, as evidenced by the low optical depth implied by the BEAGLE population synthesis models fits ($\tau_V=0.01\pm0.01$) described in \S2.2 (see Table~\ref{tab:properties}). The 
Balmer line flux ratios present a similar picture, with values ($\rm H\gamma/H\beta=0.57\pm0.17$) consistent with those expected from case B recombination (once the appropriate electron temperature and densities are adopted; see \S~\ref{sec:oxygen}). The optical line ratios also indicate the ionized gas is metal poor. We will show in \S4.2 that detection of auroral lines suggest a gas-phase metallicity of $\rm 12+\log(\rm O/H)=7.47$, corresponding to 6\% Z$_\odot$. 

The NIRSpec UV spectrum of A1703-zd6 (top panel in Figure~\ref{fig:UVspectrum}) reveals clear detection of the CIV emission seen with Keck, along with several other intense 
emission features (NIV], He II, OIII], NIII], CIII]). The  NIRSpec R=1000 data reveal more information than was possible with Keck. 
Both components of the CIV doublet are detected and resolved with NIRSpec, with an integrated strength (EW=19.4~\AA) well in excess of what is seen in nearly all lower redshift star forming galaxies \citep{Senchyna2019,Izotov2024}. As CIV is a resonant line, we expect the lines to be offset from the systemic redshift. We measure centroids of 1.2461$\mu$m ($\lambda$1548) and 1.2484$\mu$m ($\lambda$1550), both indicating the line is shifted by 240 km s$^{-1}$ from systemic. The line profile is distinctly different from RXCJ2248-ID, where the blue component was shown to be significantly attenuated by scattering effects \citep{Topping2024}. The ratio of the two components in A1703-zd6 is near unity, with a doublet ratio of $f_{\rm CIV\lambda1548}/f_{\rm CIV\lambda1550}=0.8\pm0.2$ (Table~\ref{tab:CIV}).  This is well below the intrinsic ratio (2), as expected when the blue component is preferentially scattered by outflowing gas.  Nonetheless, the mere presence of the blue component suggests A1703-zd6 faces less attenuation to the transfer of CIV photons than RXCJ2248-ID.

The detection of the  strong nitrogen line emission in A1703-zd6 follows 
the recently reported detection in RXCJ2248-ID \citep[][see also \citealt{Bunker2023, Isobe2023, Castellano2024}]{Topping2024}, providing more evidence for a connection between hard ionizing sources and nitrogen enhancement. 
Both components of the NIV] doublet are detected in A1703-zd6, revealing a total EW of $5.0\pm0.7\angstrom{}$. 
The NIV] line profile is very similar to that seen in RXCJ2248-ID, with the red component 2.1$\times$ stronger than the blue component (Figure~\ref{fig:nitrogenlines}). We will come back to discuss implications for the electron density in \S~\ref{sec:density}. The  CIV/NIV] flux ratio in A1703-zd6 (3.1) is  slightly greater than  that seen in RXCJ2248-ID (2.1). We additionally detect emission from the NIII] quintuplet (S/N=$3.2\sigma$) with an integrated EW of $3.1\pm1.0\angstrom{}$. We will use these detections to constrain the N/O ratio of A1703-zd6 in \S~\ref{sec:nabundance}.

The detection of the auroral OIII]$\lambda\lambda$1660,1666 doublet (EW=11.6~\AA) enables an additional constraint on the electron temperature which we will discuss in \S~\ref{sec:oxygen}. The [CIII], CIII]$\lambda\lambda$1907,1909 doublet is resolved with a total line strength (EW=8.8~\AA) that is comparable to what is often seen in metal poor star forming galaxies \citep{Stark2017,Hutchison2019, Tang2021, Tang2023}.  The CIV/CIII] flux ratio is well in excess of unity (3.3), suggesting a very hard spectrum. The  He II$\lambda1640$ line is also detected at $6.2\sigma$, indicating a very large EW ($6.8$\angstrom{}) rare among nearby star forming galaxies \citep{Berg2016, Berg2019, Senchyna2017}. Nevertheless, the UV line ratios appear consistent with expectations for stellar photoionization (Figure~\ref{fig:UVlineratios}). We will discuss these comparisons in more detail by Plat et al. (in prep), but here we mostly note that the hard spectrum is not clearly indicative of an underlying AGN in A1703-zd6.

\begin{figure}
    \centering
     \includegraphics[width=1.0\linewidth]{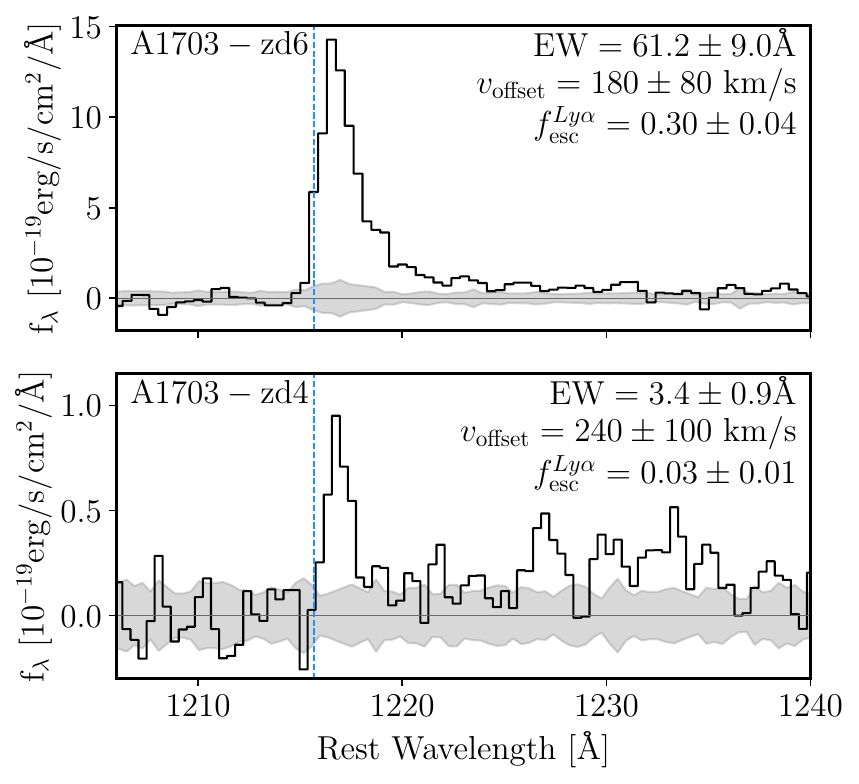}
     \caption{Ly$\alpha$ profiles of A1703-zd6 (top; \S\ref{sec:zd6}) and A1703-zd4 (bottom; \S\ref{sec:zd4}). The vacuum wavelength of Ly$\alpha$ is displayed as a vertical blue line, and the Ly$\alpha$ EW, systemic velocity offset, and escape fraction for each source are presented at the top right.}
     \label{fig:lya}
\end{figure}

The deep NIRSpec spectrum confirms the strong Ly$\alpha$ line detected from the ground by \citet{Schenker2012}.
The $R\sim1000$ grating combined with the sensitivity of NIRSpec yields a precise systemic redshift and
enables the shape of the Ly$\alpha$ emission line to be accurately characterized. The Ly$\alpha$ detection in A1703-zd6 is presented in Figure~\ref{fig:lya}.
The Ly$\alpha$ peak velocity offset measured from the $R\sim1000$ spectrum is $180\pm80$ km/s. 
This offset is consistent with the median peak velocities observed for strong line emitters at $z>6.5$, reflecting the attenuation at line center from the increasingly neutral IGM at these redshifts. The Ly$\alpha$ EW ($61.2\pm9.0\angstrom{}$) is larger what has been seen for a majority of moderately-luminous $z>7$ galaxies characterized by NIRSpec \citep[e.g.,][]{Tang2023, Saxena2023}. 
We derive the Ly$\alpha$ escape fraction for A1703-zd6 from the intrinsic Ly$\alpha$ flux inferred as $f^{int}_{\rm Ly\alpha}/f_{\rm H\beta}=24.6$, which is appropriate for the density and temperature and assuming case B recombination. 
The resulting escape fraction of $f_{\rm esc}^{\rm Ly\alpha}=0.30\pm0.04$ is comparable to typical values at $z\sim5-6$ \citep{Chen2024} implying significant transmission through the IGM.
This value is well above typical escape fractions at $z\simeq 7$ \citep[$f_{\rm esc}^{\rm Ly\alpha}\simeq0.05$][]{Jones2024, Nakane2023, Chen2024}, consistent with expectations for galaxies situated within ionized sightlines.

\begin{figure}
    \centering
     \includegraphics[width=1.0\linewidth]{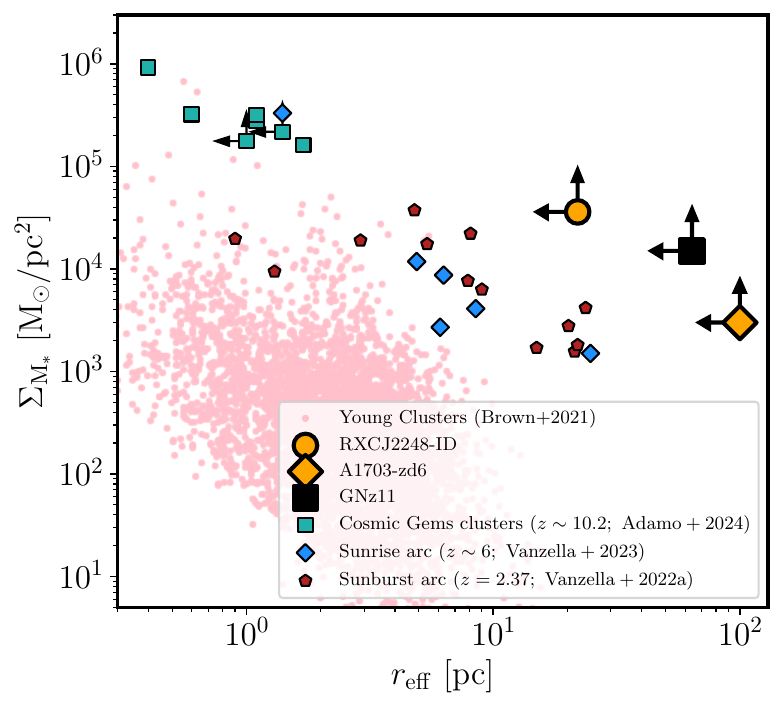}
     \caption{Stellar mass surface density versus effective radius. We display galaxies at $z>6$ including A1703-zd6 (orange diamond, This Work), RXCJ2248-ID (orange circle, \citealt{Topping2024}), and GN-z11 (black square, \citealt{Bunker2023}), in addition to individual star-forming clumps from the Cosmic Gems \citep{Adamo2024}, Sunrise arc \citep{Vanzella2023}, and the Sunburst arc \citep{Vanzella2022a}. The pink points provide measurements of individual young star clusters in the local Universe \citep{Brown2021}. Galaxies with confirmed detections in NIV] are outlined with a dotted circle. These NIV]-emitting systems have significantly higher $\Sigma_{\rm M_*}$ at fixed radius compared to local young clusters, and are even elevated compared to individual star-forming clusters at high redshift.}
     \label{fig:sizes}
\end{figure}

Finally as has been noted previously \citep{Bradley2012}, the \hst{} imaging of A1703-zd6 demonstrates that it is extremely compact (Figure~\ref{fig:images}).
It is unresolved in the image plane in the $J_{125}$ band which has a PSF FWHM of 0\secpoint12 \citep[e.g.,][]{Koekemoer2011}. This is consistent with the angular size of the rest-UV continuum trace in the 2D NIRSpec spectrum.
These constraints yield an effective radius of $<120$ pc in the source plane.
We constrain the stellar-mass surface density ($\Sigma_{\rm M_*}$) from this radius and the stellar mass from the synthesized SED, which demonstrates that A1703-zd6 has a very high concentration of stellar mass ($>3\times10^3\rm M_{\odot}/pc^2$).
In Figure~\ref{fig:sizes} we compare the mass density of A1703-zd6 to other galaxies and star clusters at high redshift \citep{Adamo2024, Vanzella2023, Vanzella2022a} and the local Universe \citep{Brown2021}.
A1703-zd6 shows similar characteristics to other high-redshift galaxies sharing several spectroscopic properties \citep[i.e., GNz11 and RXCJ2248-ID][]{Bunker2023, Topping2024}.
Specifically, all of these systems host very young stellar populations, and display high-ionization UV lines including CIV, NIV], and NIII]
potentially suggesting that high stellar density (and in particular a large density of very young O stars) may be linked to these properties.

\begin{figure*}
    \centering
     \includegraphics[width=1.0\linewidth]{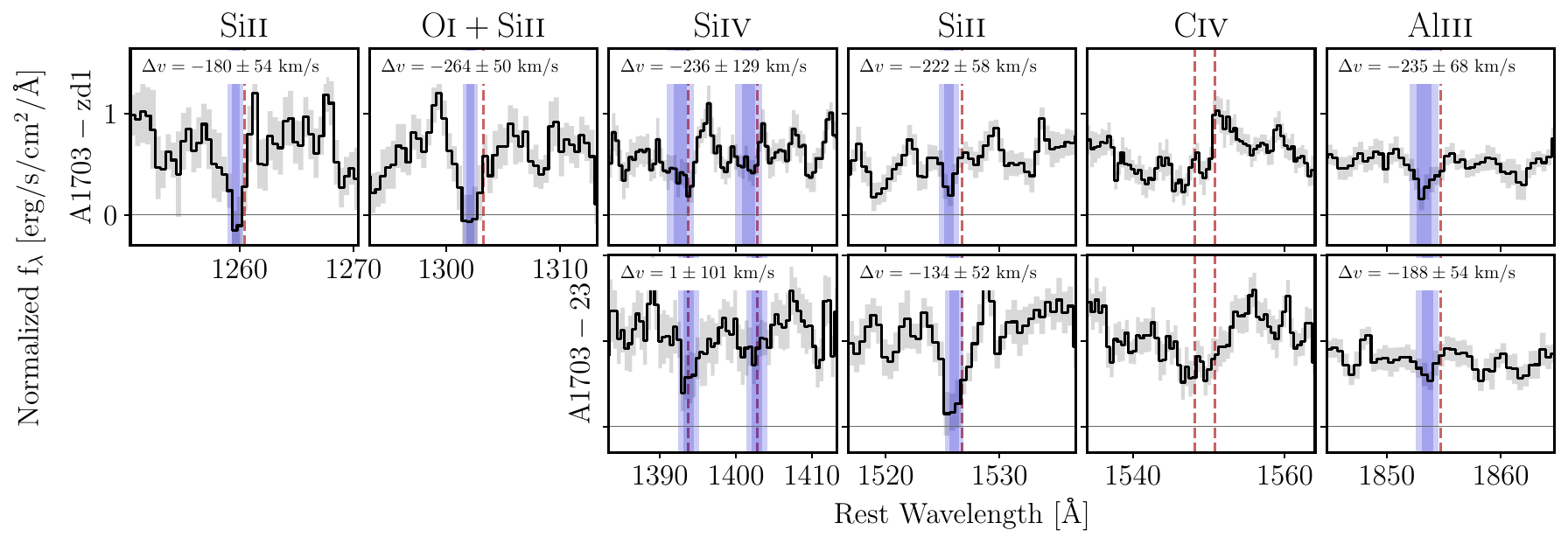}
     \caption{Rest-frame UV absorption lines in the spectrum of A1703-zd1(top row) and A1703-23(bottom row). In each panel, the systemic wavelength of the absorption feature is indicated by a red dashed line. The offset of the absorption feature is then listed at the top of each panel. The error spectra are displayed as the grey shaded regions. The vertical dark and light shaded regions illustrate the $1.5\sigma$ and $3\sigma$ uncertainties for the velocity offset of the line center from systemic.}
     \label{fig:absorption}
\end{figure*}

\subsection{A1703-zd1}
\label{sec:zd1}
A1703-zd1 is one of the brightest (H=24.0)  $z$-band dropouts known \citep{Bradley2012}, with moderate magnification ($\mu=3.1^{+0.3}_{-0.3}$) and a blue {\it Spitzer}/IRAC [3.6]-[4.5] color suggesting a photometric redshift of $z_{\rm phot}\simeq6.6-7.0$ \citep{Smit2014}. In spite of its brightness, ground-based rest-UV spectroscopy failed to confirm its redshift, likely indicating A1703-zd1 is characterized by weak Ly$\alpha$ \citep{Schenker2012} and fairly weak UV metal lines \citep{Stark2015b,Mainali2018}. The redshift of A1703-zd1 was eventually confirmed ($z=6.827$) in the far-IR, via detection of [CII]$\lambda$158$\mu$m with the Northern Extended Millimeter Array (NOEMA) \citep{Molyneux2022}. However no far-IR dust continuum was detected with NOEMA (see also \citealt{Schaerer2015}).

The NIRSpec observations of A1703-zd1 reveal a suite of strong rest-optical emission lines ([OII], [NeIII], [OIII], H$\alpha$, H$\beta$, H$\gamma$, H$\delta$), indicating a systemic redshift ($z=6.8257$) consistent with that derived from NOEMA.  We do not detect [OIII]$\lambda$4363 in this spectrum. The combined [OIII]+H$\beta$ EW is very large (1290~\AA), placing A1703-zd1 among the upper 25\% of 
the [OIII]+H$\beta$ EW distribution at $z\simeq 6-9$ \citep[e.g.,][]{Endsley2023, Matthee2023}. 
Reproducing such large EW optical lines with the synthesized SED requires relatively 
young stellar populations (37 Myr). The best fit suggests a moderate stellar mass ($\log(\rm M/M_{\odot})=9.05\pm0.22$), one of the largest in our sample. At such young stellar population ages, ionization-sensitive ratios are usually somewhat elevated \citep{Tang2019}, and indeed we find large values of O32 ($8.0^{+0.8}_{-1.0}$) and Ne3O2 ($0.62^{+0.12}_{-0.13}$) in A1703-zd1. The Balmer-line ratios are consistent with the intrinsic values 
(H$\alpha$/H$\beta=2.6^{+0.2}_{-0.3}$, H$\gamma$/H$\beta=0.48^{+0.07}_{-0.07}$; Table~\ref{tab:optlineratios}), implying minimal dust attenuation, in agreement with the upper limits on the FIR dust continuum \citep{Schaerer2015, Molyneux2022}. 

The rest-frame UV spectrum reveals confident detection of 
the continuum (S/N=5 per resolution element) without strong line emission, consistent with expectations from earlier ground-based observations. The non-detections place stringent upper limits on the rest-UV 
emission line fluxes (see Table~\ref{tab:UVlines}). At the spectroscopic redshift of A1703-zd1, the G140M spectrum covers 1250--2400\angstrom{} in the rest frame. In addition, the detector gap falls at rest-frame wavelengths of 1895--2020\angstrom{} negating our ability to constrain CIII].
Within the observed wavelength coverage, the typical EW limits ($3\sigma$) on the UV lines range from 0.7--2.2\angstrom{}, indicating much weaker emission than the relatively intense optical lines. 
One  example is OIII]$\lambda\lambda1660,1666$. The implied total EW limit ($<$2.7~\AA\ at 3$\sigma$) indicates weaker UV lines than are often seen in galaxies with such large [OIII]+H$\beta$ EWs \citep{Du2020, Tang2021}. This emphasizes that not all strong optical line emitters should be expected to power strong UV lines. The presence of weak UV lines in galaxies with very young stellar populations is often a signature of relatively elevated gas-phase metallicities, given the sensitivity of the UV transitions to electron temperature.  Strong line calibrations  (see \S4.2) provide some support for this picture, indicating a gas-phase metallicity at the upper end of this sample (12+log(O/H)=7.88). But detection of auroral lines in the UV or optical will be required to confirm this definitively.

The well-detected UV continuum in the A1703-zd1 spectrum is ideal for recovery of absorption lines. We detect both low-ionization (LIS) and high-ionization (HIS) species.
Figure~\ref{fig:absorption} presents a zoomed in view of each detected absorption feature in our spectrum, including SiII$\lambda1260$, OI+SiII$\lambda1303$, SiIV$\lambda\lambda1393,1402$, CIV$\lambda\lambda1548,1550$, and AlIII$\lambda\lambda1854,1862$. We discuss the  CIV profile in more detail later in this section, as it also includes a stellar wind component. 
We measure EWs of the LIS features in absorption that range from $-0.5$ to $-2.0$\angstrom{}, and have an average (blueshifted) velocity offset of $-220\pm80~\rm km/s$ relative to systemic. The HIS SiIV$\lambda\lambda1393,1402$ line has a lower total EW than each of the LIS lines. However, it is blue-shifted from systemic by a similar velocity ($-238\pm129$ km/s).
These features signify the presence of outflows in both phases, similar to those commonly seen at $z\sim1-5$ \citep[e.g.,][]{Shapley2003, Steidel2010, Jones2012,Pahl2020, SaldanaLopez2023}. The spectrum highlights that moderately massive galaxies at $z\simeq 7$ do present significant absorption lines with EWs approaching those seen in $z\simeq 3-4$ composites.

While most of the UV absorption lines are relatively narrow with characteristic velocity offsets of $\sim 200$~km/s, a much broader kinematic profile is evident in the CIV complex.
The absorption in CIV for this object extends to $-2000$~km/s, and is accompanied by prominent redshifted emission.
Rather than interstellar gas, a P-Cygni profile of this shape and character is suggestive of formation in the hot stellar winds of massive OB stars.
Assuming this profile is indeed stellar, we compare it to models for simple stellar populations (SSPs) from the Charlot \& Bruzual population synthesis grids \citep[see Section~\ref{sec:properties}; and description in][]{Gutkin2016,Plat2019,Senchyna2022}. 
We compare against SSP ages (2-4 Myr) that span a range of H$\beta$ EWs similar to that measured for A1703-zd1, and for which these stellar wind contributions are most prominent at a given metallicity.
These models include both stellar light and the contribution from nebular continuum in the FUV.
Since CIV is the only line with a significant potential stellar contribution at the observed SNR (consistent with model expectations), rather than detailed fitting we adopt a simple approach of qualitative comparison with the observed profile, normalized to the median flux level at 1570--1580~\AA{}.

\begin{figure}
    \centering
     \includegraphics[width=1.0\linewidth]{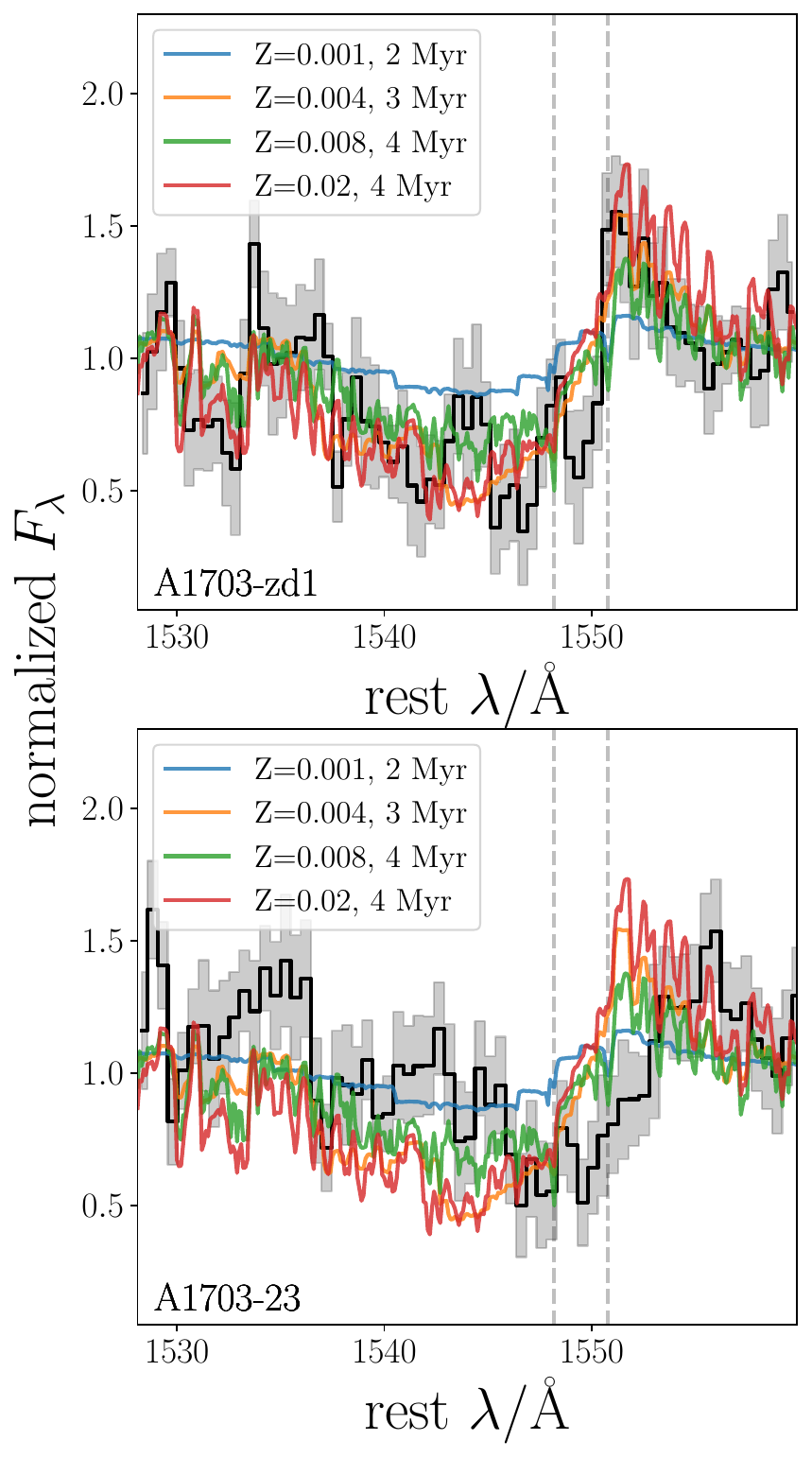}
     \caption{Prominent broad absorption in the resonant CIV 1548,1550 doublet is detected in A1703-zd1 and A1703-23 (spectra in greyscale).
     Comparison to stellar wind profiles for SSP models from Charlot \& Bruzual (colored curves) reveal reasonable agreement with the P-Cygni profile in A1703-zd1, provided a fairly high model stellar metallicity $Z\gtrsim 0.004$ ($Z/Z_\odot \gtrsim 20\%$). This suggests we may be viewing a relatively metal-rich galaxy (consistent with the weakness of the UV nebular emission lines), or one in which very luminous stars driving prominent winds are over-represented.
     The absorption in A1703-23 is of similar magnitude, but with minimal emission near systemic, in contrast to the stellar wind profiles.
     Deeper FUV spectroscopy of such galaxies probing other wind lines at similar or higher resolution promises a clearer view of the massive stars populating these systems.
     }
     \label{fig:civ_abs}
\end{figure}

The resulting comparison is displayed in Figure~\ref{fig:civ_abs} (top panel). It is clear that the young very low metallicity ($Z=0.001$) models ($Z/Z_\odot = 5\%$) substantially underestimate the observed P-Cygni profile.
Slightly older SSP ages at this low metallicity would predict even shallower profiles \citep{Vidal-Garcia2017}, even more inconsistent with the observations. At moderately low metallicities ($Z=0.004-0.008$), a range of individual SSP models at different ages are able to match both the absorption and emission profile reasonably well (underscoring the challenge of using resonant wind lines as a robust age and metallicity tracer in-isolation).
Any more complicated star formation history will suppress the strength of this P-Cygni profile relative to the SSPs where it is most prominent \citep{Senchyna2022}.
Thus, in the context of these model predictions, this comparison suggests that a massive star population with metallicities of $\sim 20\%$ solar or above is present. 
However, the strengths of these modeled stellar wind lines are highly sensitive to the uncertain treatment of winds for very luminous metal-poor massive stars, and to the uncertain proportion of such luminous stars in stellar populations.
Indeed, observations of FUV--optical spectra of metal-poor dwarf galaxies in the local Universe hosting young bursts of star formation have revealed markedly stronger \ion{C}{IV} wind profiles in comparison to models constrained in detail by other observables \citep[both nebular line emission and FUV photospheric absorption lines;][]{Senchyna2021,Senchyna2022}.
One proposed solution is that these metal poor young galaxies have an overabundance of luminous massive stars near the Eddington limit (and thus driving strong winds) not captured in the models, perhaps from a high rate of binary mass transfer and mergers occurring on very short timescales. 
These very massive stars may contribute to stronger wind lines than are captured in the existing models, allowing the wind profiles to be fit at somewhat lower metallicities than shown in Figure~\ref{fig:civ_abs}. 
A deeper FUV spectrum would reveal or place limits on additional wind features with different sensitivity to burst age, IMF, abundances, and other key properties (e.g., \citealt{Vidal-Garcia2017,Smith2016}), allowing a more robust investigation of the stellar population in A1703-zd1.

\begin{table*}
\caption{Rest-optical emission-line ratios measured for the A1703 sample. Limits are provided at the $3\sigma$ level, and `-' indicate that at least one line in the line ratio lacks wavelength coverage in the spectrum.}
\begin{adjustbox}{width=1.0\linewidth,center=1\linewidth}
\renewcommand{\arraystretch}{1.2}
\begin{tabular}{lcccccclll}
\toprule
ID & $\rm H\alpha/H\beta$ & $\rm H\gamma/H\beta$ & O3 & O32 & Ne3O2 & $12+\log(\rm O/H)_{\rm strong}$ &  $12+\log(\rm O/H)_{\rm direct}$ & $f_{\rm [CIII]}^{1907} / f_{\rm CIII]}^{1909} $ & $n_e~[\rm cm^{-3}]$ \\
\midrule
 A1703-zd1  & $2.6^{+0.2}_{-0.3}$ & $0.48^{+0.07}_{-0.07}$ & $8.1^{+0.5}_{-0.6}$ & $8.0^{+0.8}_{-1.0}$ & $0.62^{+0.12}_{-0.13}$ & $7.88\pm0.30$    & $>7.57$       & \quad$-$                    & \quad$-$\\       
 A1703-zd2  & $-$                   & $0.63^{+0.14}_{-0.16}$ & $5.2^{+0.5}_{-0.7}$ &    $-$                &  $-$                     & $7.35\pm0.30$    & $>7.27$       & \quad$1.10^{+0.21}_{-0.22}$ & $(2.0\pm1.2)\times10^4$\\           
 A1703-zd5.1& $3.4^{+1.0}_{-1.8}$ & $<1.0$                 & $3.1^{+0.9}_{-1.6}$ &   $-$                 & $-$                      & $7.19\pm0.43$    & \quad$-$      & \quad$-$                    & \quad$-$\\              
 A1703-zd5.2& $2.8^{+0.2}_{-0.3}$ & $0.57^{+0.10}_{-0.11}$ & $4.4^{+0.3}_{-0.4}$ &   $-$                 & $-$                      & $7.26\pm0.28$    & \quad$7.20\pm0.28$ & $>1.48$\hfill               & \quad$<1500$\\     
 A1703-23   & $3.6^{+0.3}_{-0.4}$ & $0.25^{+0.09}_{-0.09}$ & $7.6^{+0.6}_{-0.7}$ &   $-$                 & $-$                      & $7.92\pm0.32$    & \quad$7.84\pm0.27$       & \quad$-$                    & \quad$-$\\    
 A1703-zd4  &  $-$                  & $0.39^{+0.09}_{-0.10}$ & $7.6^{+0.7}_{-1.0}$ & $5.6^{+0.5}_{-0.7}$ & $<0.21$                & $7.98\pm0.31$    & $>7.53$       & \quad$-$                    & \quad$-$\\             
 A1703-zd6  &  $-$                  & $0.57^{+0.17}_{-0.17}$ & $6.4^{+0.3}_{-0.4}$ & $>36$               & $>3.6$                 & $7.60\pm0.30$    & \quad$7.47\pm0.19$ & \quad$0.49^{+0.19}_{-0.16}$ & $(9.4\pm4.2)\times10^4$\\
\bottomrule
\end{tabular}
\end{adjustbox}
\label{tab:optlineratios}
\end{table*}

\subsection{A1703-zd2}
\label{sec:zd2}
A1703-zd2 is a bright (H=24.9) z-band dropout  highly magnified ($\mu=19.9^{+7.1}_{-7.0}$) by the foreground cluster \citep{Bradley2012}. No previous spectroscopy of this system has been presented in the literature.  The NIRSpec rest-optical spectrum is dominated by prominent $\rm [OIII]\lambda\lambda4959,5007$ and Balmer (H$\beta$ and H$\gamma$) emission lines, implying a systemic redshift of $z_{\rm spec}=6.4267$. 
At this systemic redshift, the G395M spectrum covers wavelengths of 3900--6500\angstrom{} in the rest frame such that 
strong lines such as [NeIII], [OII], and H$\alpha$ are not observed.  The auroral [OIII]$\lambda$ 4363 line is not detected at the current S/N of the spectrum.
The [OIII]$\lambda5007$ (EW$=1096\pm412\angstrom{}$ ) and H$\beta$ (EW$=209\pm81\angstrom{}$) lines are narrow, with FWHM of 295 km/s and 320 km/s, respectively, which is consistent with the instrumental resolution at $R\sim1000$.
The total EW of the \oiiihb{} line complex is 1580\angstrom{}, which is nearly twice the average value for $z>6$ galaxies of a similar UV luminosity \citep{Endsley2023} but not as large as in A1703-zd6.
Driven by the moderately large rest-optical line EWs, the synthesized photometry is best fit by an SED model (see Section~\ref{sec:properties}) with a (CSFH) stellar population age of just $17^{+18}_{-7}$ Myr.

The H$\gamma$/H$\beta$ ratio ($0.63^{+0.14}_{-0.16}$; Table~\ref{tab:optlineratios}) implies the emission from A1703-zd2 is subject to minimal dust attenuation.
The relatively blue UV continuum ($\beta=-2.2$) provides a consistent picture of minimal dust content and low V-band optical depth ($\tau_V=0.04$).
Of the available lines in the optical spectrum, the $\rm [OIII]/H\beta$ ratio ($5.2^{+0.5}_{-0.7}$) is the best probe of the gas-phase conditions.
The measured value of $5.2^{+0.5}_{-0.7}$ is below the typical ratios observed at $z\gtrsim6$ \citep[6-8;][]{Sanders2023, Tang2023, Cameron2023}, likely suggesting low gas-phase metallicities. Using existing strong line calibrations (\S4.2), this 
suggests $\oh{}=7.4$, just $1/20Z_{\odot}$. We infer a stellar mass for A1703-zd2 of only $\log(\rm M/M_{\odot})=7.52\pm0.47$, which is the lowest value in the sample.
A low mass is consistent with the optical line properties described above, given the empirical trends between both attenuation \citep[e.g.,][]{Shapley2023} and metal content \citep[e.g.,][]{Curti2023} derived at high redshift.
We return to interpret the oxygen abundance combined with the other galaxy properties in Section~\ref{sec:oxygen}.

The rest-frame UV spectrum displays prominent detections of the high-ionization CIV and He II lines in emission.
Emission features commonly used to probe higher energies (i.e., NV and NeIV) are not covered within the wavelength range of the spectrum (1300--2350\angstrom{}).
The presence of CIV and He II in emission along with the large optical line EWs in A1703-zd2 is qualitatively similar to trends observed in low-metallicity low mass galaxies in the local Universe \citep[e.g.,][]{Senchyna2017}.
However, the observed CIV EW of $31.6\pm7.0\angstrom{}$ is considerably larger than nearly all such local systems \citep[e.g.,][]{Senchyna2019, Izotov2024}, and even exceeds the majority of objects at $z>6$. This clearly indicates this low mass galaxy is associated with a hard radiation field. 

We resolve the CIV doublet  profile (Figure~\ref{fig:UVspectrum}) into two unresolved emission lines at observed wavelengths of 1.1504 $\mu$m and 1.1521 $\mu$m. Relative to the systemic redshift, these two doublet members are offset by $+142~\rm km/s$ (Table~\ref{tab:CIV}). We find that the CIV$\lambda1548$ flux is nearly twice that of CIV$\lambda1550$, consistent with the intrinsic doublet ratio.
This indicates a low amount of scattering and relatively clear  channels for CIV to propagate through the ISM. This is likely to  contribute to the elevated CIV EW measured in this system.
The He II$\lambda1640$ line also exhibits a very large EW ($14.0\pm6.1\angstrom{}$) that exceeds values measured for even the most extreme CIV emitters locally \citep{Izotov2024} and at high-redshift \citep{Topping2024, Castellano2024}.
These two prominent emission lines characterize the hardness of the ionizing spectrum from 48--54 eV, and their ratio (CIV/He II$=3.1\pm1.0$) is  consistent with models driven by star formation or AGN activity \citep[see Figure~\ref{fig:UVlineratios};][]{Feltre2016, Gutkin2016}.
Finally, the CIII] (total EW=13.0\angstrom{}) and OIII] (total EW$<12.1\angstrom{}$) lines are found to be weaker than both He II and CIV. The 
UV line ratios implied by these features (Figure~\ref{fig:UVlineratios}) are also consistent with both stellar and AGN photoionization. Whereas other sources analyzed in this paper are more confidently in the regime of stellar photoionization, a deeper spectrum of A1703-zd2 is required to put more robust constraints on the powering mechanism of the radiation field.

\subsection{A1703-zd5.1}
\label{sec:zd5p1}
A1703-zd5.1 is one the faintest (H=25.7) galaxies in our sample and is also the most magnified ($\mu=33.6^{+14.0}_{-13.7}$) of the $z>6$ systems identified by \citet{Bradley2012}.
As such, it has the lowest luminosity of all high-redshift galaxies we observed in the Abell 1703 field.
The galaxy is a close neighbor of A1703-zd5.2 (see \S\ref{sec:zd5p2} below), situated only 0\secpoint5 away from its companion  in the image plane.
\citet{Bradley2012} interpreted their close separation as a signature of separate clumps within a single galaxy, however it is also possible that they are different galaxies currently in the process of merging.
Despite the two star forming components falling within the same slit, we leverage the high spatial resolution of \jwst{}/NIRSpec to extract separate spectra for each system. Here we present the our observations of A1703-zd5.1, while A1703-zd5.2 is described in \S\ref{sec:zd5p2}.
Despite the apparent faintness of A1703-zd5.1, NIRSpec confirms detections in H$\beta$, $\rm [OIII]\lambda5007$, and H$\alpha$ at a spectroscopic redshift of $z=6.4260$. We take this as the systemic redshift.

A1703-zd5.1 notably has very low optical-line EWs of [OIII]$\lambda5007$ (267\angstrom{}) and H$\alpha$ (157\angstrom{}). The synthesized SED is relatively flat (in f$_\nu$) across the rest-UV and rest-optical, suggesting a relatively young stellar population.  This source appears to be a spectroscopic example of the weak-line population  identified photometrically in \citet{Endsley2023}.  It is challenging to explain these SEDs with simple CSFH models, as young ages generally are accompanied by large EW rest-optical lines. 
\citet{Endsley2023_CEERS} demonstrated that such systems can be readily explained as having recently experienced a sudden decrease in their star-formation activity.
To address these complications in A1703-zd5.1, we employ a two-component star-formation history (TcSFH; \citealt{Endsley2023}) with the additional flexibility required to fit the observed emission.
These models parallel our setup described in Section~\ref{sec:properties}, except with a SFH composed of a delayed-tau model ($\textrm{SFR}\propto t \cdot e^{-t/\tau}$) combined with a decoupled CSFH episode over the most recent 20 Myr.  

The TcSFH model provides a much better fit ($\chi^2=1.6$) when compared to the CSFH model ($\chi^2=9.4$).
Furthermore, the best-fit TcSFH SED model implies a light-weighted age of 70 Myr, and stellar mass of just $10^{7.74}~\rm M_{\odot}$. The relatively low instantaneous sSFR ($3.3/\rm Gyr^{-1}$) underscores the recent downturn in star formation.
The O3 line ratio ($3.1^{+0.9}_{-1.6}$) is the lowest value in our sample, indicative of a very low oxygen abundance ($0.04Z_{\odot}$; see Section~\ref{sec:oxygen} below). 
This is perhaps not surprising given the very low luminosity of this galaxy. We note that while low metallicity will contribute to the weak [OIII], it should boost the strength of H$\alpha$ and other Balmer lines (see discussion in \citealt{Endsley2023}).  Hence low metallicity on its own cannot explain the weak emission lines in this galaxy (as evidenced by the poor $\chi^2$ for the CSFH models). The recent downturn in the SFH allowed in the TcSFH models is key for explaining the weakness of both H$\alpha$ and [OIII]. 

We note that models with ionizing photon escape can also explain the weakness of the emission lines in such SEDs. Following the approach taken in \citet{Topping2022}, we fit the SED using the density-bounded f$_{\rm{esc}}$ models of \citet{Plat2019}. In short, these models self-consistently compute the impact of density-bounded HII regions on the continuum and lines for different escape fractions (see \citealt{Plat2019} and implementation in \citealt{Topping2022}). The results indicate that reasonable solutions can be achieved for moderate escape fractions, although the reduced $\chi^2$ (3.3) is not quite as low as for the TcSFH models. 

The rest-frame UV spectrum of A1703-zd5.1 is relatively featureless, with no detections in the commonly-observed lines (e.g., CIV, He II, CIII]). We place tight upper limits on the EWs of these features ranging from 3--6\angstrom{} (Table~\ref{tab:UVlines}).
The absence of rest-UV emission lines is consistent with the low sSFR inferred from the best-fit TcSFH model described above, which implies that the massive stars that power these lines have begun to disappear following the recent downturn in SFR.
We do observe a tentative He II$\lambda1640$ emission line (S/N=2.5) that appears slightly broadened (FWHM=450 km/s) relative to the instrumental line-spread function (see Figure~\ref{fig:UVspectrum}). If confirmed, this would potentially indicate a broad stellar He II profile. This would be more challenging to explain in the TcSFH models, where recent star formation is sub-dominant, but it could be explained by the leakage models, as significant massive star populations would still be present.
Ultimately, deeper spectroscopy will be required to constrain the flux and shape of He II in order to discern its origin.

\subsection{A1703-zd5.2}
\label{sec:zd5p2}
A1703-zd5.2 is the brighter ($H$=25.3) neighbor to A1703-zd5.1, is situated only 0\secpoint5, and is magnified by a factor of $\mu=33.5^{+17.8}_{-16.9}$.
The rest-optical spectrum reveals significant detection in multiple Balmer lines (H$\alpha$, H$\beta$, H$\gamma$, H$\delta$), in addition to $\rm [OIII]\lambda\lambda4959,5007$. These lines indicate a systemic redshift of $z_{\rm spec}=6.4295$, which is very similar to its close neighbor, A1703-zd5.1.
The full extent of the optical spectrum spans from 3850--6680\angstrom{} in the rest frame, thus lacking constraints on the low-ionization [OII] and [SII] lines.
All of the emission lines detected in the rest optical are narrow, with an average FWHM (310 km/s) that is consistent with the instrumental resolution.

Despite the close proximity between A1703-zd5.1 and zd5.2, the rest-optical spectra differ significantly.
The H$\alpha$ EW ($651\pm227\angstrom{}$) is over $4\times$ that of A1703-zd5.1, while the $\rm [OIII]\lambda5007$ EW ($749\pm188\angstrom{}$) is a factor of 2.7 larger.
The observed SED is well fit by models with CSFH stellar population ages of $23^{+45}_{-17}$ Myr. 
Each of the Balmer line ratios is consistent with the absence of attenuation from dust, with H$\alpha$/H$\beta$ ($2.8^{+0.2}_{-0.3}$) being the most constraining.
We draw a similar conclusion from the relatively blue UV-continuum slope ($\beta=-2.4$).
The only strong rest-optical line ratio probed by NIRSpec is O3. We measure a low value of $4.4^{+0.3}_{-0.4}$, which suggests the strength of [OIII] is being weakened by the low gas-phase metallicity. The strong line calibration of \citet{Sanders2023} suggests this value of O3 corresponds to $12+\log(\rm O/H)=7.26\pm0.28$ (Table 3). 

The rest-UV spectrum reveals significant CIV emission  (Figure~\ref{fig:UVspectrum}). We additionally achieve detections at $\gtrsim5\sigma$ in OIII]$\lambda1666$ and [CIII]$\lambda1907$ within the observed wavelength range of 1300--2400\angstrom{}.
CIV is the strongest of the rest-UV lines in A1703-zd5.2, comprising a total EW of $10.5\pm1.0\angstrom{}$.
While not as extreme as the CIV emission observed in some other $z>6$ galaxies (e.g., A1703-zd6, RXCJ2248-ID, GHZ2), CIV EWs approaching 10\angstrom{} are rare locally \citep[e.g.,][]{Izotov2016, Izotov2024, Senchyna2017}.
Furthermore, we resolve the CIV emission into two narrow components (1.1504$\mu$m and 1.1524$\mu$m ) that are redshifted by $+80$ km/s relative to systemic (Table~\ref{tab:CIV}).
The relative fluxes of the two components ($\lambda1548/\lambda1550=2.0$) and the small velocity offset (from systemic) likely indicates the resonant CIV line may face minimal attenuation. This  may suggest the observed emission is close to the intrinsic value produced in the HII regions.
We use the OIII]$\lambda1666$ ([CIII]$\lambda1907$) detections to infer the direct gas-phase oxygen abundances (electron density) in Section~\ref{sec:oxygen} (Section~\ref{sec:density}) below.
Here, we use the CIV/He II ratio ($>5.6$) to demonstrate a significant decrease in the ionizing radiation field at energies above 47 eV. Other rest-UV line ratios provide a similar picture, with the CIV/CIII] ($>3.2$) and OIII]/He II ($>2.7$) line ratios (Figure~\ref{fig:UVlineratios})  suggesting a stellar origin for the ionizing radiation. It is striking how different A1703-zd5.2 is from its close ($\simeq $450 pc) neighbor A1703-zd5.1.  While A1703-zd5.2 is in the midst of a significant episode of star formation with intense nebular emission, A1703-zd5.1 appears to have very weak emission lines, potentially following a recent downturn in star formation.

\begin{table}
\caption{Properties of the CIV emission line for three objects in our sample with a prominent nebular component.\\
$^a$ Flux measurements are given in units of $10^{-19}~\rm erg/s/cm^2$.}
\begin{center}
\renewcommand{\arraystretch}{1.2}
\begin{tabular}{cccccccc}
\toprule
 &  A1703-zd2 &  A1703-zd5.2 &  A1703-zd6 \\
\midrule
$\rm CIV\lambda1548~Flux^a$ & $10.9\pm2.2$ & $39.0\pm3.2$ & $23.5\pm2.5$  \\ 
$\rm CIV\lambda1550~Flux^a$ & $7.0\pm1.5$ & $19.9\pm3.0$ & $30.0\pm2.8$  \\ 
$\rm CIV\lambda1548~EW$ & $17.1\pm4.5\angstrom{}$ & $7.8\pm1.4\angstrom{}$ & $8.0\pm1.8\angstrom{}$  \\ 
$\rm CIV\lambda1550~EW$ & $11.0\pm3.4\angstrom{}$ & $3.9\pm1.3\angstrom{}$ & $10.2\pm1.9\angstrom{}$  \\ 
$\textrm{CIV}\lambda1548~v_{\rm offset}$ & $157\pm48~\rm km/s$ & $67\pm24~\rm km/s$ & $220\pm26~\rm km/s$  \\ 
$\textrm{CIV}\lambda1550~v_{\rm offset}$ & $127\pm62~\rm km/s$ & $94\pm32~\rm km/s$ & $259\pm40~\rm km/s$  \\ 
\bottomrule
\end{tabular}
\end{center}

\label{tab:CIV}
\end{table}

\subsection{A1703-23}
\label{sec:23}
A1703-23 was first identified as a bright (H=23.8) $i$-band dropout \citep{Zheng2009,Richard2009}, 
in which subsequent followup reported Ly$\alpha$ emission at $z\simeq5.8$ \citep{Richard2009}.
Further spectroscopy obtained by \citet{Stark2015b} targeted the CIII]$\lambda\lambda1907,1909$ line using Keck/MOSFIRE based on the Ly$\alpha$ redshift, however no strong emission was detected.
The strongest rest-optical lines detected with \jwst{}/NIRSpec are [OIII]$\lambda5007$, H$\alpha$, and H$\beta$, at wavelengths of 3.5489$\rm \mu m$, 4.6518$\rm \mu m$, and 3.4456$\rm \mu m$, respectively. 
Based on these strong lines, we measure a redshift of $z_{\rm spec}=6.0862$, slightly different from the single-line measurement presented in \citet{Richard2009}.
Unlike the other galaxies in our sample, the Balmer-line ratios ($\rm H\alpha/H\beta=3.4^{+0.3}_{-0.4}$) imply a modest dust attenuation, 
which is consistent with the moderately red UV slope ($\beta=-1.5$) and optical depth inferred from the  SED ($\tau_V=0.17^{+0.04}_{-0.04}$).
The EWs of [OIII] ($415\pm22$\angstrom{}), H$\beta$ ($46\pm4$\angstrom{}), and H$\alpha$ ($193\pm61$\angstrom{}) are lower than typical galaxies at $z\simeq 6$ \citep{Endsley2023}. The synthesized SED fit implies a moderate age stellar population ($150^{+33}_{-61}$ Myr) and a relatively large stellar mass ($\log(\rm M/M_{\odot})=9.54\pm0.16$). The implied SFR  ($25~\rm M_{\odot}/yr$) is broadly consistent with the star forming main sequence at $z=6$ \citep{Speagle2014}. This is in contrast to most of the other galaxies we targeted in Abell 1703 which sit above the main sequence with much younger stellar populations.\footnote{Given the weak lines, we also fit the synthesized photometry with SED models assuming a TcSFH (see \S\ref{sec:zd5p1}) which yielded a consistent stellar mass, \oiiihb{} EW, and minimum $\chi^2$ to that of the best-fit CSFH model. This suggests that additional flexibility in the SFH is not required to describe A1703-23, and a CSFH is an appropriate assumption given current constraints.}

The rest-frame UV spectrum reveals CIII] emission in addition to a significantly detected continuum (4.5 per resolution element).
As a result, we detect several features in absorption (Figure~\ref{fig:absorption}).
Si~II~$\lambda1526$ is the most strongly detected absorption line, and traces low-ionization along with neutral gas \citep[e.g.,][]{Shapley2003}.
The Si~II~$\lambda1526$ absorption (EW=$-2.0\pm0.6$\angstrom{}) is moderately stronger than lines measured from stacks of typical $z\sim3-7$ galaxies \citep[e.g.,][]{Jones2012},
however the strongest absorption lines are observed in systems with the largest attenuation from dust and highest metallicity \citep{Jones2012, Du2021}.
These features are qualitatively similar to the properties of A1703-23, indicating that reasonable strong absorption lines should be present in a subset of $z\simeq 6$ galaxies.
The Si~II~$\lambda1526$ line center is blue-shifted from the systemic redshift ($\Delta v=-130\pm29$km/s) indicative of outflowing gas.
In contrast, the high-ionization HIS SiIV$\lambda1393,1402$ absorption line is detected (total EW=$1.2\pm0.4$\angstrom{}) at the systemic redshift ($\Delta v=4\pm97$km/s).
The difference in velocity structure indicates the LIS and HIS lines are kinematically distinct. Notably, this is different than what is observed in A1703-zd1, where the two phases are kinematically more similar.

As in A1703-zd1, absorption in CIV is also detected in A1703-23.
We compare this absorption profile to the same set of stellar models as described in Section~\ref{sec:zd1} in Figure~\ref{fig:civ_abs}.
Interestingly, the absorption in A1703-23 is almost absent accompanying emission near the rest wavelengths of the doublet.
This is in strong contrast with the profiles typical of the spherical outflows driven by massive star winds and thus the predictions of the SSP models we compare with.
This morphology is closer to that observed in broad absorption-line quasar spectra \citep[e.g.][]{Allen2011}, though with fairly modest $\sim 2000$~km/s velocity for such systems.
While limited by SNR, a pure absorption profile would be suggestive of interstellar gas entrained in some manner of outflow rather than stellar wind absorption; but this would stand in contrast to the near-systemic absorption in \ion{Si}{IV}.
It is also possible that the peculiar profile is a result of a relatively weak stellar wind profile combined with particularly strong interstellar absorption in \ion{C}{IV} near-systemic.
Higher-SNR and potentially higher-resolution data would help separate these contributions and enable clearer constraints to be placed on the putative stellar component directly.

\subsection{A1703-zd4}
\label{sec:zd4}
A1703-zd4 is a $z$-band dropout identified by \citet{Bradley2012} to have a photometric redshift of $z_{\rm phot}=8.4^{+0.9}_{-1.4}$. 
Multiple spectroscopic campaigns have observed A1703-zd4 using Keck/MOSFIRE targeting the Ly$\alpha$, CIV, He II, OIII], and CIII] emission lines  \citep{Stark2015b, Mainali2018}.
These observations did not yield detections of any emission lines,
however the EW limits ($<6-20$\angstrom{} at 5$\sigma$) were sufficient to distinguish A1703-zd4 from extreme UV-line emitters such as A1703-zd6.
The \jwst{}/NIRSpec spectrum provides the first spectroscopic confirmation at a redshift of $z_{\rm spec}=7.8727$ based on the [OIII]$\lambda5007$ and H$\beta$ lines.
For this spectroscopic redshift, the H-band magnitude (H=25.4) and modest magnification ($\mu=2.0^{+0.1}_{-0.1}$) correspond to an absolute UV magnitude of \muv{}$=-20.7$ \citep[i.e., $\simeq L_{\rm UV}^*$][]{Bouwens2021}.
At the stellar mass ($\log(\rm M/M_{\odot})=8.50\pm0.27$) and SFR ($\log(\rm SFR/M_{\odot}yr^{-1})=1.06^{+0.52}_{-0.45}$) inferred from the SED fit, we find that A1703-zd4 is elevated above the star-forming main sequence by $0.4$ dex \citep{Speagle2014}.
The best-fit SED implies a non-zero impact from dust attenuation ($\tau_V=0.07^{+0.04}_{-0.04}$), which is consistent with the UV slope measured using NIRSpec ($\beta=-2.0$) and relative fluxes of the strongest observed Balmer lines ($\rm H\gamma/H\beta=0.39^{+0.09}_{-0.10}$; Table~\ref{tab:optlineratios}).

We measure large [OIII]$\lambda5007$ and H$\beta$ EWs of $1317\pm432$\angstrom{} and $148\pm46$\angstrom{}, respectively, which are both consistent with the high sSFR of the SED.
A1703-zd4 is one of the two objects in our sample with an [OII] detection enabling the O32 ratio to be calculated.
As the best-fit SED and Balmer-line ratios imply the presence of dust, we correct the [OIII] and [OII] fluxes for attenuation using the \citet{Cardelli1989} attenuation curve and assuming an intrinsic $\rm H\gamma/H\beta=0.47$.
We derive a resulting O32 ratio of $4.0^{+0.9}_{-0.9}$, while we calculate a line ratio prior to the dust correction of $5.6^{+0.5}_{-0.7}$.
Star-forming galaxies in this epoch commonly have O32 in excess of the A1703-zd4 value \citep[$\rm O32\simeq10$;][]{Tang2023, Sanders2023, Cameron2023}, with a clear trend between [OIII] EW and O32 \citep{Tang2023}. A1703-zd4 appears broadly consistent with this relation.

In the rest-frame UV we detect clear Ly$\alpha$ emission as well as [CIII]$\lambda1907$ at $\rm S/N>3$.
The [CIII]$\lambda1907$ EW is just $3.6\pm0.7$\AA, which is easily matched by EELGs at $z\sim2$ with comparable optical line properties \citep[e.g.,][]{Tang2019}.
While no other rest-UV metal line is detected (EWs$<3-6\angstrom{}$), we compute limits on the CIV/CIII] ($<0.3$) and CIII]/He II ($>4.8$) line ratios.
Both of these constraints support a population of massive stars as source of ionizing radiation in A1703-zd4 (Figure~\ref{fig:UVlineratios}).
We also detect a weak Ly$\alpha$ line with an EW of $3.4\pm0.9$\angstrom{}. %
We calculate a velocity offset of $+240\rm km/s$ for Ly$\alpha$ relative to the systemic redshift set by the optical lines.
The Ly$\alpha$ EW in A1703-zd4 is relatively small, and the typical EWs of galaxies at lower redshift with the same optical line EWs suggest a significant attenuation of Ly$\alpha$ photons from A1703-zd4. 
We compute the Ly$\alpha$ escape fraction following the same method outlined in Section~\ref{sec:zd6}, and recover a value of just $f_{\rm esc}^{\rm Ly\alpha}=0.03\pm0.01$, indeed suggesting the observed line is a small portion of the intrinsic emission.

\begin{figure}
    \centering
     \includegraphics[width=1.0\linewidth]{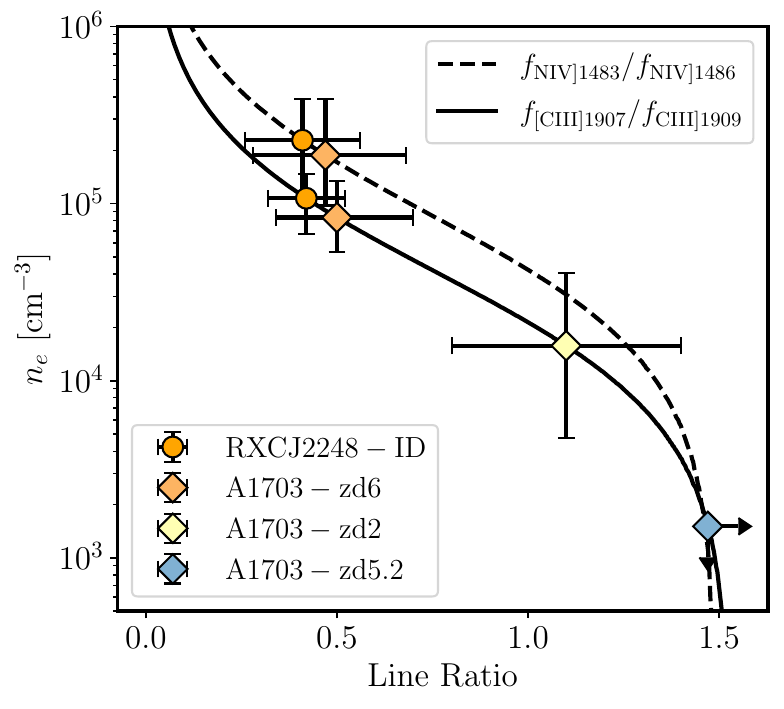}
     \caption{Electron densities inferred from the CIII] and NIV] doublet ratios. Theoretical CIII] and NIV] line ratios calculated for different densities and assuming $\rm T_e=20000K$ are displayed as the solid and dashed lines, respectively. Measured line ratios and inferred densities for the high-redshift galaxies are shown as colored markers. Densities inferred from the CIII] doublet are placed along the solid line, while densities inferred from NIV] are plotted on the dashed line. A1703-zd2 and A1703-zd5.2 are only detected using the CIII] line, while A1703-zd6 and RXCJ2248-ID are detected in CIII] and NIV].} 
     \label{fig:densityratios}
\end{figure}

\begin{figure}
    \centering
     \includegraphics[width=1.0\linewidth]{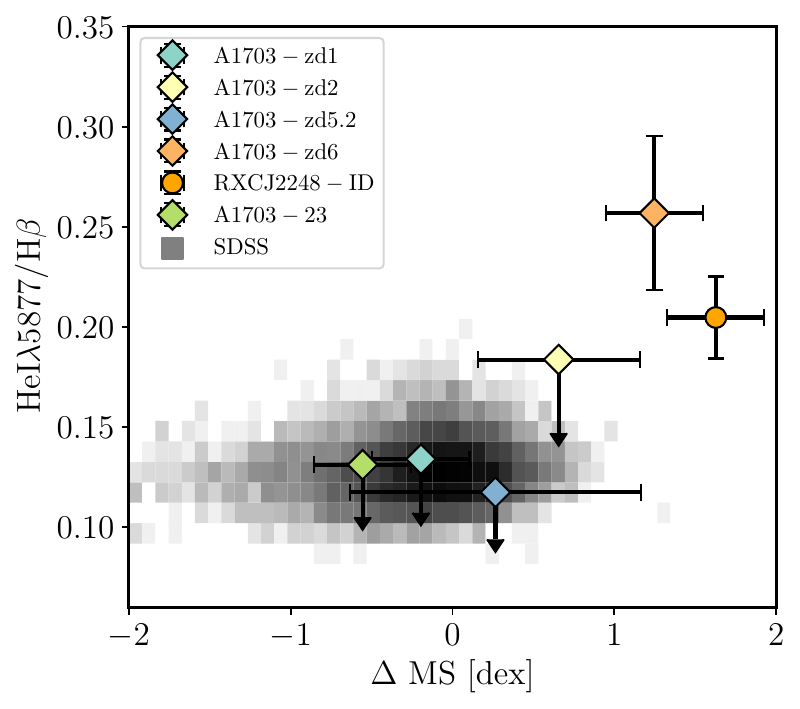}
     \caption{Relative HeI$\lambda5877$/H$\beta$ flux as a function of distance from the star-forming main sequence. Galaxies at $z>6$ are displayed as coloured symbols, while the locus traced by SDSS galaxies is shown as the grey histogram. Typical galaxies in the local universe have an average HeI/H$\beta$ line flux of $0.12$, regardless of their location in the SFR versus M$_*$ plane. While the objects in our sample within 1 dex of the main sequence are consistent with this line ratio, A1703-zd6 and RXCJ2248-ID are significantly He I enhanced, resulting from the high densities preferentially boosting He I relative to H$\beta$ from collisional effects.} 
     \label{fig:heiratio}
\end{figure}

\begin{figure}
    \centering
     \includegraphics[width=1.0\linewidth]{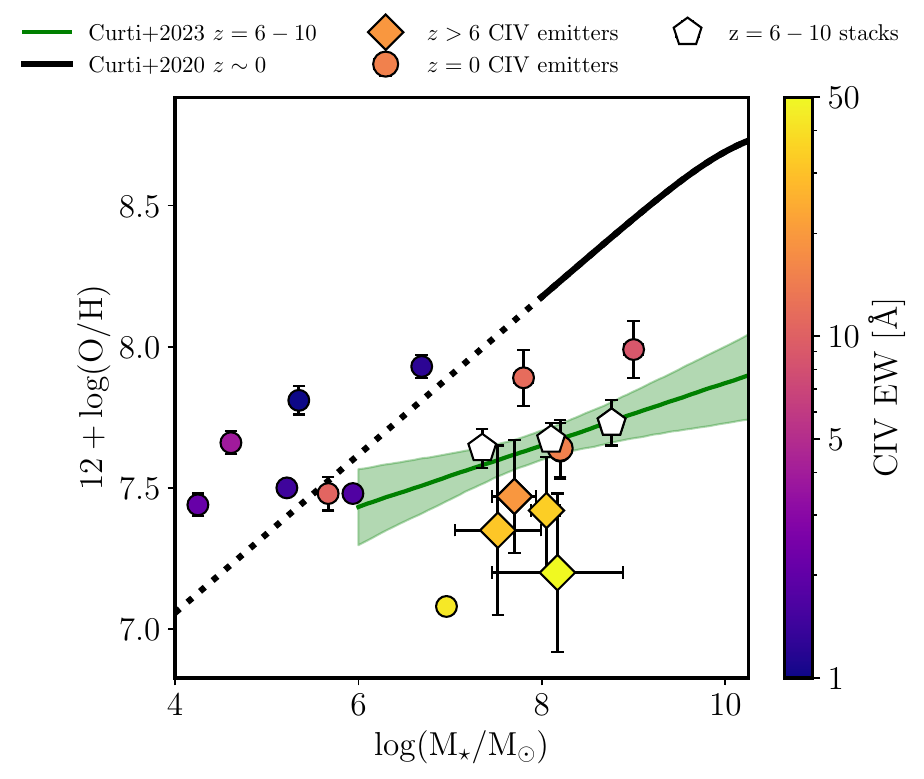}
     \caption{Inferred oxygen abundance as a function of stellar mass for CIV emitters at $z>6$ (diamonds) from the A1703 field in addition to RXCJ2248-ID \citep{Topping2024}, compared to local CIV emitters (squares; \citealt{Izotov2016, Izotov2016b, Izotov2018, Izotov2018b, Senchyna2017, Senchyna2022, Wofford2021, Schaerer2022}). Objects are color-coded based on their CIV EWs as indicated by the colorbar on the right. We display the mass-metallicity relation derived locally (black line; \citealt{Curti2020}). The dotted black line represents an extrapolation of the local MZR from \citet{Curti2020} below the mass range probed by their sample.  The green line and white pentagons are the best-fit MZR and composite measurements of $z\sim6-10$ galaxies from \citet{Curti2023}. The local systems scatter about the MZR, with the smallest and largest CIV emitters lying above and below the MZR, respectively. All of the $z>6$ CIV emitters have below-average metallicities for their stellar mass.}
     \label{fig:MZR}
\end{figure}

\section{Sample Properties in Abell 1703}
\label{sec:sec4}

In the previous section, we presented the spectroscopic properties of our 
$z\gtrsim 6$ sample in the Abell 1703 field, describing each source individually. Here we describe the  electron densities (\S4.1), gas-phase oxygen abundances (\S4.2),  carbon and nitrogen abundances (\S4.3 and \S4.4). Finally we discuss existing constraints on  aluminum enhancements, with the goal of identifying an additional anomalous abundance pattern in globular clusters  (\S4.5). In all subsections below, our broad goal is using the new data to improve our understanding of the properties of  galaxies with hard ionizing sources and nitrogen-enhanced ionized gas.

\subsection{High electron densities in ionized gas}
\label{sec:density}

Recent work has revealed ionized gas with extremely high electron densities ($n_e\simeq 10^5$ cm$^{-3}$) in a subset of reionization era galaxies, several orders of magnitude larger than typical densities at lower redshifts \citep{Bunker2023, Senchyna2023, Topping2024}. This was first inferred for GNz11 based on the NIV] doublet ratio \citep{Bunker2023, Senchyna2023, Maiolino2023} and was then demonstrated  in RXCJ2248-ID based on three density-sensitive doublets in the UV \citep{Topping2024}. In GNz11 it has been suggested the densities may be related to the broad line region of an AGN \citep{Maiolino2023}, but in RXCJ2248-ID the high ionized gas densities appear likely related to dense conditions during a strong burst of star formation \citep{Topping2024}.

A1703-zd6 has a similar spectrum as RXCJ2248-ID, with strong CIV emission and extremely powerful rest-optical emission lines ([OIII]+H$\beta$ EW = 4116~\AA) suggesting a very young stellar population (1.6 Myr). We detect two density-sensitive doublets in the NIRSpec observations presented in \S3.1. We infer densities using the \textsc{getTemDen} procedure from the \textsc{pyneb} software package \citep{Luridiana2015}.
The NIV] doublet ratio ($f^{1483}/f^{1486}=0.47^{+0.19}_{-0.21}$) suggests the ionized gas is very dense ($19^{+20}_{-9}\times10^4\rm/cm^3$), comparable to that seen in RXCJ2248-ID. The CIII] doublet ratio ($f^{1907}/f^{1909}=0.49^{+0.19}_{-0.16}$) reveals a similar picture 
($9.4\pm4.2\times10^4$ cm$^{-3}$). Given the higher ionization potential of NIV] (47.4 eV), its emission originates nearer to the ionizing source when compared to the CIII] emission (24.4 eV). The electron density follows a similar trend and increases toward the inner regions of the H II regions, leading higher-ionization probes to yield higher densities \citep[e.g.,][]{Kennicutt1984, Kewley2019, Berg2021}.

There are several potential physical explanations for the elevated densities. We may expect  high ionized gas densities ($\simeq 10^5$ cm$^{-3}$)  to arise in a short window after a strong burst of star formation, when  gas surface densities are sufficiently high to create an over-pressurized environment for the HII regions \citep[e.g.,][]{Lehnert2009}.
In this picture, we would only observe very dense gas in the subset of galaxies with the largest sSFRs (and hence the largest optical line EWs).  
On the other hand, the UV-based tracers of electron densities (CIII], NIV]) have long been known to probe denser gas than the more typically-used doublets in the optical ([OII], [SII]) \citep{James2014, James2018, Mainali2023}. It is thus possible that the UV lines may reveal elevated densities similar to RXCJ2248-ID and A1703-zd6 in all reionization era galaxies, not just those undergoing strong bursts.  

We can attempt to distinguish between these two scenarios using our Abell 1703 sample.  In addition to A1703-zd6, we detect the density-sensitive CIII] doublet in A1703-zd2 and zd5.2, both with lower H$\beta$ EWs than A1703-zd6. The spectra of the three galaxies reveal distinctly different doublet flux ratios and hence different densities (Figure~\ref{fig:densityratios}). A1703-zd5.2 has the lowest H$\beta$ EW  (170~\AA) and has a 1907/1909 flux ratio ($>1.48$) implying moderately low ionized gas densities  ($<$1500 cm$^{-3}$). A1703-zd2 has a larger H$\beta$ EW (209~\AA) and a 1907/1909 flux ratio ($1.10^{+0.21}_{-0.22}$) indicating moderately large densities (2$\times$10$^4$ cm$^{-3}$). Taking these results together with the spectra of A1703-zd6 and RXCJ2248-ID, we find that the UV-based densities do increase with H$\beta$ EW (albeit in small samples), as would be expected if densities are elevated in the strongest bursts. The highest densities ($>$10$^5$ cm$^{-3}$) appear limited to subset of galaxies with H$\beta$ EWs in excess of 300~\AA, corresponding to extremely young stellar populations ($\lesssim 2$ Myr) formed in a recent upturn of star formation.

Signatures of very high densities are not limited to the rest-UV. In galaxies with ionized densities of  10$^{5}$ cm$^{-3}$, we expect the [OII] emission line to be collisionally de-excited, leading to high O32 indices ($184$ and $>36$ in RXCJ2248-ID and A1703-zd6, respectively). We also expect to see strong He I$\lambda5876$ emission owing to enhanced collisional excitation of this transition \citep[see e.g.,][]{Yanagisawa2024}. In RXCJ2248-ID, He I$\lambda5876$  was found to be  21\% as strong as H$\beta$, roughly 2$\times$ larger than is typically found in local star forming galaxies \citep{Abazajian2009}. The enhancement in the line ratio can be fully explained by the elevated gas densities (see \citealt{Topping2024}). We find similarly enhanced He I~$\lambda5876$ emission in A1703-zd6, with a line flux that is 26\% that of H$\beta$. The other galaxies in our sample are not detected in He I~$\lambda5876$, as expected given the lower ionized gas densities. The upper limits we are able to place on the He I$~\lambda5876$/H$\beta$ flux ratios in these systems ($\lesssim$0.10-0.15) are typically 2$\times$ lower than that found in A1703-zd6 and RXCJ2248-ID (Figure~\ref{fig:heiratio}), consistent with flux ratios found in nearby galaxies. These results suggest that deep rest-optical spectroscopy may be able to identify high density sources via elevated He I~$\lambda5876$ emission (see also \citealt{Yanagisawa2024}).

\begin{figure}
    \centering
     \includegraphics[width=1.0\linewidth]{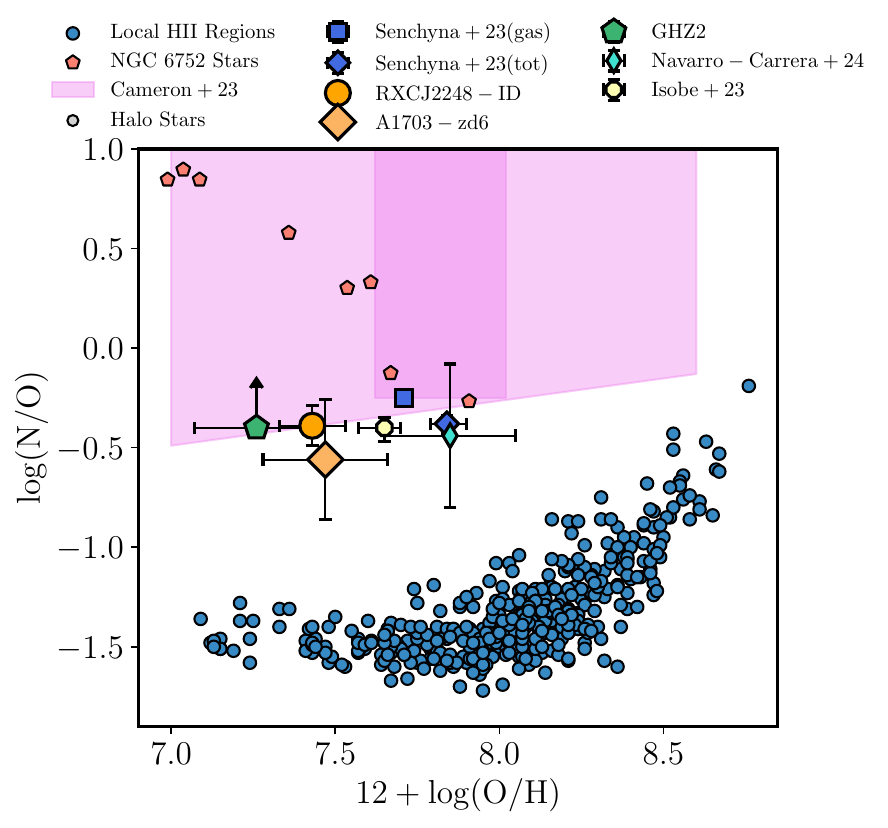}
     \caption{N/O as a function of O/H for stars and HII regions in the local Universe and galaxies at high redshift. Local HII regions trace out a tight locus that asymptotes at $\log(\rm N/O)\simeq-1.4$ at low metallicities. We display measurements from a sample of N-enriched high-redshift galaxies including A1703-zd6 (orange diamond) with nitrogen abundances 10$\times$ higher than HII regions at fixed O/H. This parameter space is shared by individual stars in the globular cluster NGC 6752 indicating that both types of objects may undergo common formation processes.} 
     \label{fig:no}
\end{figure}

\subsection{Gas-phase oxygen abundances}
\label{sec:oxygen}

One of the primary explanations of nebular CIV emission at $z\gtrsim 6$ is hard radiation from a population of very metal poor massive stars. In this case, we would expect the CIV emitting galaxies to be linked to metal poor gas, as is often found in nearby galaxies \citep{Senchyna2017, Senchyna2022, Wofford2021, Schaerer2022}. Photoionization modeling in \citet{Stark2015b} suggested that reproducing the strong CIV emission in A1703-zd6 may require metallicities approaching 2\% of the Solar value ($\oh{}=7.05$), but these inferences were based on very limited information available from ground-based observations. With the NIRSpec observations presented in this paper, we can now derive a direct gas-phase metallicity for A1703-zd6, while also investigating implications for the metal content in the other lensed systems in our sample.

We quantify gas-phase oxygen abundances for each object in our sample using the available emission-line constraints.
The $11\sigma$ $\rm [OIII]\lambda4363$ detection in A1703-zd6 provides the best O/H constraint in our sample, while A1703-zd5.2 and A1703-23 are detected in the auroral $\rm [OIII]\lambda1666$ line. For these systems we derive direct metallicities following the methods described in \citet[][see also \citealt{Izotov2006}]{Topping2024}.
For the remaining galaxies in our sample, we infer the oxygen abundance using the O3 and O32 strong-line calibrations derived from high-redshift galaxies \citep{Sanders2023b}.
Results are listed in Table~\ref{tab:optlineratios}. Gas-phase oxygen oxygen abundances are generally found to be low, ranging from $\oh{}=7.20$ to $7.98$. 

We start with our primary target, the strong CIV emitting galaxy A1703-zd6. Adopting the density found in Section~\ref{sec:density}, we recover an electron temperature of $\rm T_e(\rm O^{++})=23000\pm3200$K, which is consistent with low metallicity galaxies in this epoch \citep[e.g.,][]{Laseter2023, Katz2023}.
For this electron temperature and the  $\rm [OIII]\lambda5007$ and H$\beta$ fluxes, we derive an oxygen abundance of $\oh=7.47\pm0.19$.
This value represents oxygen only in the doubly-ionized state, which we expect to dominate the abundance in this system given the very large O32 ratio ($>36$). 
We derive a consistent metallicity when considering the O3 ratio ($6.4^{+0.3}_{-0.4}$), which yields an abundance of $\oh=7.60\pm0.30$ using the \citet{Sanders2023b} calibrations. Calibrations using the O32 and Ne3O2 ratios indicate much lower metallicities, however these line ratios are likely severely impacted by the density of the system (see Section \ref{sec:density}). Overall, these results suggest the gas in A1703-zd6 is metal poor ($6\%~Z_\odot$), consistent with lower redshift results \citep[e.g.,][]{Senchyna2017, Senchyna2022, Wofford2021, Schaerer2022} and as would be expected if there is a low metallicity stellar population contributing to the hard radiation field. However what stands out most about A1703-zd6 in our sample is not its metallicity, but rather its very young inferred stellar population age ($1.6^{+0.5}_{-0.4}$ Myr) and correspondingly large H$\beta$ EW ($424\pm112$~\AA). Both are at the extremes of what has been seen by {\it JWST} in the $z\gtrsim 6-8$ galaxy population. Presumably the coupling of the extremely young stellar population and low gas phase metallicity are mutually required for strong CIV emission (e.g., \citealt{Senchyna2019}).

We have also detected strong CIV emission ($>10~\angstrom{}$) in two additional galaxies in the Abell 1703 field (A1703-zd2 and A1703-zd5.2). As expected, the new CIV emitters are among the most metal poor in the sample. The O3 ratios of these two systems ($5.2^{+0.5}_{-0.7}$ and $4.4^{+0.3}_{-0.4}$) are well below average (O3=7--10)  at $z\simeq 6-8$ \citep[e.g.,][]{Tang2023, Sanders2023, Cameron2023}. Using the calibrations described above, these likely point to low metallicities: $\oh{}=7.35\pm0.30$ (A1703-zd2) and $7.26\pm0.28$ (A1703-zd5.2). We additionally detect the OIII] UV-based auroral line in A1703-zd5.2, suggesting a direct method metallicity ($\oh{}=7.20\pm0.28$)  consistent with the O3-based value described above. The inferred stellar populations ages of A1703-zd2 (17 Myr) and A1703-zd5.2 (23 Myr) are the next two youngest in our sample after A1703-zd6. 

We note that galaxies in Abell 1703 with metal poor gas but without young stellar populations do not show strong CIV emission. A1703-zd5.1 provides the best example of such a source, with ionized gas that appears metal poor ($\oh{}=7.19\pm0.43$) 
but comparatively weak CIV emission ($<$4.3~\AA). As pointed out in \S~\ref{sec:zd5p2}, this source appears best fit by a recent downturn in star formation (or ionizing photon leakage), resulting in a spectrum without strong CIV emission in spite of the low metallicity.  In contrast, A1703-zd4 has a relatively young stellar population age (23 Myr) but gas-phase metallicity at the upper end of our sample ($\oh{}=7.98\pm0.31$). Here we see no strong CIV emission ($<4.9$~\AA).  Collectively these results  emphasize that the presence of strong CIV emission is limited to sources that show both young stellar populations ($\lesssim~$25 Myr) and low metallicity gas ($\lesssim$ 0.1 Z$_\odot$), consistent with trends seen at lower redshift \citep[e.g.,][]{Senchyna2017, Senchyna2022, Wofford2021, Schaerer2022}. 
We will come back to discuss the demographics of the CIV population through investigating the full public spectroscopic database in \S5.

To place the $z\gtrsim 6$ CIV-emitting galaxies in context, we consider their position with respect to the  mass-metallicity relationship (MZR; \citealt{Curti2023}), as derived from {\it JWST} spectroscopy of $z\simeq 6-10$ galaxies (Figure~\ref{fig:MZR}). We also include several additional strong $z\gtrsim 6$ CIV emitters  \citep{Topping2024} as well the local star forming galaxies with CIV detections \citep{Izotov2016, Izotov2016b, Izotov2018, Izotov2018b, Senchyna2017, Senchyna2022, Wofford2021, Schaerer2022}. It is clear in the figure that the galaxies with strong CIV emission at $z\gtrsim 6$ typically fall below the MZR, implying that they are both younger and more metal poor than the parent sample of early galaxies. The low redshift galaxies tend to have much weaker CIV EWs ($\simeq 1-3$~\AA; c.f. \citealt{Izotov2024}) and lower stellar masses ($\lesssim 10^6$ M$_\odot$) than existing detections at $z\gtrsim 6$. The mismatch in the mass of the CIV emitters at low and high redshift arises because the low metallicities required to power hard radiation fields ($\lesssim$ 0.1 Z$_\odot$) are possible in moderately massive galaxies in the reionization era ($\simeq 10^8$ M$_\odot$). In the local universe, these low metallicities are only found in much lower mass galaxies. As a result, there is potentially a significant difference between the $z\gtrsim6$ CIV emitters and the local reference sample.  Future work is required to investigate whether these differences lead to distinctly different spectra at fixed metallicity.

\begin{figure}
    \centering
     \includegraphics[width=1.0\linewidth]{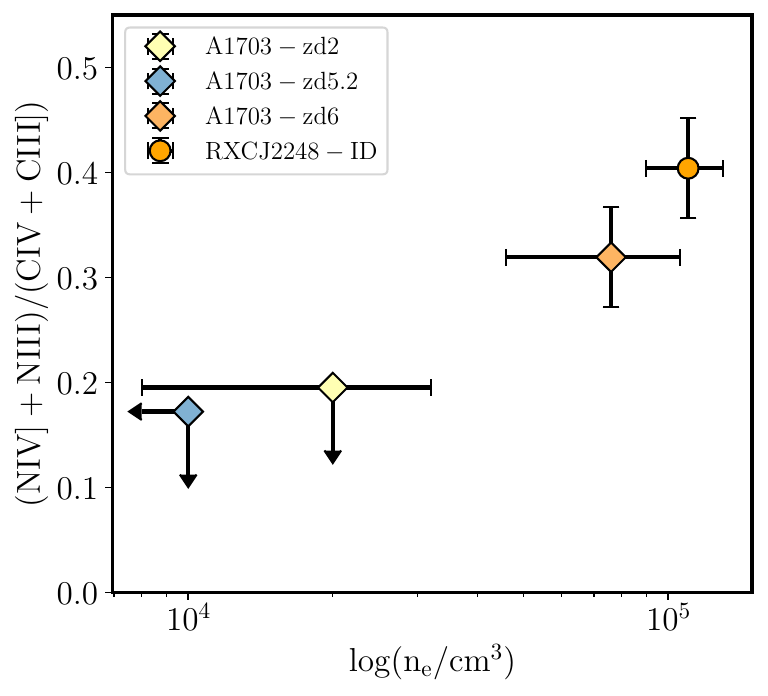}
     \caption{Total NIV]+NIII flux compared to the total CIV+CIII] flux as a function of electron density. We display all objects from our sample for which the density can be constrained from the CIII] ratio. The two galaxies with nitrogen-line detections (and thus higher N/C line ratios) have the highest densities, while the galaxies with more typical CIII] electron densities have (NIV]+NIII)/(CIV+CIII]) ratios that are $<0.2$.}
     \label{fig:densities}
\end{figure}

\subsection{C/O abundances} 
\label{sec:abundances}

The  abundance of carbon (relative to oxygen) encodes  insights into the past star formation history \citep{Henry2000, Esteban2014, Berg2016, Berg2019} and integrated yields of earlier generations of stars \citep[e.g.,][]{Garnett1995, Yin2011, Berg2019}. Early {\it JWST} spectra have constrained carbon-to-oxygen (C/O) ratios in a small number of early galaxies with deep UV constraints on emission from carbon and oxygen lines \citep{Arellano2022, Jones2023, DEugenio2023, Topping2024}. The detections (and upper limits) of the UV carbon and oxygen lines in our sample allow us to build on these studies. Here, we focus on galaxies in the Abell 1703 field with detections in CIV, CIII], and OIII] (A1703-zd6 and A1703-zd5.2) which will provide the most robust constraints on C/O.

Briefly, we infer the C/O abundance ratio by first computing $\rm C^{++}/O^{++}$ from the observed CIII] and OIII] emission lines using \textsc{pyneb} \citep{Luridiana2015}.
In calculating this abundance ratio, we assume an electron temperature and density inferred from OIII]$\lambda1666$/[OIII]$\lambda5007$ and the CIII] doublet, respectively (see Table~\ref{tab:optlineratios}.
The CIII] doublet in A1703-zd5.2 is in the low-density regime, providing only an upper limit on the electron density. For this object, we assume a density of $1000~\rm cm^{-3}$, which is consistent with typical densities observed in this epoch \citep[e.g.,][]{Sanders2023b, Laseter2023, Katz2023}. We note that the inferred abundance ratios vary minimally if a different assumption on density is imposed.
These emission lines yield $\log(\rm C^{++}/O^{++})$ abundance ratios of $-1.20\pm0.12$ and $-1.13\pm0.15$ for A1703-zd6 and A1703-zd5.2, respectively.

Both systems show strong CIV emission ($19.4\pm1.4$~\AA\ in A1703-zd6 and $10.5\pm1.0$ ~\AA\ in A1703-zd5.2), implying a significant amount of the carbon budget exists in the triply-ionized state.
The CIV emission is subject to  resonant scattering, and the observed line fluxes may underestimate the true values. As discussed in \S3, the line profiles of both galaxies give indications of modest scattering. 
Here we consider two bounding cases. On one hand, we compute ionic ratios assuming the observed CIV EWs are the intrinsic values. For the other limiting case, we assume that 50\% of the CIV flux has been scattered out of the line of sight, providing a conservative bound on the fraction of carbon in the triply-ionized state. 
We obtain $\rm C^{3+}/C^{++}$ ionization fractions using \textsc{pyneb} by computing the $\rm C^{3+}$ and $\rm C^{++}$ emissivities at the densities and temperatures defined above.
Using the observed CIV emission as intrinsic we obtain lower-bound $\rm C^{3+}/C^{++}$ of $0.08\pm0.07$ (A1703-zd6) and $0.06\pm0.08$ (A1703-zd5.2) and upper-bound values of $0.38\pm0.12$ (A1703-zd6) and $0.36\pm 0.14$ (A1703-zd5.2). These give the likely range of ionization fractions. To select a value within this range to use for our calculations, we consider an approach used in \citet{Jones2024}. Given that the redder component of the CIV doublet is less susceptible to attenuation (and that both profiles appear to show only modest signs of scattering), we can adopt its flux as intrinsic and compute the total CIV flux using the theoretical doublet ratio ($\lambda$1548/$\lambda$1550=2). This suggests 
$\rm C^{3+}/C^{++}$ values of $0.27\pm0.08$ (A1703-zd6) and $0.13\pm0.11$ (A1703-zd5.2), between the two boundary cases quoted above. We will use these values in our calculations below.

Using the ionic abundances computed above, we derive a total abundance ratio of $\log(\rm C/O)=-0.74\pm0.18$ and $\log(\rm C/O)=-0.79\pm0.17$ for A1703-zd6 and A1703-zd5.2, respectively. Both suggest ionized gas that is not  substantially enriched in carbon relative to oxygen,  well below ($\simeq4\times$) the solar value ($\co_{\odot} = -0.23$; \citealt{Asplund2009}).
Similarly low C/O ratios are commonly inferred for metal poor galaxies locally (\citealt{Berg2016, Berg2018, Berg2020, Ravindranath2020}; c.f., \citealt{Dofour1988}) and at $z\simeq 2$ \citep{Erb2010,Christensen2012,Stark2014,Berg2019}.
Similarly, many systems in the reionization era have significantly sub-solar C/O \citep[e.g.,][]{Jones2023, Arellano2022, Stiavelli2023, Hu2024}, consistent with our results in this paper.  The low C/O observed in these systems are consistent with  enrichment from supernovae, without significant pollution 
from AGB stars \citep[e.g.,][]{Yin2011, Berg2019}.
The absence of such evolved stellar populations are consistent with the young ages and massive stars implied by the rest-UV and optical spectra described in Section~\ref{sec:results}. We note that C/O ratios approaching the solar value are also expected at extremely low metallicities, possibly reflecting yields from population III stars \citep[e.g.,][]{Akerman2004, Carigi2011}. We find no evidence of these in the {\it JWST} spectra analyzed above.

\subsection{N/O abundances}
\label{sec:nabundance}

In \citet{Topping2024}, we demonstrated that the strong CIV emitting galaxy RXCJ2248-ID has an anomalously large N/O ratio, similar to that reported by \citealt{Bunker2023} in GNz11 (see also \citealt{Isobe2023,Marques-Chaves2024,NavarroCarrera2024}).  
We detect similar nitrogen line emission (NIV] and [NIII]) in the CIV emitting galaxy A1703-zd6 presented in this paper.  We have showed that this galaxy has metal poor gas ($\oh{}=7.47\pm0.19$) and a sub-solar C/O ratio ($\log(\rm C/O)=-0.74$). Here we first constrain the N/O ratio in A1703-zd6, before commenting briefly on the relative nitrogen abundances in the other systems in our sample. As we also detect CIV in two additional galaxies, our measurements allow us to further explore the potential link between galaxies with hard radiation fields and significant nitrogen enhancements.

Our methodology is similar to that discussed in  detail in \citet{Topping2024} (and largely parallels our derivation of C/O described above). We will expand on our abundance calculations in Plat et al. 2024 in prep. We infer N/O in A1703-zd6 and place limits on N/O for systems with OIII] detections and nitrogen line non-detections (A1703-zd5.2, A1703-23) using the $\frac{\rm NIV]+NIII}{\rm OIII]}$ line ratio \citep[see e.g.,][]{Senchyna2023, Cameron2023}.
Briefly, in deriving the N/O abundances we assume the electron temperature inferred using the highest S/N auroral emission line ($\rm [OIII]\lambda4363$ or $\rm OIII]\lambda1666$). 
For A1703-zd6 and A1703-zd5.2, we fix the electron density to the value inferred from the NIV] and CIII] doublets, respectively.  Lacking a density-sensitive measurement in A1703-23, we assume a  value of $1000~\rm cm^{-3}$ in our abundance calculations, motivated by typical densities in similar sources at these redshifts \citep[e.g.,][]{Sanders2023b, Laseter2023, Katz2023}. We stress that adopting different densities does not have a large systematic effect on our abundance estimates.

We find that the NIV], NIII], and OIII] line fluxes in A1703-zd6 imply a super-solar nitrogen abundance, with a relative abundance of $\no=-0.6\pm0.3$.
Figure~\ref{fig:no} demonstrates that  this is  moderately higher than the solar value ($\no_{\odot}=-0.86)$, \citealt{Asplund2009}) and nearly $10\times$ higher than typical values seen in HII regions at the metallicity of A1703-zd6.
For A1703-zd5.2 and A1703-23 we follow the methodology described above using the $3\sigma$ upper limits on NIV] and NIII] and find constraints of $\no<-0.9$ and $\no<-0.8$, respectively. 
While we cannot rule out the possibility that these two systems have a minor nitrogen enhancement relative to the local HII regions, their N/O constraints are  consistent with the local locus.

These measurements offer new insight into the nature of the nitrogen-emitting galaxies that have been found in the reionization era with {\it JWST}.  Our Cycle 1 program has now confirmed  nitrogen-enhanced gas in both of our primary targets (A1703-zd6 and RXCJ2248-ID), each originally selected only on strong CIV emission in ground-based UV spectroscopy. This potentially hints at a connection between the processes leading to the formation of hard ionizing sources and those leading to the overabundance of nitrogen. We have shown that both galaxies are likely in the midst of strong bursts of star formation with spectra dominated by the light from very young stellar populations ($\lesssim2$  Myr). This results in rest-optical emission lines that are among the most prominent (upper 1\% of [OIII]+H$\beta$ EW) seen in existing samples of  reionization era galaxies \citep[e.g.,][]{Endsley2023, Matthee2023}. 
Both galaxies also appear to be host  compact star clusters with very dense star formation conditions, the latter revealed by very high electron densities ($\simeq 10^5$ cm$^{-3}$) associated with the density-sensitive emission line doublets in the UV. These high densities might correspond to a short period ($\simeq 1-3$ Myr) after strong bursts where HII regions remain confined by very high gas densities that led to the strong burst. At later evolutionary stages, a variety of feedback processes will disperse the birth cloud gas, and the HII regions will acquire more typically-observed densities ($\simeq 10^2-3$ cm$^{-3}$). 

In this picture, the observed link between nitrogen enrichment and hard ionizing sources (the latter responsible for CIV or He II emission) may have its  origin in the extremely dense conditions that arise in a brief window ($\lesssim$ 1-3 Myr) following strong bursts in the reionization era. The high stellar densities ($>3.6\times10^{4}~\rm M_{\odot}/pc^2$  in RXCJ2248-ID; \citealt{Topping2024}) that arise in the clusters formed during these bursts  will enhance  dynamical interactions (i.e., runaway collisions, binary mass transfer) that may lead to the formation of very massive stars and hot stripped stars, boosting the  population of  hard ionizing sources relative to less dense environments. Such dynamical processes may also help explain how the ISM is polluted with its overabundance of nitrogen relative to oxygen \citep[e.g.,][]{Gieles2018, Martins2020, Charbonnel2023}.

Our observations also unveiled two new CIV emitting galaxies (A1703-zd5.2, A1703-zd2), with CIV EWs similar to A1703-zd6. According to the picture described above, we may expect both to show nitrogen lines, but our observations reveal no  NIII] or NIV] emission in either galaxy.
In Figure~\ref{fig:densities}, we plot the upper limit on the flux ratio of nitrogen (NIV] and NIII) and carbon lines (CIII] and CIV) for A1703-zd5.2 ($<0.17$) and A1703-zd2 ($<0.19$). The  flux limits confirm that these two CIV emitters are significantly weaker nitrogen-line emitters than  RXCJ2248-ID and A1703-zd6, where this ratio was found to be 1.5--2.0$\times$ greater. Clearly the nitrogen line emission is not present in all strong CIV emitters. While both A1703-zd5.2 and A1703-zd2 are dominated by relatively young stellar populations ($23^{+45}_{-17}$ and $17^{+18}_{-7}$ Myr, respectively), they are not as young as A1703-zd6 ($1.6^{+0.5}_{-0.4}$ Myr) or RXCJ2248-ID ($1.8^{+0.7}_{-0.4}$ Myr). Similarly, the ionized gas densities appear elevated ($\simeq 1-2\times$ 10$^4$ cm$^{-3}$) but are 5-10$\times$ lower than the densities in A1703-zd6 and RXCJ2248-ID. It is plausible that these new systems correspond to a slightly later evolutionary state, or that they simply experienced less intense bursts where densities were not elevated sufficiently to initiate the conditions required for seeding the ISM with the nitrogen enhancement. Larger samples will be required to confirm this picture in greater detail. We will come back to this in \S5, investigating the nitrogen line emission in a large public spectroscopic sample.

\subsection{Signatures of additional abundance anomalies}

The $z\gtrsim 6$ nitrogen line emitters may provide our first opportunity to directly study the nature of the polluters responsible for the anomalous abundances in globular clusters \citep[see e.g.,][ for a recent review]{Bastian2018}. If the N/O and N/C ratios are truly related to those found in globular clusters, we should also see evidence of the additional abundance anomalies associated with hydrogen burning. \citet{Marques-Chaves2024} have suggested that aluminum enhancements (from the MgAl cycle)  are likely to be the most readily detectable in early galaxies.
The most accessible probes of aluminum in these galaxies are the 
$\rm Al III\lambda1670$ and  
$\rm Al III\lambda\lambda1854,1862$ lines. These features are most commonly observed in absorption in high redshift star forming galaxies \citep[e.g.,][]{Steidel2016}, but are occasionally seen in emission toward some AGNs \citep{Alexandroff2013}. If aluminum is overabundant in the ISM of the nitrogen line emitters, we may expect to see the Al III lines in emission. 

In RXCJ2248-ID, we presented an upper limit on the strength of the $\rm Al III\lambda\lambda1854,1862$ doublet, with each individual component likely having an EW below $2.0$~\AA. Here we extend this analysis to A1703-zd6. We search for Al III emission near the expected observed wavelengths of the doublet. 
The spectrum of A1703-zd6 displays an emission feature at the expected wavelengths of the $\rm Al III\lambda 1862$ emission line.  We closely inspect the spectrum surrounding this  feature and find that one exposure suffers from a large residual resulting from the subtraction from a snowball event (see Section~\ref{sec:data}). 
To remedy this, we repeat our data reduction procedures while removing the affected exposures. The resulting spectrum still exhibits an emission feature (EW=2.7~\AA) at the expected wavelength of the $\rm Al III\lambda1854$ line. As the S/N is low (2.6) we consider this feature to be tentative. We do not detect any emission at the location of the $\rm Al III\lambda 1862$ line, consistent with expectations given the predicted flux ratio ($\lambda$1854/$\lambda$1862=1.5) of the doublet.

To quantify the Al abundance implied by this tentative feature in A1703-zd6, we consider the $\rm Al III\lambda1854/OIII]\lambda1666$ line ratio, for which we measure $f_{\rm Al III}/f_{\rm OIII]}=0.36$.  For the metallicity and ionization parameter we infer for A1703-zd6, the observed $f_{\rm Al III}/f_{\rm OIII]}$ indicates an aluminum enhancement that is 30--100$\times$ the solar value ($\log(\rm Al/O)_{\odot}=-2.24$, \citealt{Asplund2009}). A more detailed treatment of the abundance ratios will be presented by Plat et al. (in prep). But we note that this level of aluminum enhancement is consistent with the abundance pattern seen in a subset of individual globular cluster stars \citep[e.g.,][]{Carretta2018, Carlos2023}.  Deeper spectroscopy of A1703-zd6 will ultimately be required to verify this tentative feature.

\begin{figure}
    \centering
     \includegraphics[width=1.0\linewidth]{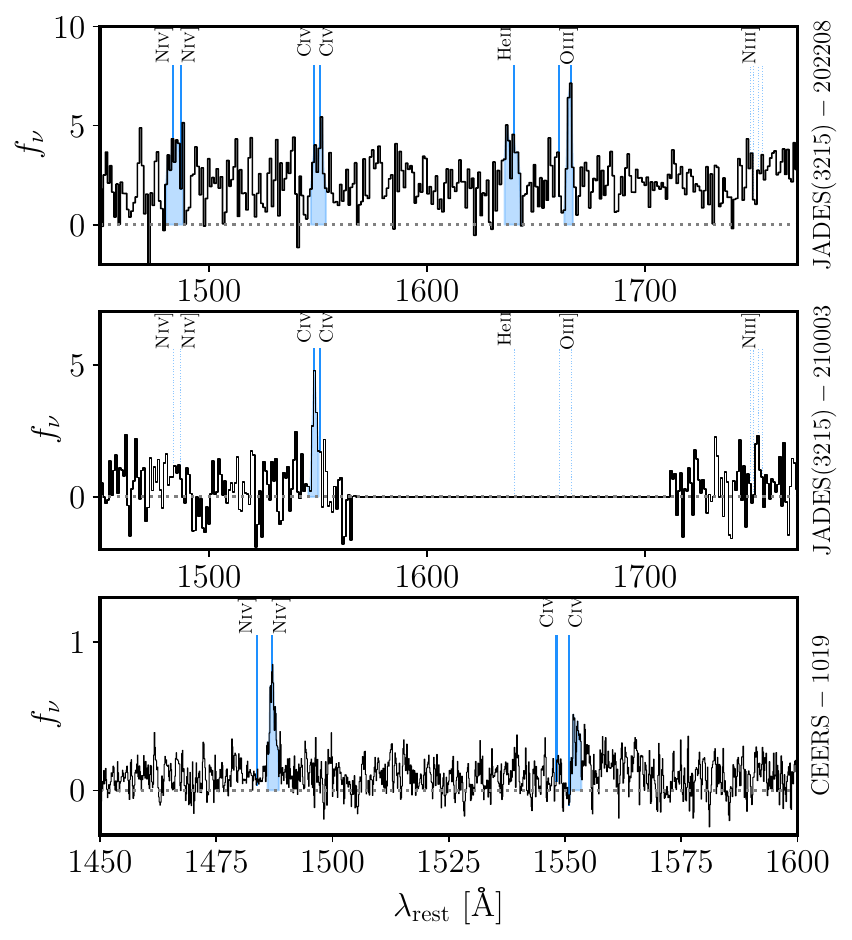}
     \caption{Spectra for galaxies with previously-unreported CIV or NIV] detections presented in the rest frame. The top, center, and bottom panels shows the spectra for JADES-202208, JADES-210003, and CEERS-1019, respectively, as indicated by their labels on the right (see also Table~\ref{tab:literatureemitters}). The spectra of JADES-202208 and JADES-210003 were obtained using the $R\sim1000$ G140M grating, and CEERS-1019 was obtained using the $R\sim2700$ G140H grating. The spectrum of JADES-210003 between 1565--1710\angstrom{} is covered by the NIRSpec detector gap. The wavelength range of the CEERS-1019 spectrum is narrowed to better show the emission lines at high resolution, and none of the redder rest-UV lines (i.e., He II$\lambda1640$, OIII]$\lambda\lambda1660,1666$, NIII]$\lambda1750$) are detected at $>3\sigma$. Vacuum wavelengths of emission lines in the rest-UV are indicated by vertical blue lines that are labelled at the top of each panel. Emission lines that are detected have their fluxes shaded blue. A 5$\sigma$ detection of NIV]$\lambda1486$ in CEERS-1019 was previously reported by \citet{Larson2023} and \citet{Isobe2023},  now detected at 17$\sigma$ in the $R~\sim2700$ spectrum. This spectrum also presents the first detection of CIV in CEERS-1019 using the NIRSpec grating.}
     \label{fig:database}
\end{figure}

%
%
%
%
\section{Demographics of $z>4$ UV line emitters}
\label{sec:disc}

The discovery of enhanced N/O and N/C ratios in  early galaxies 
has been one of the more surprising early findings  of {\it JWST} \citep{Bunker2023,
Isobe2023,Marques-Chaves2024,Topping2024,Castellano2024}, potentially providing a signpost of globular clusters in formation 
\citep{Senchyna2023,Charbonnel2023,Vink2023}.
In this paper, we have presented more evidence suggesting that the nitrogen emitters  also tend to have a population of hard ionizing sources and very dense ionized gas  (see \S4.3 and \citealt{Topping2024}).

In this section, we attempt to place the strong nitrogen and carbon line emitters in a broader context, establishing how common these systems are at very high redshift and comparing their basic properties to that of the full galaxy population. 
To achieve this goal in a self-consistent manner, we assemble a large spectroscopic sample of $z\gtrsim 4$ galaxies with NIRSpec observations and characterize the incidence of strong UV line emission. 
While the nitrogen emitters observed as part of this program (i.e., A1703-zd6 and RXCJ2248-ID) were observed due to their strong CIV emission, this large spectroscopic sample was not constructed with the same prerequisite.
The data reduction and construction of the NIRSpec spectroscopic catalog are briefly described in \S5.1. We then discuss the statistics and properties of the CIV-emitters in \S5.2 and that of the nitrogen emitters in \S5.3.

\subsection{Public \jwst{}/NIRSpec spectra at $z>4$}
\label{sec:database}

We assemble the NIRSpec MOS observations from the Cosmic Evolution Early Release Science (CEERS\footnote{\url{https://ceers.github.io/}}; ERS-1345, PI: S. Finkelstein; Finkelstein et al. (in prep), also see \citealt{Finkelstein2022a}), the Director’s Discretionary Time (DDT) observations (DDT-2750, PI: P. Arrabal Haro; \citealt{ArrabalHaro2023a,ArrabalHaro2023b}) in the CEERS field, the JWST Advanced Deep Extragalactic Survey (JADES\footnote{\url{https://jades-survey.github.io/}}; GTO-1180, GTO-1181, and GTO-1210, PI: Eisenstein; \citealt{Eisenstein2023}), the JADES Origins Field (JOF; GO-3215, PI:Eisenstein; \citealt{Eisenstein2023_JOF}), and the Ultradeep NIRSpec and NIRCam ObserVations before the Epoch of Reionization (UNCOVER\footnote{\url{https://jwst-uncover.github.io/}}; GO-2561, PIs: I. Labb\'e \& R. Bezanson; \citealt{Bezanson2022}) and GO-4287 (PI: Mason).
A full description of the sample selection and analysis methods applied to these public spectroscopic samples will be provided in Tang et al. (in prep), however we provide a brief description in this section.
We obtained final 2D and 1D spectra following the reduction procedure described in \S \ref{sec:observations}.
We also collect the {\it HST}/ACS+{\it JWST}/NIRCam photometry available from \cite{Endsley2023_CEERS} in the CEERS field, the JADES DR2 photometric catalog in the GOODS-S field (corresponding to sources observed in programs GTO-1180 and 1210, and GO-3215; \citealt{Rieke2023}), photometry from the DAWN JWST Archive \citep{Brammer2023,Valentino2023} in the GOODS-N field (for GTO-1181), and the UNCOVER DR2 catalog from \cite{Weaver2024}.

We follow the procedure outlined in \citet{Tang2023} to identify spectroscopic redshifts based on the H$\alpha$ (for $z\lesssim6.7$) in addition to $\rm [OIII]+H\beta$ (for $z\lesssim9.4$) emission lines.
For the small number of systems at yet higher redshift, we utilized the strongest emission line (typically [NeIII], [OII], or CIII]) if present, otherwise 
we adopted the redshift implied by the Ly$\alpha$ continuum break. Break redshifts typically have larger uncertainties ($\sigma_z\sim 0.1$) than those measured from emission lines \citep[e.g.,][]{Jones2024, Hainline2024}, however this does not significantly impact our results.
In total, these efforts yield a parent sample of 737 galaxies with spectroscopic redshifts spanning $z=3.90$ to $13.36$ (median $5.29$). This sample does not include the CIV emitter targeted in RXCJ2248-ID \citep{Topping2024} or any of the Abell 1703 sources presented in this paper. We will comment on how the detections in these samples compare to the overall population trends.

Following the methods outlined in \citet{Tang2023, Tang2024, Chen2024}, we fit the {\it HST}/ACS+{\it JWST}/NIRCam photometry of each of the 737 galaxies using {\sc BEAGLE} following the model setup described in \S \ref{sec:properties} with the redshifts fixed to the spectroscopic value. We utilize the best-fit SEDs to define the absolute UV magnitudes ($M_{\rm UV}$) as well as to infer stellar population properties including stellar mass, SFR, and age from their best-fit SEDs.
In addition, we measure $\rm [OIII]\lambda5007$ and H$\beta$ EWs for each galaxy in the sample. In systems where the rest-optical continuum is detected at $>3\sigma$, the optical-line EWs are measured directly from the spectrum. For these objects, we compare the measured EWs to the values inferred from the best-fit SED and achieve an excellent agreement. 
In the absence of a significant rest-optical continuum detection in the spectrum, we adopt the $\rm [OIII]\lambda5007$ and H$\beta$ EWs derived from the best-fit SED. 

Figure~\ref{fig:database} presents the \muv{} and combined \oiiihb{} EWs for the 737 galaxies comprising the spectroscopic sample.
The absolute UV magnitudes of the spectroscopic sample range from \muv{}$=-22.6$ to $-16.1$, with a median value of $-19.6$.
This \muv{} distribution is consistent with the UV-bright end of the UV luminosity distribution of photometrically-selected objects \citep[e.g.,][]{Endsley2023}, the latter also including fainter systems (down to \muv{}$=-16.0$) and having a luminous median \muv{} ($-18.0$).
Both the photometric and spectroscopic samples span the similar ranges of \oiiihb{} EW of 150--6000\angstrom{} and 45--8300\angstrom{} respectively, and the median \oiiihb{} of the spectroscopic sample ($710\angstrom{}$) is well-matched to that of the photometric sample at fixed \muv{}.

We select galaxies from the spectroscopic sample displaying CIV or NIV] emission lines using a combination of visual inspection and S/N criteria.
First, we measure emission-line fluxes by shifting the PRISM and grating spectrum into the rest frame, and fitting Gaussian profiles at the vacuum wavelength of each line. 
The centroid of the profiles were allowed to vary by one resolution element in the PRISM spectra ($\simeq10000$ km/s), and 1000 km/s in the grating spectra to allow for small variations in the wavelength solution in addition to ensure capturing of resonance lines that do not necessarily trace the systemic redshift (e.g., CIV).
Uncertainties are set by width of the flux distribution derived by repeatedly fitting profiles to 1000 instances of the spectrum perturbed by its error.
Each $>5\sigma$ CIV and NIV] emission line was visually inspected to remove artifacts and poor fits to the data, and we verified that each emission line yielded consistent fluxes when covered by multiple observing modes. 
The resulting robust sample comprises 11 galaxies detected in either CIV (10 galaxies; Section~\ref{sec:civdemographics}) or NIV] (3 galaxies; Section~\ref{sec:nivdemographics}) spanning from $z=5.449$ up to $10.621$, which are presented in Table \ref{tab:literatureemitters}. The spectroscopic redshifts for each of these systems are measured from rest-optical emission lines, rather than the continuum break.
Among the 9 galaxies detected in CIV, the measured equivalent widths range from 4.9 \AA{} up to 47.6 \AA{} with a median of 25.4 \AA{}, comparable to what is observed in A1703-zd2 and A1703-zd6. 
When present, the NIV] emission lines have EWs that span a smaller range (5.1--8.3\angstrom{}), which may contribute to their smaller sample size. 
Finally, we additionally select two (four) galaxies with tentative NIV] (CIV) detections, which satisfy only one of our S/N and visual inspection criteria (Table~\ref{tab:literatureemitters}).
The spectra of several of our robust sources have been previously presented in the literature, as indicated in the Table. Additionally, we identify two new galaxies (JADES-202208 and 210003) with either a CIV or NIV] detection (Figure~\ref{fig:database}).

In addition, we present a deep, high-resolution ($R\simeq2700$) NIRSpec spectrum of CEERS-1019 \citep[i.e., EGSY8p7;][]{Zitrin2015, Roberts-Borsani2016} in the rest-frame UV that reveals CIV emission (Whitler et al. in prep). Previous spectroscopy using the $R\sim1000$ grating from CEERS was not sufficiently deep to detect the CIV feature \citep[e.g.,][]{Larson2023, Tang2023}.\footnote{CEERS observations targeting EGSY8p7 using the NIRSpec PRISM suffered from an electrical short resulting in significant excess emission throughout the detector. However, \citet{Marques-Chaves2024} note the presence of a feature at the expected wavelength of CIV in the contaminated spectrum}
We measure a total EW for the CIV emission in EGSY8p7 of $4.6^{+0.5}_{-0.6}\angstrom{}$. The CIV profile of EGSY8p7 is asymmetric with no sign of the blue doublet component. This is reminiscent of the line profile of RXCJ2248-ID and easily explained by resonant scattering effects  (see \citealt{Topping2024}). It is well-fit by a skewed Gaussian profile redshifted from the systemic wavelength. The properties of the fit are also similar to the RXCJ2248-ID line; we measure a FWHM of 310 km/s after correcting for the instrumental resolution ($\sim100$ km/s), and a velocity offset from CIV$\lambda1550$ of the emission peak of $+190$ km/s. Finally, the emission extends to yet higher velocities, with the best-fit profile extending above the error spectrum up to velocities of +800 km/s.  

In the following subsections, our primary goal is quantifying the incidence of CIV and NIV], independent of the powering mechanism of the lines.  We do briefly consider  whether any of our sources clearly stand out as having lines powered by AGN photoionization. Here we use the  UV diagnostics discussed in \S \ref{sec:zd6}  \citep[e.g.,][]{Feltre2016,Mingozzi2022,Scholtz2023}. Our measurements reveal CIV/He II ratios ($\gtrsim1$) and CIII]/He II ratios ($\gtrsim2$) that are are  consistent with stellar photoionization, as we also found for RXCJ2248-ID and A1703-zd6. We will discuss these in more detail in Plat et al. (2024, in prep). Of course, very high redshift AGN may have different spectra than those predicted in these models, and some level of mixing between AGN and stellar photoionization is also plausible. 
We therefore do not rule out contribution from AGN photoionization in these sources. We note that one source (JADES-954) has been confirmed as a broad line AGN in \citet{Maiolino2023_infantBH}. We will discuss this source in more detail below. One additional source (JADES-58975) was listed in the narrow line AGN sample of \citet{Scholtz2023}. We find that this spectrum has UV lines  that are plausibly consistent with a stellar origin (OIII]/HeII$=1.4$, [Ne IV] EW $<$ 30\angstrom{}), but this does not significantly impact our demographic investigation below \citep[see also,][]{Curti2024}.
\begin{figure*}
    \centering
     \includegraphics[width=1.0\linewidth]{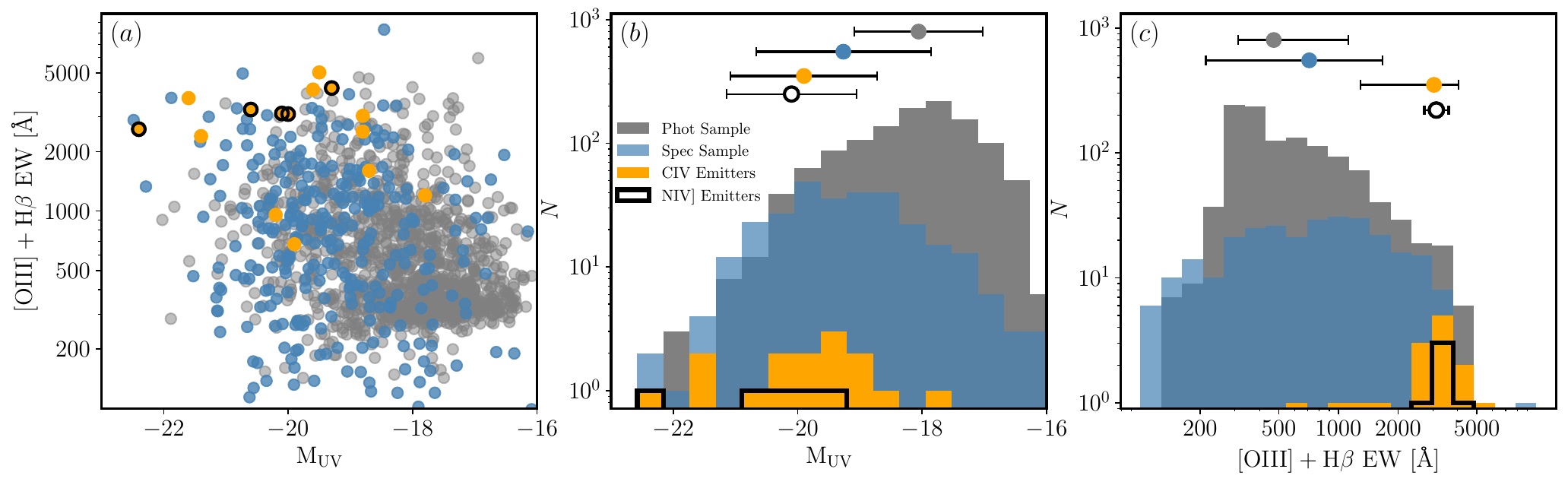}
     \caption{$\rm [OIII]+H\beta$ EW versus \muv{} of robust NIV] (black) and CIV (orange) emitters compared to the full sample of spectroscopic (teal) and photometric (grey) measurements (panel a). Panels (b) and (c) provide the distributions of \muv{} and \oiiihb{}, respectively for the four samples. In these panels, the points and errorbars above each histogram display the median and standard deviation, respectively, of each sample. The CIV and NIV] emitters appear to have UV luminosities at or above the median of the spectroscopic sample, however nearly all of these objects have EWs elevated significantly above the sample average. The two CIV emitters with $\rm [OIII]+H\beta$ EWs that are less than 1000\angstrom{} have been shown to display evidence of AGN activity \citep{Maiolino2023_infantBH, Scholtz2023}. }
     \label{fig:distributions}
\end{figure*}

\subsection{The Demographics of CIV Emitters at $z\gtrsim 4$}
\label{sec:civdemographics}

We have presented new evidence in this paper that links nitrogen line emitters to those with strong CIV emission (see also \citealt{Topping2024,Castellano2024}). 
In this subsection, we attempt to quantify the fraction of $z\gtrsim 4$ galaxies with strong CIV emission, establishing how commonly hard ionizing sources are present in early galaxies.  This builds on earlier efforts to establish the CIV emitting fraction in the galaxy population with ground based spectroscopy \citep{Mainali2018}.

We have identified 11 galaxies with robust CIV detections among the 737 systems in our $z>4$ spectroscopic sample.
Table~\ref{tab:literatureemitters} presents the properties of each of these systems.
Seven of these CIV detections have been previously reported, including CEERS-1019 \citep[EGSY8p7][]{Larson2023, Isobe2023, Tang2023, Marques-Chaves2024}, JADES-3991 \citep[GNz11][]{Bunker2023, Maiolino2023}, CEERS-397 \citep{Harikane2023}, JADES-954 \citep{Maiolino2023_infantBH}, JADES-1899  \citep{Witstok2024, Tang2024}, JADES-13176 \citep{Cameron2023b}, JADES-58975 \citep{Scholtz2023, Curti2024}, and UNCOVER-10646 and UNCOVER-22223 \citep{Fujimoto2023_uncover}.
The CIV emission in JADES-210003 and JADES-202208 have not been previously reported, and we present their spectra in Figure~\ref{fig:database}.  

The  detection rate of CIV emitting galaxies (11 among 737 systems) is extremely low, as would be expected if the hard ionizing sources required to power strong CIV are only present in a small subset of the parent population. Of course, not all 737 galaxies have sufficiently deep spectra to place a meaningful limit on the CIV EW. To achieve a more robust CIV emitter fraction, we identify spectra with the depth to detect CIV with an EW$>10\angstrom{}$ at the $>5\sigma$ level.
Of the 737 $z\gtrsim4$ galaxies with spectra described in Section~\ref{sec:database}, we identify 99 systems with sufficient CIV sensitivity. With this more refined sample, we still find that a very small fraction ($8\%$) of $z>4$ galaxies show robust CIV emission with an EW in excess of $10\angstrom{}$.

To understand the low success rate in detecting CIV emission, we compare the \muv{} and $\rm [OIII]+H\beta$ EWs of the CIV emitters to the full spectroscopic sample, as well as to a larger NIRCam-based photometric sample of $z\sim6-9$ galaxies from \citet{Endsley2023} in Figure~\ref{fig:distributions}. The eight CIV emitters span a fairly broad range of UV luminosities (\muv{} of $-21.6$ to $-17.7$), with a median absolute UV magnitude of \muv{}$=-19.6$). 
In contrast, the [OIII]+H$\beta$ EWs of the CIV emitters are far from typical. Nearly all of the CIV emitters have very large $\rm [OIII]+H\beta$ EWs (median 3000\angstrom{}), indicating they are on the  tail of the full distribution of galaxies.\footnote{One of the CIV emitters (UNCOVER-22223) lacks an \oiiihb{} EW measurement as the lines are redshifted out of the NIRSpec wavelength coverage.}   According to the $z\simeq 6-9$ [OIII]+H$\beta$ EW distribution in \citet{Endsley2023}, the median [OIII]+H$\beta$ EW exhibited by the CIV emitters corresponds to the upper 2\% of the population.  
 There is one CIV emitter in our sample (JADES 1181-954) that shows relatively small \oiiihb{} EW (679~\AA), distinguishing it from the rest of the sample. This source is  a broad line AGN with a very red optical continuum  \citep{Maiolino2023_infantBH}. It is likely that the AGN is contributing significantly to the continuum flux density in the rest-optical, decreasing the \oiiihb{} EW from what would have been observed from the host galaxy.

The large optical line EWs 
are consistent with expectations for extremely young stellar populations, formed in a recent strong burst of star formation. Indeed, the BEAGLE SEDs that describe these systems show a median constant star formation age of just 5 Myr. These results are consistent with our findings for A1703-zd6 and RXCJ2248-ID, where [OIII]+H$\beta$ EWs are large (4120 and 3100~\AA, respectively) and stellar population ages are young ($1.6^{+0.5}_{-0.4}$ and $1.8^{+0.7}_{-0.4}$ Myr, respectively).
Based on these results, we suggest that the rarity of strong CIV emission is likely due to the [OIII]+H$\beta$ EW (or age) threshold required to power the high ionization lines. Only a small fraction of the galaxy population is found with the extremely young stellar population ages that are linked to the large [OIII]+H$\beta$ EWs associated with strong CIV emission. If we consider only those objects in 
the parent spectroscopic sample with  \oiiihb{} EWs larger than 2000\angstrom{} (and spectra sufficiently deep to detect CIV with EW$>$10~\AA), we find that 40\% (5 of 12) show strong  CIV emission ($>$10~\AA). This suggests that hard ionizing sources capable of powering CIV are likely fairly common in galaxies in a brief (few Myr) window following a strong burst of star formation. By selecting galaxies with very high [OIII]+H$\beta$ EWs (via photometric flux excesses or spectroscopy), it should be possible to build a larger sample of systems with high-ionization lines in the rest-UV.

\begin{figure}
    \centering
     \includegraphics[width=1.0\linewidth]{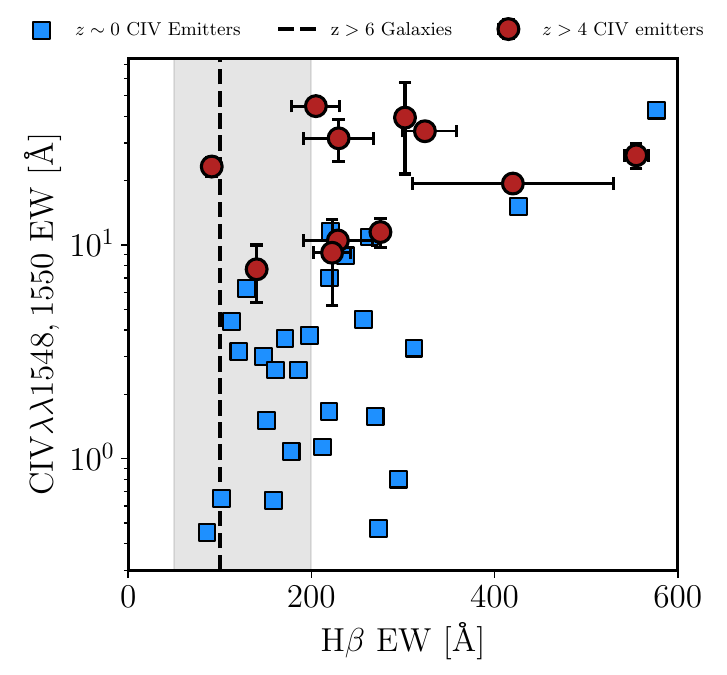}
     \caption{CIV EW versus H$\beta$ EW for galaxies in the local universe (blue squares) and at high redshift (red circles). The high-redshift CIV-emitter sample includes objects in the A1703 field in addition to RXCJ2248-ID \citep{Topping2024} and the CIV emitters identified in Table~\ref{tab:literatureemitters}.  The local comparison samples include objects presented by \citet{Senchyna2019}, \citet{Izotov2016, Izotov2016b, Izotov2018, Izotov2018b}, \citet{Berg2016, Berg2019},  \citet{Senchyna2017}, and \citep{Izotov2024}. We indicate the average H$\beta$ EW and the $1\sigma$ distribution width at $z>6$ as the vertical dashed line and shaded region, respectively, from the photometric sample of \citet{Endsley2023}. The high-redshift galaxies nearly universally have higher CIV EWs than the local samples. For H$\beta$ EWs of 100-300\angstrom{}, the highest CIV EWs locally only reach $\simeq10\angstrom{}$, while several galaxies at $z>4$ exceed 30\angstrom{}. At the highest H$\beta$ EWs ($>400\angstrom{}$), the local and high-redshift galaxies have comparable CIV EWs, implying they may be reaching the maximum EW that can be produced by photoionization from massive stars. }
     \label{fig:CIVEW}
\end{figure}

We compare the CIV EWs of the $z\gtrsim 4$ galaxies to local star forming systems \citep[e.g.,][]{Berg2016, Berg2019, Izotov2016, Izotov2016b, Izotov2018, Izotov2018b, Senchyna2017, Senchyna2019, Izotov2024} with similar H$\beta$ EW in Figure~\ref{fig:CIVEW}. These two quantities provide insight into how the hard radiation field varies with the stellar population age. We see that local metal poor galaxies often power relatively weak CIV emission (1-10~\AA) at H$\beta$ EW = 100-300~\AA. In contrast, the early galaxies with CIV detections extend to much higher CIV EWs ($\gtrsim$ 20-50~\AA) than are seen locally 
at similar H$\beta$ EWs. While several newly-discovered local systems are beginning to approach the intense CIV EWs found in the reionization era \citep[e.g.,][]{Izotov2024}, it appears that these very large CIV EWs are not the norm in strong bursts of star formation at low redshift. This may indicate that the stellar populations in reionization era galaxies host harder ionizing sources, or it could hint at some contribution from AGN photoionization. Alternatively a larger fraction of the resonant CIV emission may be escaping from very high redshift galaxies (see \citealt{Senchyna2022}), with observed EWs approaching the maximum CIV EW achievable by stellar photoionization models (Plat et al. in prep). Deeper spectra of the CIV emitters should reveal which of these is most likely to be driving the stronger high ionization line emission seen at $z\gtrsim 4$.

\subsection{The Demographics of NIV] Emitters at $z\gtrsim 4$}
\label{sec:nivdemographics}

We now quantify the prevalence of the NIV] emitters in the spectroscopic sample. As described in \S5.1, we recover three systems with significant NIV] detections (Table~\ref{tab:literatureemitters}) including GN-z11 \citep{Bunker2023} and CEERS-1019 \citep{Larson2023, Isobe2023, Marques-Chaves2024}, as well as one new source (MSA ID: JADES-202208). 
We display the spectrum of JADES-202208 ($z=5.449$) in Figure~\ref{fig:database}. The NIV] emission is detected at S/N=5.0, along with CIV (S/N=5.1), He II (S/N=4.4), and a strong detection of the OIII] auroral line (S/N=4.9). The NIV] EW ($5.1^{+1.6}_{-2.6}$~\AA) and CIV EW ($5.6^{+2.0}_{-2.0}$~\AA) are comparable, while the [OIII]+H$\beta$ EW (3268\AA) is among the largest in the parent catalog.

The recovery of only 3 NIV]  lines in a redshift catalog of 747 galaxies highlights the  challenges in blindly increasing the nitrogen line sample with non-targeted spectroscopy. One reason for the low detection rate is that the nitrogen lines are typically weak (EW$\simeq 5-10$~\AA), requiring extremely deep rest-UV spectroscopy for meaningful limits in fainter sources. To quantify the 
fraction of nitrogen line emitters in the $z>4$ galaxy sample, we only consider spectra where a NIV] line with an EW larger than 5\angstrom{} can be constrained at $>5\sigma$.
This limiting EW is more strict than the one imposed for CIV in Section~\ref{sec:civdemographics} reflecting the typically-weaker NIV] line. Of the 737 galaxies across the full sample, only 38 are of sufficient depth to reach this limit. Given the 3 NIV] detections, this suggests that 7\% of $z\gtrsim 4$ galaxies with sufficiently deep {\it JWST} spectra exhibit NIV] with EW $>$ 5~\AA. This fraction is comparably low to that found for strong CIV emitting galaxies, suggesting that the majority of early galaxies are not observed in a phase with very hard ionizing sources and enhanced nitrogen. 

To understand what makes nitrogen line emitters rare at $z\gtrsim 4$, we compare the absolute magnitude and [OIII]+$H\beta$ EWs of the NIV] emitters to the parent spectroscopic sample in  Figure~\ref{fig:distributions}. Currently the NIV] detections are primarily limited to the bright galaxies (median \muv{}$=-21.1$) which is not surprising given the sensitivity to detect NIV]. It is conceivable that deeper spectroscopy (or highly-magnified sources in cluster fields) may reveal strong nitrogen line emission in fainter galaxies.  What does appear to be clear is that 
the NIV] emitters exclusively have very large \oiiihb{} EWs (2600-4200\angstrom{}), placing them in the upper 2\% of the EW distribution at similar redshifts and indicating extremely young stellar populations \citep{Endsley2023, Matthee2023}. Of course, very large  \oiiihb{} EWs are also found in RXCJ2248-ID and A1703-zd6, neither of which is included in the sample described above. If we consider galaxies with  high \oiiihb{} EWs ($>2000\angstrom{}$) and spectra deep enough to recover NIV] emission at our EW limit, we find that nitrogen emitters comprise a much larger fraction ($30\%$) of the population.  Spectroscopic programs targeting the highest equivalent width optical line emitters may provide a more efficient means of identifying additional nitrogen emitters.

\begin{figure*}
    \centering
     \includegraphics[width=1.0\linewidth]{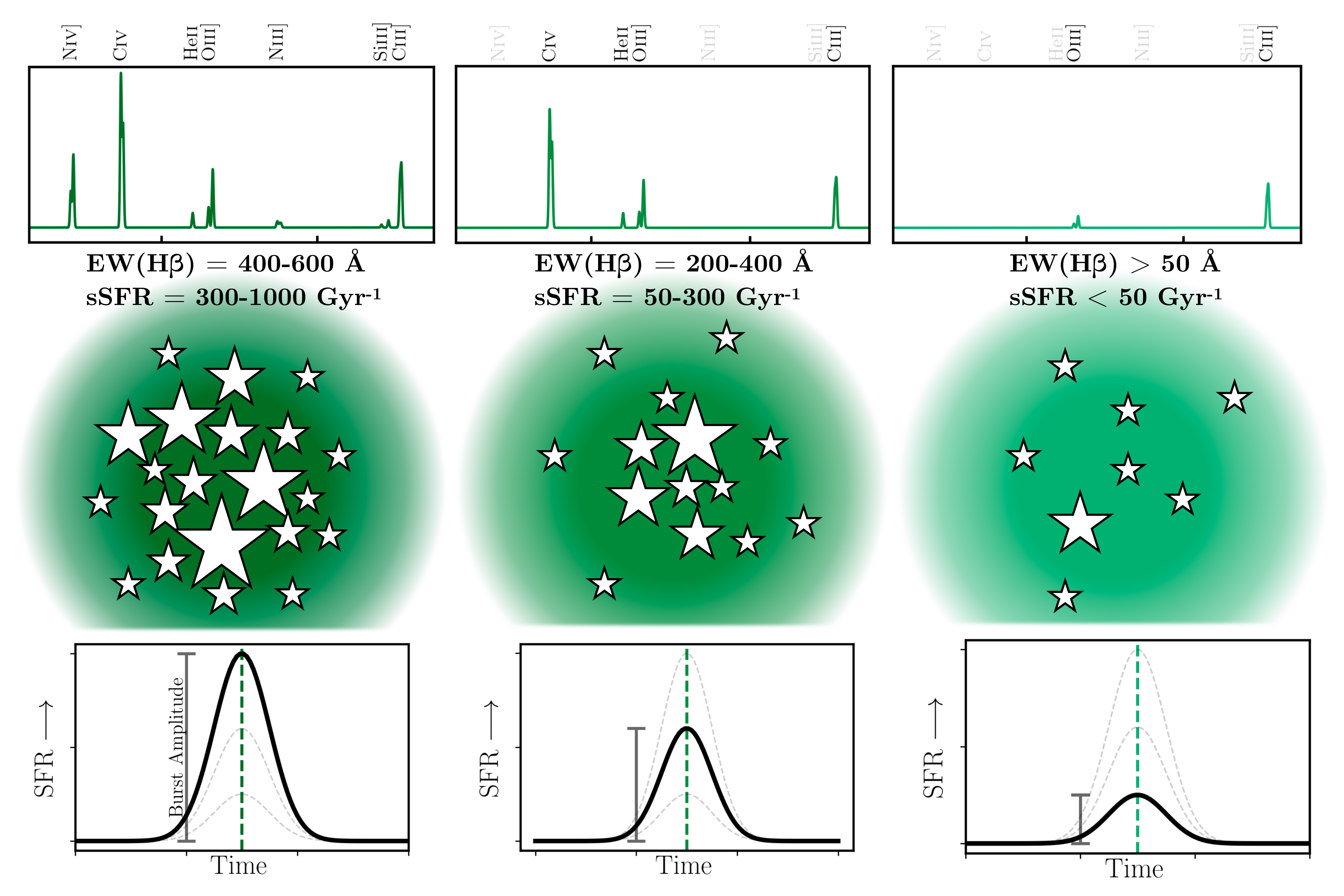}
     \caption{Schematic demonstrating the physical scenarios that result in a given rest-frame UV spectrum. The left panels present a UV spectrum that contains high-ionization lines (e.g., He II and CIV) along with strong nitrogen lines (e.g., NIV] and NIII]) such as those seen in A1703-zd6 (This Work) and RXCJ2248-ID \citep{Topping2024}. In this scenario, the UV lines are driven a significant burst of star formation (sSFR$=300-1000~\rm Gyr^{-1}$) which also drives very strong rest-optical emission lines (H$\beta$ EW$=400-600\angstrom{}$). The middle panels reflect rest-UV spectra such as those of A1703-zd2 and A1703-zd5.2, which display strong high-ionization lines (CIV and He II), but do not show emission from nitrogen. These systems are driven by a moderate burst of star formation (sSFR$=50-300~\rm Gyr^{-1}$) and have above-average rest-optical emission line EWs (H$\beta$ EW$=200-400\angstrom{}$). The right panels demonstrate a scenario where the most massive stars are absent, representing recent star-formation activity corresponding to an sSFR of $<50~\rm Gyr^{-1}$. These systems do not have the ionizing radiation field capable of driving CIV and He II emission, and do not have an ISM enriched in nitrogen, and display H$\beta$ EWs down to $\simeq50\angstrom{}$. The most common line seen in the rest-frame UV for this scenario is CIII].
     }
     \label{fig:schematic}
\end{figure*}

\section{Discussion}
\label{sec:discussion}

Over the past several years, a distinct phase in the early evolution of reionization-era galaxies has been uncovered by deep spectroscopy.  The first glimpse of galaxies in or near this phase was via discovery of  prominent \ion{C}{iv} emission detected pre-JWST in a handful of luminous lensed systems, a transition rarely encountered in nebular emission at such strengths in lower-redshift star-forming systems \citep[][]{Stark2015b,Mainali2017,Schmidt2017}.
More recently, the detection of strong lines of ionized nitrogen including \ion{N}{iv}] in GN-z11 clearly established that some high-redshift galaxies harbor gas in a remarkable nitrogen-enriched and high-density phase that is not yet fully understood \citep{Bunker2023,Cameron2023b,Senchyna2023,Maiolino2023}.
In \citet{Topping2024} and this paper, we have shown that both prominent   $z>6$ \ion{C}{iv} emitters known  before {\it JWST} launch also reveal signatures of dense, nitrogen-enriched gas, suggesting that these two phenomena are potentially linked.

With improved statistics on the demographics of nitrogen emitters, we can begin to place some constraints on the nature of this observational phase.
The clear association of strong \ion{C}{iv} and \ion{N}{iv} detections with very intense optical nebular emission (H$\beta$ EW ~$\gtrsim 400$~\AA{}) in a sample of over 700 $z\gtrsim 4$ galaxies establishes that this is likely a short-lived phase during bursts of star formation.  
The elevated H$\beta$ EWs correspond to galaxies caught with very large instantaneous sSFR.  
In the context of bursty star formation histories likely to be present at $z\gtrsim 6$ \citep[e.g.,][]{Ma2018, Furlanetto2022, Dome2023, Endsley2023, Strait2023}, the largest sSFRs will be found in galaxies with the highest amplitude bursts (i.e., the highest star formation rate at fixed stellar mass) observed near the peak of this star formation activity.
The H$\beta$ EWs thus depend both on the light-weighted stellar population age and on the relative strength of the burst.  The demographic trends we found in \S5 may therefore suggest that the nitrogen enrichment is limited to  the highest peak intensity bursts of star formation. Specifically, while galaxies with reasonably intense line emission (H$\beta$ EW $\simeq$ 200\AA) are likely in the midst of recent upturns of star formation,  they appear not to power the nitrogen emission seen in more intense bursts.
We illustrate this schematically in Figure~\ref{fig:schematic}. 

The fact that these nitrogen signatures are only observed at the highest-sSFRs is an important clue as to their nature.
It is natural to expect that 
galaxies undergoing the strongest bursts are likely to have unique ISM conditions. In particular, we expect  the cold  gas densities to reach extremely high values in these systems \citep[e.g.,][]{Kim2021, Grudic2021} where the cluster formation efficiency may be elevated \citep[e.g.,][]{Adamo2011, Adamo2015, Ma2020}. 
Hence the nitrogen line emitters appear to be  preferentially found in galaxies that are likely to be forming very dense star cluster complexes, potentially similar to the conditions under which globular clusters formed.
This fact also suggests that the abundances are short-lived in the ionized gas phase, as otherwise we would expect to find similar evidence for nitrogen over-abundances (in these or lower-ionization nitrogen lines) in galaxies with somewhat less extreme optical equivalent widths, representing the same bursts at somewhat older ages.
Some combination of processes must alter this situation on a short (of-order Myrs; \S5.2) timescale: for instance, ongoing star formation involving this gas (as is the case if some of this material is locked-up in stars with globular cluster-like abundance patterns), stellar feedback heating and evacuating this gas, yields from Type II supernovae or time-integrated normal massive star winds polluting this material \citep[which should boost oxygen and carbon relative to nitrogen; e.g.][]{Nomoto2013,Charbonnel2023}, or inflows of pristine gas diluting these abundances.

However, high-sSFR alone is clearly not sufficient on its own to produce the conditions responsible for this enrichment.
Strong bursts with similar H$\beta$ EWs in the local universe do not appear to show the prominent nitrogen lines being recovered at $z\gtrsim 6$ \citep[e.g.,][]{Berg2019, Senchyna2017, Senchyna2019, Izotov2016, Izotov2016b, Izotov2018, Izotov2024}. This suggests that the  conditions required to produce the anomalous nitrogen abundance pattern may be uniquely present in  high redshift galaxies, perhaps reflecting different star formation processes. One possibility is that the combination of high gas densities and low metallicities (thought to be  common at $z\gtrsim 6$) may lead to rapid cooling times and less effective feedback, resulting in  elevated star formation efficiencies in 
a subset of early galaxies \citep{Dekel2023, Li2023ffb}. If present in early galaxies, these so-called `feedback-free starbursts' (FFBs) would likely be the highest sSFR galaxies at a given epoch, the same population that appears to often show \ion{C}{iv} and \ion{N}{iv} detections.  We note that the compact star forming complexes associated with the nitrogen-enhancements do appear to reach stellar mass surface densities \citep{Topping2022} where feedback-free starburst conditions may be expected \citep{Grudic2021}.  Hence while FFBs have recently been  
 discussed as a means of explaining the excess of galaxies at $z\gtrsim 10$,  they may also play an important role in driving the nitrogen-enhanced abundance pattern that is being seen in the strongest bursts of the reionization era and in the fossil record of globular clusters \citep[see also e.g.][]{Renzini2023}. 

The ionizing sources responsible for driving the nitrogen enhancement have been discussed at length in the recent literature \citep[e.g.,][]{Senchyna2023, Nandal2024b, Topping2024, Maiolino2023, Castellano2024, Marques-Chaves2024}. With improved constraints on the demographics of this population, we can gain new insights into the nature of this population. 
The fact that this abundance pattern is only found in very strong bursts where densities are large is consistent with a picture where dynamical interactions  play an important role in forming the polluting sources \citep{Gieles2018, Martins2020, Charbonnel2023}, as pointed out previously \citep{Senchyna2023}. Runaway collisions may alter the stellar initial mass function in these dense stellar systems, creating a population of very massive stars (VMSs; $\simeq 10^2$-$10^3$ M$_\odot$) or supermassive stars (SMSs; $\simeq 10^3$-10$^5$ M$_\odot$). 
The contribution of both populations to the nitrogen-enhanced abundance pattern has been recently investigated in some detail \citep[e.g.,][]{Nagele2023, Charbonnel2023, Senchyna2023, Topping2024, Nandal2024b, Marques-Chaves2024}. If low metallicity VMSs altered by interactions and mergers are present in nitrogen emitters, we would expect a hard radiation field, perhaps even more intense than is commonly found in metal poor galaxies locally. This may help explain the intense CIV emission in these systems, with EWs in excess of most local samples (Figure 17). In contrast, the SMSs are not expected to power hard radiation fields \citep[e.g.,][]{Gieles2018, Martins2020}, so if such a population was responsible for expelling the nitrogen-enriched material, we would likely also need there to be a separate population of hard ionizing sources to power the high ionization lines. Or we may
be witnessing a later evolution stage (following the SMS phase) during a second burst of star formation, where massive stars forming from already-enriched material now photoionize their nitrogen-enhanced surroundings.
Along these lines, the potentially important role of multiple bursts in explaining the abundance pattern has been discussed in \citet{Kobayashi2024}. In  multiple burst scenarios, a key prediction is that there should be a  population of strong bursts without significant nitrogen emission lines (i.e., the first burst in such scenarios). The demographics presented in \S5 are to first order consistent with this possibility, as not all galaxies with H$\beta$ EW$\gtrsim 400$~\AA\ support strong nitrogen line emission. 
More detailed comparison of formation models to these demographics may aid in differentiating between these scenarios.

It has been recently noted  (e.g., \citealt{Marques-Chaves2024}) that the dense conditions in the nitrogen emitters also appear optimized for creating intermediate mass black holes (IMBHs). If VMSs or SMSs are common in the star clusters of nitrogen emitters (see discussion above), they may provide an efficient path for the formation of black holes, with subsequent growth from tidal disruption events and black hole mergers \citep[e.g.,][]{Quinlan1990, PortegiesZwart2004, Rantala2024}. In cases where this picture holds, the nitrogen emitters at $z\gtrsim 6$ may correspond to an early phase where relatively massive black holes are efficiently formed following intense and dense bursts of star formation. 
In multiple burst scenarios (e.g. \citealt{Kobayashi2024}), it is possible that IMBHs could have already assembled in earlier star formation episodes. In this case, the nitrogen emitters may already harbor relatively massive black holes which may be fueled during the current burst of star formation. Here we may expect to see AGN signatures, as is plausibly the case for GNz11 \citep{Maiolino2023}. Distinguishing between these cases requires confirmation of the presence of VMSs or SMSs in the nitrogen emitters or their progenitors.

Many of the outstanding questions about these objects are connected to the uncertain nature of the stellar populations and ionizing sources powering them.
An important first step in characterizing such massive star populations was taken in \citet{Rivera-Thorsen2024}, using deep \textit{JWST} spectroscopy of the $z=2.37$ Sunburst Arc, probing the rest-frame optical of this object.
The Sunburst Arc is a crucial lower-redshift laboratory for understanding these $z>6$ spectra, as the rest-UV spectrum of the highly-lensed LyC-leaking cluster in this galaxy reveals similarly prominent \ion{N}{iii}] emission requiring significant nitrogen enrichment: $\mathrm{N/O}= 0.6$ \citep{Pascale2023}, albeit at a much smaller mass scale (a lensed single cluster) and likely higher metallicity (plausibly consistent with the lack of \ion{N}{iv}] emission).
The new \textit{JWST} spectra reveal strong broadened stellar wind emission in both the blue and red `bumps' typically associated with Wolf-Rayet (WR) stars \citep{Rivera-Thorsen2024}.
These bumps are composed of blended lines of (predominantly) ionized nitrogen/helium and carbon, respectively; and their detection supports the presence of a prominent population of luminous and stripped (or otherwise processed to reveal layers exposed to hot H-burning) massive stars driving dense, highly-ionized winds.
This may be critical to understanding the source of the nitrogen-enriched material seen in the ionized gas phase in this object.
The strong winds of very massive stars ($\sim 100$--$1000$~$\mathrm{M_\odot}$) near the Eddington limit are a natural source of nitrogen-enriched material themselves \citep[e.g.][]{Vink2023}, and are expected to be visible in both optical WR bumps \citep[e.g.][]{Martins2022}.
However, the wind signatures of these VMSs are highly sensitive to mass loss rates and to detailed evolution and wind modeling, and challenging to disentangle from `classical' WR stars in integrated light \citep[e.g.][]{Martins2023}.
If VMS or WR winds are directly responsible for the pollution observed in these $z>6$ sources, deep UV spectroscopy promises to readily reveal their continuum signatures even at low metallicities \citep[e.g.][]{Senchyna2021,Martins2023}; and in any case, detailed comparisons both with lower-redshift samples and with population synthesis predictions including these sources would enable the first characterization of their constituent massive stars and a critical test for the various enrichment scenarios.

%
%
%
%
\section{Summary}

We present new {\it JWST}/NIRSpec observations of galaxies in the Abell 1703 field. Our primary target is A1703-zd6, a bright (H=25.9) $z=7.0435$ galaxy known to power strong CIV emission from ground-based observations \citep{Stark2015a}. The new NIRSpec spectrum spans the rest-UV to optical, constraining the metal content, abundance pattern, and ionizing sources linked with such strong CIV emission. We also present spectra of seven additional bright (H=23.8--25.9) $z\simeq 6-8$ galaxies in the Abell 1703 field. To place these results in context, we investigate the prevalence of CIV and nitrogen lines in a newly-constructed redshift catalog of 737 $z\gtrsim 4$ galaxies with public NIRSpec spectra. 

One of our central goals is to understand the physical conditions that lead to the nitrogen enhancements seen in recent metal poor galaxies at high redshift \citep{Bunker2023,Marques-Chaves2024,Topping2024}, testing whether there is a link between the presence of hard radiation fields and the anomalous abundance pattern.  We summarize our primary findings below.

1. The rest-UV spectrum of A1703-zd6 confirms the strong CIV emission (EW=19.4~\AA) seen with Keck, while also revealing strong emission from a suite of other emission lines (i.e., OIII], CIII], He II).
The CIV EW is larger than that seen in most metal poor star forming galaxies found in nearby ($z<0.02$) galaxies. 
The line ratios indicate a hard radiation field (CIV/CIII]=3.3), albeit consistent with expectations for stellar photoionization.

2. The rest-optical spectrum of A1703-zd6 exhibits a suite of strong emission lines. The [OIII]+H$\beta$ EW (4116~\AA) is extremely large, corresponding to the upper 0.5\%  of EWs seen in the reionization-era. The ionized gas is found to be metal poor ($\oh=7.47$) and very high in electron density ($8-19\times10^4 ~\rm cm^{-3}$). The elevated gas densities result in collisional de-excitation of [OII] (boosting the O32 ratio) and strong He I lines (He I$\lambda$5876/H$\beta$=0.26). The spectral features are similar to those of RXCJ2248-ID \citep{Topping2024}, suggesting high densities are a feature of HII regions in  reionization-era galaxies with the largest [OIII] and H$\beta$ EWs.

3. We find that the emission line spectrum of A1703-zd6 can be explained by a low metallicity young (1.6 Myr) stellar population 
formed in a recent burst of star formation. The stellar mass (10$^{7.70}$ M$_\odot$) associated with the burst appears concentrated in a compact (unresolved) region less than 120 pc in size.  NIRCam imaging will be required to put a more stringent limit on the size. Nonetheless the current measurement suggests that A1703-zd6 hosts a high density of metal poor massive stars.

4. The rest-UV spectrum of A1703-zd6 reveals strong NIV] and NIII] emission lines, indicating that the ionized gas is both elevated in nitrogen and deficient in oxygen ($\no=-0.6$). The abundance pattern is similar to that seen in GNz11 \citet{Bunker2023} and RXCJ2248-ID \citep{Topping2024}. The discovery of nitrogen enhancements in both A1703-zd6 and RXCJ2248-ID, two galaxies identified based on strong CIV emission, suggests that there may be a physical link between the presence of hard ionizing sources and the processes leading to the elevated nitrogen content in the ISM.  We suggest both may be related to a dense clusters of massive stars formed after a strong burst of star formation.

5. We also find strong CIV emission (EW=10.5 and 31.6~\AA) in A1703-zd2 and A1703-zd5.2, two $z\gtrsim 6$ galaxies with rest-optical emission lines suggesting metal poor gas ($\oh{}=7.35\pm0.30$ and $7.26\pm0.28$). Both galaxies are also found to lie above the star forming main sequence at $z\simeq 6-8$ with large specific star formation rates ($59$ and $32$ Gyr$^{-1}$ assuming a CSFH). The  H$\beta$ EWs are very large (209~\angstrom{} and 170~\angstrom{}), yet they are $\gtrsim 2\times$ less than A1703-zd6.  Neither source shows evidence for nitrogen enhancements. Doublet ratios suggest electron densities are large ($1-2\times10^4$ cm$^{-3}$) but not as elevated as the sources showing the nitrogen emission. We suggest that the H$\beta$ EW is likely to provide the best indicator of whether a galaxy exhibits nitrogen enhancements seen in a handful of early {\it JWST} spectra.

6. To better test this physical picture, 
we investigate the spectra of 737 $z\gtrsim 4$ galaxies available in our reductions of the public {\it JWST} database. Strong CIV ($>$10~\AA) and NIV] ($>$5~\AA) emission is  extremely rare, found in only 8 and 7\% of $z\gtrsim 4$ galaxies, where we only consider those spectra sufficiently deep to reach the respective EW limits. We find that the strong UV lines are primarily seen in galaxies with extremely large [OIII]+H$\beta$ EWs ($>$2500~\AA), corresponding to extremely young stellar populations ($<$5 Myr) formed in a recent strong burst of star formation. Such extremely young ages are rare in early star forming systems.  Targeted observations of galaxies with these large optical line EWs should yield larger samples of nitrogen-enhanced galaxies with hard ionizing sources.

\label{sec:summary}

\begin{table*}
\caption{Additional CIV and NIV] emitters at $z\gtrsim4$ identified from public NIRSpec datasets.
We list the JWST program where the NIRSpec observation is taken, their IDs, coordinates, spectroscopic redshifts, absolute UV magnitudes ($M_{\rm UV}$), [OIII]+H$\beta$ equivalent widths, the CIV and NIV] equivalent widths, the mode of the NIRSpec observations (i.e., prism or $R\sim1000$ medium grating), and literature references when available.
Here, the $M_{\rm UV}$ for UNCOVER sources in the Abell 2744 cluster have been corrected for gravitational lensing with the \protect\cite{Furtak2023} lensing map.
For [OIII]+H$\beta$ EWs, we adopt the values derived directly from the prism spectra when the rest-optical continuum is detected, and values inferred from NIRCam SED fitting in other cases. Emission-line EWs marked with a `*' are designated as tentative.\\
$^a$ Broad-line AGNs reported in \protect\cite{Harikane2023_BH} and \protect\cite{Maiolino2023_infantBH}.
}
\begin{adjustbox}{width=1.0\textwidth,center=1\textwidth}
\renewcommand{\arraystretch}{1.2}
\hskip-0.45cm\begin{tabular}{lrrcrcccccc}
\toprule
Program & Object ID & RA & DEC & $z_{\rm spec}$ & $\rm M_{\rm UV}$ & $\rm [OIII]+H\beta$ EW & CIV EW & NIV] EW & NRS Mode & Ref.\\
        &           &    &     & &    (AB Mag)   & (\angstrom{}) & (\angstrom{}) & (\angstrom{}) & & \\

\midrule
   \multicolumn{11}{c}{Galaxies selected on NIV] emission} \\
\midrule
CEERS(1345,4287)    &   1019     & 215.0353910 & +52.8906620 &  8.678 & $-22.4_{-0.1}^{+0.1}$ & $2599_{- 459}^{+ 642}$ & $ 4.6_{- 0.6}^{+ 0.5}$ & $ 10.1_{- 0.8}^{+ 0.7}$  & R1000,R2700 & [1],[2],[3],[4] \\
JADES(1181)    &   3991     & 189.1060540 & +62.2420490 & 10.621 & $-21.5_{-0.1}^{+0.1}$ &                     -- & $ 5.4_{- 2.9}^{+ 1.5}$ & $ 5.1_{- 1.7}^{+ 1.7}$  & PRISM,R1000 & [5],[6] \\
JADES(3215)    & 202208     &  53.1640684 & -27.7997202 &  5.449 & $-20.6_{-0.1}^{+0.1}$ & $3268_{- 102}^{+ 112}$ & $ 5.1_{- 2.6}^{+ 1.6}$ & $ 5.6_{- 2.0}^{+ 2.0}$  & PRISM,R1000 & -- \\
\midrule
   \multicolumn{11}{c}{Galaxies selected on CIV emission without a robust NIV] detection} \\
\midrule
CEERS(1345)    &    397$^a$ & 214.8361970 & +52.8826930 &  6.000 & $-21.4_{-0.1}^{+0.1}$ & $2396_{-  92}^{+  95}$ & $ 7.7_{- 2.4}^{+ 2.2}$ & --                      & PRISM,R1000 & [5],[7] \\
JADES(1181)    &    954$^a$ & 189.1519660 & +62.2596350 &  6.761 & $-19.9_{-0.1}^{+0.1}$ & $ 679_{-  36}^{+  35}$ & $23.3_{- 3.7}^{+ 3.8}$ &                     --  & PRISM,R1000 & [5],[8] \\
JADES(1181)    &   1899     & 189.1977400 & +62.2569640 &  8.279 & $-19.5_{-0.1}^{+0.1}$ & $5039_{- 951}^{+ 702}$ & $44.7_{-14.7}^{+22.1}$ & $*27.5_{-14.6}^{+21.6}$ & PRISM,R1000 & [5] [9],[10],[11] \\
JADES(1210)    &  13176     &  53.1217573 & -27.7976379 &  5.941 & $-19.6_{-0.1}^{+0.1}$ & $4127_{- 392}^{+ 448}$ & $39.5_{- 4.7}^{+ 3.5}$ &                     --  & PRISM,R1000 & [5],[12] \\
JADES(1210)    &  58975     &  53.1124340 & -27.7746090 &  9.436 & $-20.2_{-0.2}^{+0.1}$ & $ 957_{- 101}^{+ 121}$ & $9.2_{-7.6}^{+1.1}$ &                     --  & PRISM,R1000 & [5],[13],[14] \\
JADES(3215)    & 210003     &  53.1318414 & -27.7737748 &  5.780 & $-18.8_{-0.1}^{+0.1}$ & $3037_{- 212}^{+ 261}$ & $26.3_{- 3.3}^{+ 3.8}$ &                     --  & PRISM,R1000 & [5] \\
UNCOVER(2561)  &  10646     &   3.6369600 & -30.4063620 &  8.511 & $-21.6_{-0.1}^{+0.1}$ & $3733_{- 335}^{+ 341}$ & $11.5_{-1.7}^{+1.9}$ &                     --  & PRISM       & [5],[15] \\
UNCOVER(2561)  &  22223     &   3.5681150 & -30.3830520 &  9.570 & $-17.7_{-0.1}^{+0.1}$ &                     -- & $22.8_{-6.4}^{+7.2}$ &                     --  & PRISM       & [5],[15] \\
 \midrule
 \midrule
    \multicolumn{11}{c}{Tentative NIV]-Emitters} \\
 \midrule
 JADES(1181)    &   3990     & 189.0169950 & +62.2415820 &  9.374 & $-20.7_{-0.1}^{+0.1}$ &                     -- &                     -- & $*12.4_{-5.9}^{+10.0}$  & PRISM,R1000 & [5],[16] \\
 UNCOVER(2561)  &  34265     &   3.6071810 & -30.3648170 &  6.354 & $-18.7_{-0.1}^{+0.1}$ & $1115_{- 225}^{+ 275}$ &                     -- & $*10.0_{- 2.8}^{+4.2}$  & PRISM       & [5] \\
 \midrule
    \multicolumn{11}{c}{Tentative CIV-Emitters} \\
 \midrule
 JADES(1210)    &   5173     &  53.1568262 & -27.7671606 &  7.980 & $-18.9_{-0.1}^{+0.2}$ & $1021_{- 288}^{+ 261}$ & $*27.1_{-4.3}^{+4.0}$ &                     --  & PRISM,R1000 & [5] \\
 JADES(1210)    &   9903     &  53.1690468 & -27.7788335 &  6.631 & $-18.6_{-0.1}^{+0.1}$ & $2564_{- 380}^{+ 437}$ & $*19.5_{-2.4}^{+2.2}$ &                     --  & PRISM,R1000 & [5] \\
 UNCOVER(2561)  &  10155     &   3.5822060 & -30.4071140 &  5.660 & $-16.4_{-0.1}^{+0.1}$ & $2671_{-1307}^{+1291}$ & $*73.7_{-16.2}^{+25.5}$ &                     --  & PRISM       & [5] \\
 UNCOVER(2561)  &  23604     &   3.6052470 & -30.3805840 &  7.882 & $-18.3_{-0.1}^{+0.1}$ & $ 668_{- 190}^{+ 313}$ & $*36.8_{-6.0}^{+7.3}$ &                     --  & PRISM       & [5],[17] \\
\bottomrule
\multicolumn{11}{l}{[1] \citet{Larson2023}; [2] \citet{Isobe2023}; [3] \citealt{Tang2023}; [4] \citet{Marques-Chaves2024}; [5] \citet{Heintz2024}; [6] \citet{Bunker2023}; [7] \citet{Harikane2023_BH};  }\\
\multicolumn{11}{l}{[8] \citealt{Maiolino2023_infantBH}; [9] \citealt{Witstok2024}; [10] \citealt{Tang2024}; [11] \cite{NavarroCarrera2024}; [12] \cite{Cameron2023c}; [13] \citealt{Scholtz2023}; [14] \citealt{Curti2024};  }\\
\multicolumn{11}{l}{[15] \citealt{Fujimoto2023_uncover};  [16] \citealt{Schaerer2024}; [17] \citealt{Chen2024}}\\

\end{tabular}
\end{adjustbox}
\label{tab:literatureemitters}
\end{table*}

\section*{Acknowledgements}
MWT acknowledges support from the NASA ADAP program through the grant number 80NSSC23K0467.
DPS acknowledges support from the National Science Foundation through the grant AST-2109066. 
AZ acknowledges support by Grant No. 2020750 from the United States-Israel Binational Science Foundation (BSF) and Grant No. 2109066 from the United States National Science Foundation (NSF); by the Ministry of Science \& Technology, Israel; and by the Israel Science Foundation Grant No. 864/23.

This work is based in part on observations made with the NASA/ESA/CSA James Webb Space Telescope. The data were obtained from the Mikulski Archive for Space Telescopes at the Space Telescope Science Institute, which is operated by the Association of Universities for Research in Astronomy, Inc., under NASA contract NAS 5-03127 for JWST. The authors acknowledge the CEERS, DDT-2750, and UNCOVER teams led by Steven L. Finkelstein, Pablo Arrabal Haro, and I. Labb\'e \& R. Bezanson for developing their observing program with a zero-exclusive-access period.
Some of the data products presented herein were retrieved from the Dawn JWST Archive (DJA). DJA is an initiative of the Cosmic Dawn Center (DAWN), which is funded by the Danish National Research Foundation under grant DNRF140.

 \section*{Data Availability}
The data underlying this article may be presented upon reasonable request to the corresponding author.

\bibliographystyle{mnras}
\bibliography{main}

\appendix
\section{Emission line fluxes and equivalent widths}
\label{sec:appendix}
Tables~\ref{tab:UVlines} and \ref{tab:opticallines} provide the full set of emission lines in the rest-frame UV and optical, respectively, that we measured for the spectroscopic sample.

\begin{table*}
\caption{Emission-line fluxes and equivalent widths of rest-frame UV features for our JWST/NIRSpec sample of galaxies in Abell 1703. Upper limits are provided at the $3\sigma$ level. Fluxes and equivalent widths of emission lines lacking wavelength coverage are indicated by a `-'. Equivalent widths are calculated using the continuum level measured from the spectrum if the continuum is detected at $>3\sigma$.}
\begin{center}
\renewcommand{\arraystretch}{1.2}
\begin{tabular}{llllllll}
\toprule
 & A1703-zd1 & A1703-zd2 & A1703-zd5.1 & A1703-zd5.2 & A1703-23 & A1703-zd4 & A1703-zd6 \\
\midrule
   \multicolumn{8}{c}{$\textrm{Line Flux}~[10^{-19}~\rm erg/s/cm^2]$} \\
\midrule
$\rm Ly\alpha$ & $\hfill-\hfill$ & $\hfill-\hfill$ & $\hfill-\hfill$ & $\hfill-\hfill$ & $\hfill-\hfill$ & $\quad7.9\pm1.6$ & $\quad249.5\pm9.3$ \\ 
$\rm NIV]\rm \lambda1483$ & $<4.2\hfill$ & $<4.1\hfill$ & $<11.2\hfill$ & $<8.6\hfill$ & $<6.8\hfill$ & $<4.2\hfill$ & $\quad5.8\pm1.9$ \\ 
$\rm NIV]\lambda1486$ & $<4.4\hfill$ & $<4.8\hfill$ & $<9.8\hfill$ & $<10.9\hfill$ & $<7.3\hfill$ & $<3.9\hfill$ & $\quad12.3\pm1.6$ \\ 
$\rm CIV\lambda1549$ & $-$ & $\quad20.1\pm2.4$ & $<14.6\hfill$ & $\quad51.9\pm4.7$ & $<10.6\hfill$ & $<6.1\hfill$ & $\quad57.0\pm3.1$ \\ 
$\rm HeII\lambda1640$ & $<3.9\hfill$ & $\quad6.5\pm1.5$ & $<9.5\hfill$ & $<9.5\hfill$ & $<4.9\hfill$ & $<3.7\hfill$ & $\quad6.8\pm1.1$ \\ 
$\rm OIII]\lambda1660$ & $<4.0\hfill$ & $<4.4\hfill$ & $<10.1\hfill$ & $<9.4\hfill$ & $\quad7.0\pm1.7$ & $<4.6\hfill$ & $\quad10.7\pm1.7$ \\ 
$\rm OIII]\lambda1666$ & $<3.5\hfill$ & $<6.1\hfill$ & $<9.9\hfill$ & $\quad16.0\pm2.9$ & $\quad7.2\pm1.8$ & $<4.3\hfill$ & $\quad19.8\pm1.8$ \\ 
$\rm NIII1750$ & $<5.1\hfill$ & $<6.3\hfill$ & $<13.8\hfill$ & $<13.7\hfill$ & $<6.0\hfill$ & $<5.3\hfill$ & $\quad7.0\pm2.2$ \\ 
$\rm SiIII]\lambda1883$ & $<8.9\hfill$ & $<4.1\hfill$ & $<9.2\hfill$ & $<7.9\hfill$ & $<4.8\hfill$ & $<4.2\hfill$ & $<5.0\hfill$ \\ 
$\rm SiIII]\lambda1892$ & $<7.4\hfill$ & $<3.4\hfill$ & $<7.7\hfill$ & $<8.5\hfill$ & $<3.8\hfill$ & $<3.8\hfill$ & $<3.6\hfill$ \\ 
$\rm [CIII]\lambda1907$ & $\hfill-\hfill$ & $\quad4.4\pm1.3$ & $<8.5\hfill$ & $\quad10.7\pm2.2$ & $\quad9.5\pm1.5$ & $\quad9.6\pm1.9$ & $\quad5.8\pm1.5$ \\ 
$\rm CIII]\lambda1909$ & $\hfill-\hfill$ & $\quad3.8\pm1.7$ & $<8.5\hfill$ & $<5.4\hfill$ & $\hfill-\hfill$ & $<8.4\hfill$ & $\quad11.7\pm1.8$ \\
\toprule
   \multicolumn{8}{c}{$\textrm{Equivalent Width}~[\angstrom{}]$} \\
 \midrule
$\rm Ly\alpha$ & $\hfill-\hfill$ & $\hfill-\hfill$ & $\hfill-\hfill$ & $\hfill-\hfill$ & $\hfill-\hfill$ & $\quad3.4\pm0.9$ & $\quad61.2\pm9.0$ \\ 
$\rm NIV]\rm \lambda1483$ & $<1.4\hfill$ & $<4.1\hfill$ & $<3.1\hfill$ & $<1.7\hfill$ & $<1.2\hfill$ & $<3.0\hfill$ & $\quad1.6\pm0.5$ \\ 
$\rm NIV]\lambda1486$ & $<1.5\hfill$ & $<4.9\hfill$ & $<2.7\hfill$ & $<2.1\hfill$ & $<1.3\hfill$ & $<2.8\hfill$ & $\quad3.4\pm0.5$ \\ 
$\rm CIV\lambda1549$ & $-$ & $\quad31.6\pm7.0$ & $<4.3\hfill$ & $\quad10.5\pm1.0$ & $<1.9\hfill$ & $<4.9\hfill$ & $\quad19.4\pm1.4$ \\ 
$\rm HeII\lambda1640$ & $<1.5\hfill$ & $\quad14.0\pm6.1$ & $<3.1\hfill$ & $<2.2\hfill$ & $<1.0\hfill$ & $<4.8\hfill$ & $\quad2.2\pm0.4$ \\ 
$\rm OIII]\lambda1660$ & $<1.4\hfill$ & $<7.1\hfill$ & $<3.6\hfill$ & $<2.3\hfill$ & $\quad1.4\pm0.4$ & $<5.5\hfill$ & $\quad4.0\pm0.7$ \\ 
$\rm OIII]\lambda1666$ & $<1.3\hfill$ & $<9.1\hfill$ & $<3.5\hfill$ & $\quad3.9\pm0.7$ & $\quad1.5\pm0.5$ & $<5.0\hfill$ & $\quad7.6\pm0.9$ \\ 
$\rm NIII1750$ & $<2.2\hfill$ & $<11.3\hfill$ & $<5.2\hfill$ & $<3.5\hfill$ & $<1.3\hfill$ & $<10.0\hfill$ & $\quad3.1\pm1.0$ \\ 
$\rm SiIII]\lambda1883$ & $<0.9\hfill$ & $<6.4\hfill$ & $<4.1\hfill$ & $<2.5\hfill$ & $<1.4\hfill$ & $<5.0\hfill$ & $<2.6\hfill$ \\ 
$\rm SiIII]\lambda1892$ & $<0.7\hfill$ & $<7.3\hfill$ & $<3.5\hfill$ & $<2.7\hfill$ & $<1.1\hfill$ & $<4.5\hfill$ & $<1.9\hfill$ \\ 
$\rm [CIII]\lambda1907$ & $\hfill-\hfill$ & $\quad6.5\pm2.7$ & $<3.6\hfill$ & $\quad3.1\pm0.6$ & $\quad2.8\pm0.5$ & $\quad3.6\pm0.7$ & $\quad2.9\pm0.7$ \\ 
$\rm CIII]\lambda1909$ & $\hfill-\hfill$ & $\quad5.5\pm2.2\hfill$ & $<5.5\hfill$ & $<2.4\hfill$ & $\hfill-\hfill$ & $<4.9\hfill$ & $\quad5.9\pm1.1$ \\

\bottomrule
\end{tabular}
\end{center}
\label{tab:UVlines}
\end{table*}

\begin{table*}
\caption{Emission-line fluxes and equivalent widths of rest-frame optical features for our JWST/NIRSpec sample of galaxies in Abell1703. Upper limits are provided at the $3\sigma$ level. Emission lines indicated by a `-' do not have wavelength coverage in the spectra.}
\begin{center}
\renewcommand{\arraystretch}{1.2}
\begin{tabular}{llllllll}
\toprule
Line & A1703-zd1 & A1703-zd2 & A1703-zd5.1 & A1703-zd5.2 & A1703-23 & A1703-zd4 & A1703-zd6 \\
   \multicolumn{8}{c}{$\textrm{Line Flux}~[10^{-19}~\rm erg/s/cm^2]$} \\
\midrule
$\rm [OII]\lambda3728$ & $\quad23.1\pm2.2$ & $\hfill-\hfill$ & $\hfill-\hfill$ & $\hfill-\hfill$ & $\hfill-\hfill$ & $\quad15.1\pm1.3$ & $<5.8\hfill$ \\ 
$\rm [NeIII]\lambda3869$ & $\quad14.4\pm2.0$ & $<13.0\hfill$ & $<17.4\hfill$ & $<19.3\hfill$ & $\hfill-\hfill$ & $<3.2\hfill$ & $\quad19.5\pm1.6$ \\ 
$\rm HeI\lambda3890+H8$ & $<5.1\hfill$ & $<9.9\hfill$ & $<14.8\hfill$ & $<11.8\hfill$ & $\hfill-\hfill$ & $<3.6\hfill$ & $\quad6.5\pm1.6$ \\ 
$\rm H\epsilon$ & $<5.6\hfill$ & $<6.7\hfill$ & $<9.9\hfill$ & $<9.7\hfill$ & $\hfill-\hfill$ & $<3.6\hfill$ & $\quad13.0\pm1.7$ \\ 
$\rm H\delta$ & $\quad10.8\pm1.2$ & $<4.7\hfill$ & $<6.9\hfill$ & $\quad8.4\pm2.3$ & $<17.2\hfill$ & $<3.0\hfill$ & $\quad9.2\pm1.2$ \\ 
$\rm H\gamma$ & $\quad11.0\pm1.2$ & $\quad7.8\pm1.4$ & $<6.1\hfill$ & $\quad14.2\pm2.0$ & $\quad4.8\pm1.4$ & $\quad4.3\pm0.8$ & $\quad20.0\pm1.4$ \\ 
$\rm [OIII]\lambda4363$ & $<3.3\hfill$ & $<4.2\hfill$ & $<4.8\hfill$ & $<5.1\hfill$ & $<4.2\hfill$ & $<3.4\hfill$ & $\quad11.3\pm1.0$ \\ 
$\rm HeII\lambda4686$ & $<3.7\hfill$ & $<4.4\hfill$ & $<6.1\hfill$ & $<5.8\hfill$ & $<4.8\hfill$ & $<3.2\hfill$ & $<3.5\hfill$ \\ 
$\rm H\beta$ & $\quad23.0\pm1.3$ & $\quad12.4\pm1.1$ & $\quad5.7\pm1.6$ & $\quad25.1\pm1.6$ & $\quad19.4\pm1.3$ & $\quad11.1\pm1.0$ & $\quad32.7\pm1.3$ \\ 
$\rm [OIII]\lambda4959$ & $\quad60.4\pm1.6$ & $\quad16.2\pm1.2$ & $<5.4\hfill$ & $\quad32.4\pm1.8$ & $\quad46.8\pm1.4$ & $\quad28.4\pm1.3$ & $\quad77.1\pm1.8$ \\ 
$\rm [OIII]\lambda5007$ & $\quad185.5\pm2.4$ & $\quad64.9\pm1.7$ & $\quad17.8\pm2.3$ & $\quad110.8\pm2.1$ & $\quad147.9\pm2.2$ & $\quad84.1\pm1.6$ & $\quad207.8\pm2.4$ \\ 
$\rm HeI\lambda5876$ & $<4.6\hfill$ & $<3.4\hfill$ & $<4.3\hfill$ & $<4.4\hfill$ & $<3.8\hfill$ & $\hfill-\hfill$ & $\quad8.4\pm1.4$ \\ 
$\rm H\alpha$ & $\quad60.7\pm3.1$ & $\hfill-\hfill$ & $\quad19.3\pm2.8$ & $\quad70.1\pm2.8$ & $\quad69.8\pm2.3$ & $\hfill-\hfill$ & $\hfill-\hfill$ \\ 
$\rm [NII]\lambda6585$ & $<6.4\hfill$ & $\hfill-\hfill$ & $<5.5\hfill$ & $<5.7\hfill$ & $<5.4\hfill$ & $\hfill-\hfill$ & $\hfill-\hfill$ \\ 
$\rm [SII]\lambda6717$ & $\hfill-\hfill$ & $\hfill-\hfill$ & $\hfill-\hfill$ & $\hfill-\hfill$ & $\quad7.5\pm2.2$ & $\hfill-\hfill$ & $\hfill-\hfill$ \\ 
$\rm [SII]\lambda6730$ & $\hfill-\hfill$ & $\hfill-\hfill$ & $\hfill-\hfill$ & $\hfill-\hfill$ & $\quad7.1\pm2.1$ & $\hfill-\hfill$ & $\hfill-\hfill$ \\ 
\toprule
\multicolumn{8}{c}{$\textrm{Equivalent Width}~[\angstrom{}]$} \\
\midrule
$\rm [OII]\lambda3728$ & $\quad9.9\pm1.4$ & $\hfill-\hfill$ & $\hfill-\hfill$ & $\hfill-\hfill$ & $\hfill-\hfill$ & $\quad214.4\pm56.8$ & $<15.5\hfill$ \\ 
$\rm [NeIII]\lambda3869$ & $\quad34.5\pm7.3$ & $<41.1$ & $<79.8$ & $<8.5\hfill$ & $\hfill-\hfill$ & $<27.7\hfill$ & $\quad58.3\pm9.7$ \\ 
$\rm HeI\lambda3890+H8$ & $<12.0\hfill$ & $<34.5$ & $<63.9$ & $<7.2\hfill$ & $\hfill-\hfill$ & $<34.4\hfill$ & $\quad20.0\pm6.3$ \\ 
$\rm H\epsilon$ & $<17.6\hfill$ & $<35.7$ & $<30.5\hfill$ & $<17.0\hfill$ & $\hfill-\hfill$ & $\quad67.1\pm23.8$ & $\quad67.1\pm17.6$ \\ 
$\rm H\delta$ & $\quad50.2\pm9.7$ & $<37.9\hfill$ & $\quad34.0\pm14.4$ & $\quad17.5\pm5.4$ & $<101.2$ & $<30.8$ & $\quad80.3\pm25.2$ \\ 
$\rm H\gamma$ & $\quad69.5\pm18.3$ & $\quad131.9\pm55.0$ & $<18.7$ & $\quad91.2\pm35.3$ & $\quad9.8\pm3.2$ & $\quad61.7\pm19.4$ & $\quad133.8\pm33.0$ \\ 
$\rm [OIII]\lambda4363$ & $<22.2\hfill$ & $<39.3$ & $<33.9$ & $<5.7$ & $<8.7\hfill$ & $<32.9$ & $\quad73.1\pm18.5$ \\ 
$\rm HeII\lambda4686$ & $<23.8\hfill$ & $<48.7$ & $<48.0$ & $<25.6$ & $<9.9\hfill$ & $<38.3\hfill$ & $<28.9\hfill$ \\ 
$\rm H\beta$ & $\quad116.9\pm14.5$ & $\quad208.8\pm80.6$ & $\quad86.3\pm53.6$ & $\quad169.7\pm43.7$ & $\quad46.0\pm4.2$ & $\quad147.9\pm46.4$ & $\quad423.8\pm112.4$ \\ 
$\rm [OIII]\lambda4959$ & $\quad292.7\pm26.8$ & $\quad274.4\pm104.8$ & $<42.8$ & $\quad218.6\pm56.0$ & $\quad123.5\pm7.3$ & $\quad431.4\pm133.2$ & $\quad1006.1\pm219.7$ \\ 
$\rm [OIII]\lambda5007$ & $\quad880.2\pm82.7$ & $\quad1096.4\pm412.2$ & $\quad267.3\pm152.3$ & $\quad748.9\pm187.8$ & $\quad414.8\pm22.1$ & $\quad1316.7\pm432.4$ & $\quad2686.6\pm632.8$ \\ 
$\rm HeI\lambda5876$ & $<30.0\hfill$ & $<43.3$ & $<58.0$ & $<10.7$ & $<15.7\hfill$ & $\hfill-\hfill$ & $\quad43.7\pm8.2$ \\ 
$\rm H\alpha$ & $\quad210.1\pm44.4$ & $\hfill-\hfill$ & $\quad157.7\pm57.6$ & $\quad651.5\pm227.3$ & $\quad192.9\pm40.8$ & $\hfill-\hfill$ & $\hfill-\hfill$ \\ 
$\rm [NII]\lambda6585$ & $<22.5\hfill$ & $\hfill-\hfill$ & $<42.1\hfill$ & $<48.1\hfill$ & $<28.6\hfill$ & $\hfill-\hfill$ & $\hfill-\hfill$ \\ 
$\rm [SII]\lambda6717$ & $\hfill-\hfill$ & $\hfill-\hfill$ & $\hfill-\hfill$ & $\hfill-\hfill$ & $\quad15.0\pm4.5$ & $\hfill-\hfill$ & $\hfill-\hfill$ \\ 
$\rm [SII]\lambda6730$ & $\hfill-\hfill$ & $\hfill-\hfill$ & $\hfill-\hfill$ & $\hfill-\hfill$ & $\quad14.2\pm4.3$ & $\hfill-\hfill$ & $\hfill-\hfill$ \\ 
\bottomrule
\end{tabular}
\end{center}

\label{tab:opticallines}
\end{table*}

\section{Tentative CIV and NIV] emitting sources}
\label{sec:tentative}
In searching for CIV and NIV] emitters in the public \jwst{} spectra, we identified several objects that appeared to display emission in these UV lines, yet we were unable to classify the lines as `robust'.
These tentative sources are presented at the bottom of Table~\ref{tab:literatureemitters}.
We identify tentative NIV] lines in the spectra of JADES-3990 and UNCOVER-34265. \citet{Schaerer2024} recently presented a detection of NIV] from the PRISM spectrum of JADES-3990. While we also note an apparent emission feature at the rest-frame wavelength of NIV], we do not measure it at the $5\sigma$ significance we require to consider it in our robust sample.
The spectrum of UNCOVER-34265 presents an excess of emission coincident with NIV]. However in addition to being measured at $<5\sigma$, the line comprises only one pixel in the spectrum, increasing the likelihood of being either an artifact or the result of scatter from noise.
In addition to these tentative NIV] emitters, we identify several objects with tentative CIV emission lines.
All of JADES-5173, UNCOVER-10155, and UNCOVER-23604 show emission features in their PRISM spectra coincident with the wavelength of CIV. However in each of these cases the S/N of the line does not reach the necessary $5\sigma$ limit. In addition, the tentative CIV lines in JADES-5173 spectra do not appear in its corresponding $R\sim1000$ NIRSpec spectra.
The apparent CIV emission line in JADES-9903 is detected at a higher S/N than the other tentative sources, however the line centroid is offset from systemic by $\sim5000$ km/s. While this is comparable to the instrumental resolution of the NIRSpec PRISM at this wavelength, this offset is significant enough to preclude JADES-9903 from being included in the robust sample.

\end{document}